\documentclass[12pt,vi]{mitthesis}
\usepackage{lgrind}
\usepackage{graphicx}
\usepackage{hyperref}
\newcommand{\eq}[1]{eq.~\ref{#1}}
\newcommand{\be}{\begin{equation}}
\newcommand{\ee}{\end{equation}}
\newcommand{\bea}{\begin{eqnarray}}
\newcommand{\eea}{\end{eqnarray}}
\newcommand{\half}{\frac{1}{2}}
\newcommand{\Tr}{\mbox{Tr}}
\newcommand{\tr}{\mbox{tr}}

\newcommand{\classicalVacuum}{vacuum}
\newcommand{\ClassicalVacuum}{Vacuum}
\newcommand{\HiggsGauge}{Higgs-gauge }
\newcommand{\ID}{\mbox{{\sf 1}\zr{-0.16}\rule{0.04em}{1.55ex}\zr{0.1}}}

\newcommand{\oneF}{_{\rm eff}^{(1)}}
\newcommand{\zeroF}{_{\rm eff}^{(0)}}
\newcommand{\zr}[1]{\mbox{\hspace*{#1em}}}
\newcommand{\fourier}[1]{\tilde{#1}}

\newcommand{\smallhalf}{{\scriptstyle \frac{1}{2}}}
\newcommand{\Aslash}{A \hskip -0.7em /}
\newcommand{\dslash}{\partial \hskip -0.7em /}
\newcommand{\Qt}{{QED$_{2+1}$\ }}
\newcommand{\Qf}{{QED$_{3+1}$\ }}

\newcommand\res[1]{\mathcal{#1}}

\newcommand{\flux}{F }
\newcommand{\gaussSub}{{\rm G}}
\newcommand{\EclSub}{{\rm cl}}
\newcommand{\EvacSub}{{\rm vac}}
\newcommand{\EctSub}{{\rm CT}}
\newcommand{\EdeltaSub}{{\delta}}
\newcommand{\EfdSub}{{\rm FD}}
\newcommand{\Det}{\mbox{Det}} 

\newcommand{\azAngle}{\varphi}

\begin{document}

%
%
%
%
%
%
%
\title{Searching for Novel Objects in the Electroweak Theory}

\author{Vishesh Khemani}
\department{Department of Physics}
\degree{Doctor of Philosophy}
\degreemonth{June}
\degreeyear{2004}
\thesisdate{April 30, 2004}


\supervisor{Edward H. Farhi}{Professor}

\chairman{Thomas J. Greytak}{Associate Department Head for Education}

\maketitle



\cleardoublepage
\setcounter{savepage}{\thepage}
\begin{abstractpage}
We explore the Higgs-Gauge configuration space in the standard electroweak theory.  We outline a general prescription that uses the non-trivial topology associated with the gauge group of the theory, to find known solutions of the Euclidean classical equations of motion and motivate the existence of novel ones.  In Minkowski spacetime we present evidence for the existence of approximate breathers -- long-lived, spatially localized, temporally periodic configurations.  We consider heavy fermion quantum fluctuations about static Higgs-Gauge configurations, and argue for the existence of stable fermionic solitons.  These could resolve the fermion decoupling puzzle in chiral gauge theories.  We describe our search for a fermionic soliton within a spherical ansatz, and discuss the quantum corrected sphaleron and the emergence of new barriers suppressing the decay of heavy fermions.  Finally, we consider electroweak strings and how they could give rise to stable multi-quark objects.  

\end{abstractpage}


\cleardoublepage

\section*{Acknowledgments}

I would like to thank my advisor, Edward Farhi, for his infectious enthusiasm that has made work seem like play, and for sharing his eclectic tastes that have shaped my interests in Physics.  I am indebted to Robert Jaffe for being an inspirational teacher, and for being so generous in his encouragement and support.  I am grateful to Noah Graham and Herbert Weigel for being my oracles through my first project, and for being such engaging collaborators ever since.  I have enjoyed my many stimulating conversations with Oliver Schroeder and Markus Quandt.  And finally, I thank Victoria for...well, everything.      


\pagestyle{plain}
\tableofcontents

\chapter{Introduction}

The Standard Model of particle physics has several well-known configurations of Higgs and gauge fields that are solutions of the classical equations of motion.  These have rich phenomenology associated with them.  For example, the electroweak instanton \cite{Belavin:fg,'tHooft} in Euclidean spacetime mediates non-perturbative fermion number violation through quantum tunneling.  There are compelling reasons to expect the existence of novel configurations that drive physics ranging from decoupling of heavy fermions to electroweak baryogenesis.  I explore these possibilities in this work.

I begin by examining the space of solutions of the classical equations of motion in the electroweak theory, in Chap.~\ref{chap:ClassicalSolutions}.  I describe my work with E.~Farhi and N.~Graham, in which we use topologically non-trivial maps into the gauge group to construct and motivate solutions in Euclidean spacetime.  This method is a generalization of ideas introduced by Manton \cite{Manton:1983nd} and Klinkhamer \cite{Klinkhamer:2003hz}.  It encapsulates the various known solutions (electroweak instanton, sphaleron, strings, etc.) into a unified framework.  Moreover, the use of all the topological properties of the theory leads to the possibility of new, unstable solutions.

I then consider Minkowski spacetime, and argue for the existence of approximate breathers -- long-lived, spatially localized, temporally periodic configurations -- in the electroweak theory.  E.~Farhi, N.~Graham and I have used the mechanism that creates intrinsic  localized modes in anharmonic crystals \cite{SieversTakeno,CampbellFlachKivshar} to demonstrate the presence of approximate breathers in the 1+1 dimensional Abelian Higgs model.  We are currently looking for such objects within a spherical ansatz in the electroweak theory.  These breathers have lifetimes that are orders of magnitude larger than all scales in the problem and challenge the notion of naturalness in field theories.  On a more phenomenological front, electroweak breathers could create out-of-equilibrium regions in space during the electroweak phase transition and drive baryogenesis. 

All known (and expected) spatially varying, static solutions in the electroweak theory are unstable and are generically called {\it sphalerons} (to distinguish them from stable solutions or {\it solitons}).  They do not have any associated quantum extended-particle states.  The discovery of a stable configuration would result in a soliton sector in the Hilbert space of states, in addition to the familiar vacuum sector.  There is no topological reason for stability of a static configuration in the electroweak theory, but a non-topological soliton (corresponding to a local minimum of the energy) may still exist.  However, in the absence of a topological beacon, it is difficult to search for such an object.  If we consider quantum fluctuations around classical configurations, then there are compelling reasons to expect the existence of {\it quantum solitons}, and well-understood mechanisms to guide the search for them.  Such objects could be stabilized by virtue of carrying a conserved quantum number, in analogy with topological solitons that carry a topological charge.  

In Chap.~\ref{chap:QuantumSolitons} I explain how the quarks and leptons in the Standard Model could be strongly bound by certain configurations of Higgs and gauge fields, giving rise to the possibility of fermionic solitons.  These would allow heavy fermions to decouple from the theory because the lower energy fermionic solitons would carry their quantum numbers and maintain anomaly cancellation.  Since these solitons are quantum-stabilized, we have to include quantum corrections to their energy when analyzing their stability.  I review an efficient method based on scattering theory that allows an exact computation of the one-loop quantum corrections to the energy non-perturbatively, with physical on-shell renormalization carried out in the perturbative sector \cite{PhaseshiftsGeneral,PhaseshiftsFermions}.  This makes it feasible to carry out a variational search for fermionic solitons.  I describe my search (with E.~Farhi, N.~Graham, R.~L.~Jaffe and H.~Weigel \cite{KhemaniDecoupling}) within a spherical ansatz and discuss  the quantum-corrected sphaleron and the emergence of new barriers that suppress heavy fermion decay. We find no evidence for a spherical fermionic soliton.  However, this does not preclude the existence of such objects outside the ansatz.

In Chap.~\ref{chap:ElectroweakStrings}, I briefly review the family of electroweak strings \cite{VachaspatiReview}.  These are static, unstable solutions of the classical equations of motion that localize energy within a cylindrical region of space.  The Higgs condensate is suppressed in the core of a string, and so heavy quarks are bound along its length and resist its disintegration.  If this mechanism could stabilize electroweak strings, then they would constitute a crucial ingredient in a viable scenario for electroweak baryogenesis without requiring a first-order phase transition \cite{Brandenberger}.  Furthermore, a gas of electroweak strings would have negative pressure and could contribute to the dark energy required for the observed cosmic acceleration \cite{Perlmutter:1998np,Riess:1998cb}.  An analysis of the stability of these objects requires a computation of the fermion quantum correction to the energy.  This is most efficiently done using scattering theory methods and there are technical challenges associated with the long-range nature of the potential generated by such configurations.  So, as a stepping stone to the calculation I investigate the similar, but simpler, problem of one-loop quantum corrections to the energies of magnetic flux tubes in QED (in collaboration with N.~Graham, M.~Quandt, O.~Schroeder and H.~Weigel).  I comment on the puzzles and challenges associated with the presence of fermions in such backgrounds.  We find that when we include a region of return flux (which is the truly physical case) the scattering potential becomes short-range and the calculation becomes tractable.  As the return flux is made more diffuse, the quantum-corrected energy of the configuration has a well-defined limit and gives the energy associated with an isolated flux tube.  We also discover that when the energy is properly renormalized, the results in two and three spatial dimensions become similar.  These findings suggest that a natural next step of research would be to examine electroweak vortices in two spatial dimensions by including a region of spread-out return flux.  A fermionic soliton in that theory would be indicative of a stable electroweak string in the Standard Model.

\chapter{Classical Solutions}
\label{chap:ClassicalSolutions}
We briefly review the bosonic sector of the electroweak theory.  We use topologically non-trivial maps into the gauge group to explore the space of solutions to the classical equations of motion, in Euclidean spacetime.  We show how the method can be used to construct well-known solutions and how it suggests the existence of novel, unstable solutions.  We also discuss the possibility of the existence of approximate breathers (long-lived, spatially localized, temporally periodic configurations) in Minkowski spacetime.

\section{The \HiggsGauge Sector}
\label{sec:HiggsGauge}
The bosonic sector of the electroweak theory is an $SU(2) \times U(1)$ gauged Higgs model.  The Abelian coupling constant ($g'$) is known to be much smaller than the non-Abelian coupling constant ($g$).  For simplicity and clarity we choose $g'=0$ and ignore the dynamics of the $U(1)$ hypercharge gauge fields.  The three $SU(2)$ gauge bosons are denoted by $W_\mu^a$.  These may be expressed as a matrix valued field, using the group generators $\tau^a$,
\be
W_\mu = W_{\mu}^a \frac{\tau^a}{2} \, .
\ee
There are two complex scalar fields that form the Higgs doublet
\be
\phi =  \Biggl( \begin{array}{c} \phi_+ \\[-0.1in] \phi_0 \end{array} \Biggr) \, ,
\ee
where the subscripts denote the electric charge (had the U(1) fields been included).
The Higgs may be written as a $2 \times 2$ matrix field
\begin{equation}
\Phi =  \left( \begin{array}{cc} \phi_0^{*} & \phi_+ \\ -\phi_+^{*} & 
\phi_0 \end{array} \right) \, .
\end{equation}
Under a gauge transformation, $U(x) \in SU(2)$, the fields transform as
\be
W_\mu \rightarrow U \left( W_\mu + \frac{i}{g} \partial_\mu \right) U^\dag \, , \; \Phi \rightarrow U \Phi \, .
\ee  
The \HiggsGauge sector is defined by the action
\bea
\label{eq:HiggsAction}
S_H[\Phi, W_\mu] =  \int d^4 x \left[ -\frac{1}{2} \tr \left(W^{\mu\nu}W_{\mu\nu}\right) 
+ \frac{1}{2}\tr\left(\left(D^{\mu}\Phi \right)^{\dag} D_{\mu}\Phi\right) \right. \nonumber \\
\left. - \frac{\lambda}{4}\left( \tr\left(\Phi^{\dag}\Phi\right) - 2v^2 \right)^2 \right] \, ,
\eea
where the field-strength tensor and the covariant derivative are defined as follows:
\begin{eqnarray}
\label{covariant}
W_{\mu\nu} & = & \partial_\mu W_\nu - \partial_\nu W_\mu - i g\left[ W_\mu 
, W_\nu \right] \, , \nonumber \\
D_\mu \Phi & = & \left( \partial_\mu - i g W_\mu \right)\Phi \, ,
\end{eqnarray}  
and $\lambda$ is the Higgs self-interaction coupling constant.  

The gauge symmetry is spontaneously broken because the Higgs doublet has a non-zero vacuum expectation value.  This results in the following particle spectrum: a single scalar Higgs particle with mass $m_h^{(0)}=2v\sqrt{\lambda}$ and three degenerate gauge bosons with mass $m_w^{(0)} = gv/\sqrt{2}$.  The superscript `(0)' denotes that the masses are at tree level, and later in Chap.~\ref{chap:QuantumSolitons} we will discuss one-loop quantum corrections.
\section{Solutions in Euclidean Spacetime}
In Euclidean spacetime, the \HiggsGauge action is positive definite and is given by
\begin{equation}
\label{eq:HiggsActionEuclidean}
S_H^{(E)}[\Phi, W_\mu] =  \int d^4 x \left[ \frac{1}{2} \tr \left(W_{\mu\nu}W_{\mu\nu}\right) 
+ \frac{1}{2}\tr\left(\left(D_{\mu}\Phi \right)^{\dag} D_{\mu}\Phi\right) +
\frac{\lambda}{4}\left( 
\tr\left(\Phi^{\dag}\Phi\right) - 2v^2 \right)^2 \right] \, .
\end{equation}
Since the metric is $\delta_{\mu\nu}$, there is no difference between upper and lower indices.  Also, for notational simplicity, we use the same Greek letters to denote spacetime indices in both Euclidean and Minkowski space.

The classical equations of motion, obtained by extremizing the action with respect to the Higgs and gauge fields, are
\bea
D_\mu D_\mu \Phi & = & \lambda \left[ \tr\left(\Phi^\dag \Phi \right) - 2 v^2 \right] \Phi \, , \nonumber \\
D_\mu F_{\mu \nu} & = & \frac{i g}{4} \left[ D_\nu \Phi \Phi^\dag - \Phi \left( D_\nu \Phi \right)^\dag \right] \, .
\label{eq:EOM}
\eea  
The covariant derivatives for the Higgs and gauge fields are
\bea
D_\mu \Phi & = & \left( \partial_\mu - i g W_\mu \right) \Phi \, , \nonumber \\
D^\mu F_{\mu \nu} & = & \partial^\mu F_{\mu \nu} - i g \left[ W^\mu , F_{\mu \nu} \right] \, .
\eea
\subsection{\ClassicalVacuum{} Solutions}
The totally-trivial configuration, $\Phi=v\ID , W_\mu=0$, is obviously a solution at which the action has a global minimum of zero.  All configurations gauge-equivalent to the totally-trivial configuration are also global minimum solutions.  We refer to these as \classicalVacuum{} configurations.  Any $U(x) \in SU(2)$ completely specifies such solutions as pure-gauge configurations
\be
\Phi = v U \, , \;\;\; W_\mu = \frac{i}{g} U \partial_\mu U^\dag \, .
\ee 
\subsection{Non-\classicalVacuum{} Solutions and Topology}
\label{sec:ClassicalSolutions}
In addition to the \classicalVacuum{} solutions with zero action, there are several solutions corresponding to local minima and saddle-points of the action.  There are various strategies that enable us to find such solutions.  One way is to embed known solutions of simpler theories into the electroweak theory.  For example, the kink solution in 1+1 dimensional, real $\phi^4$ theory becomes a {\em domain wall} solution in the electroweak theory.  Also, Nielsen-Olesen vortices \cite{Nielsen:1973cs} in the Abelian Higgs model become a family of {\em string} solutions in the electroweak theory \cite{Vachaspati:1992fi}.  A second strategy is to use topology, and this has lead to the discovery of the {\em instanton} \cite{Belavin:fg} and the {\em sphaleron} \cite{Manton:1983nd,KlinkhamerManton}.

We now present a general prescription, which uses the topology associated with maps into the gauge group of the theory, to find known solutions in the theory and motivate the existence of novel ones.  The basic idea, due to Manton, is to construct non-contractible loops in configuration space, and the ``top'' of the tightest loop should correspond to a solution.  This is how the well-known electroweak sphaleron was constructed.  Klinkhamer has motivated the existence of other solutions by constructing different non-contractible loops \cite{Klinkhamer:2003hz}.  We generalize this procedure and demonstrate how to construct {\em all} possible non-contractible loops in a gauge theory, and list the solutions indicated by these.
\subsection{Topological Prescription}
A finite Euclidean-action configuration is pure gauge at spacetime infinity:
\begin{equation}
\Phi^{(\infty)} = v U \, , \;\;\; W_\mu^{(\infty)} = \frac{i}{g} U \partial_\mu U^\dag \, , 
\label{eq:asymptotic}
\end{equation}
where $U$ is a map from the boundary of spacetime to the gauge group, $SU(2)$.  We allow the configuration to have trivial dimensions, in which case it has finite action per unit volume of the trivial dimensions, and the domain of $U$ is the appropriate subspace of the boundary of spacetime.  For example, a static configuration has time as a trivial dimension and in order to have finite energy (action per unit time), it must be pure gauge at spatial infinity.  

Now the third and fourth homotopy groups of $SU(2)$ are non-trivial:
\begin{equation}
\Pi_3(SU(2)) = {\mathcal Z} \, , \Pi_4(SU(2)) = {\mathcal Z}_2 \, .
\end{equation}        
So each map from $S_3$ into $SU(2)$ belongs to a homotopic class labeled by an integer winding number, and it cannot be continuously deformed into any map in a distinct class.  Similarly, each map from $S_4$ into $SU(2)$ belongs either to the trivial class (which contains the trivial map) or the non-trivial class.  Consider any topologically non-trivial map into $SU(2)$ from $S_3$ or $S_4$.  Identify a subspace of the domain with the boundary of spacetime spanned by the non-trivial dimensions.  Any remaining coordinates in the domain are interpolation parameters that define a sequence of pure-gauge, asymptotic configurations.  These asymptotic configurations may be smoothly continued into the bulk to obtain a sequence of configurations.  The sequence becomes a loop when we restrict all configurations on the boundary of the interpolation space to be the totally-trivial configuration ($\Phi=v\ID, W_\mu = 0$).  If the loop can be shrunk to a point, then the topologically non-trivial map into the gauge group can be continuously deformed to the identity, which is impossible.  The top of the tightest non-contractible loop, obtained by minimizing the action (per unit volume of any trivial dimensions) for each point on the loop, should be an unstable solution.

We now present detailed examples of the above construction and list the solutions indicated when all possible non-contractible loops are considered.
\subsection{Winding 1 Map from $S_3$ to $SU(2)$}
\label{sec:winding1}
Let $\beta_1, \beta_2, \alpha$ denote the angular coordinates that describe a three-dimensional sphere, $S_3$, with $0 \le \beta_i \le \pi$ and $0 \le \alpha < 2\pi$.  The $S_3$ can be embedded as a unit sphere in a four-dimensional Euclidean space and is described by unit vectors
\be
\hat{n} (\beta_1, \beta_2, \alpha) = \left( \begin{array}{c}
  \cos \beta_1 \\
  \sin\beta_1 \sin\beta_2 \cos\alpha \\
  \sin\beta_1 \sin\beta_2 \sin\alpha \\
  \sin\beta_1 \cos\beta_2 
  \end{array} \right) \, .
\ee
Then, the canonical winding 1 map from $S_3$ to $SU(2)$ is given by the one-to-one map
\be
U^{(1)} (\beta_1, \beta_2, \alpha) = \hat{n}_0 \ID + i \hat{n}_i \tau_i \, .
\label{eq:S3_1}
\ee    
Now we will identify different subspaces of the domain $S_3$ with the boundary of spacetime, in accordance with the topological prescription, to find non-\classicalVacuum{} solutions to the classical equations of motion.  It turns out that all the solutions found using this map are well-known, and our method simply encapsulates them into a unified framework.  However, as we shall see later, other non-trivial maps lead to novel solutions.
\subsubsection{The Weak Instanton}
The entire $S_3$ is identified with the boundary of spacetime by choosing the unit vectors $\hat{n}$ to be the unit position vectors $\hat{x}$ in Euclidean spacetime.  This example is an exception because there are no parameters remaining in the domain to construct a loop.  The asymptotic configuration is determined by
\be
U(\hat{x}) = \hat{x}_0 \ID + i \hat{x}_i \tau_i \, , 
\ee
using \eq{eq:asymptotic}.  Now we extend the configuration to the interior of spacetime, without introducing any singularities, using the ansatz
\bea
\Phi & = & f_H(r) v U \, , \nonumber \\
W_\mu & = & f_W(r) \frac{i}{g} U \partial_\mu U^\dag \, , 
\eea
where $r$ denotes the radius vector in Euclidean spacetime.  The radial functions $f_H, f_W$ go from 0 at $r=0$ (for regularity at the origin) to 1 as $r$ goes to $\infty$ (for finite-action).  This configuration cannot be continuously deformed to the totally-trivial configuration because that would imply that $U$ can be continuously deformed to $\ID$.  This indicates the existence of a topologically stable solution.  Indeed, in the absence of the Higgs fields, the choice 
\be
f_W(r) = \frac{r^2}{r^2 + w^2} 
\ee    
gives a stable solution to the equations of motion for every choice of the width $w$.  This is the well-known weak instanton \cite{Belavin:fg}.  It mediates non-perturbative fermion number violating processes via tunneling \cite{'tHooft}.  The action has a local minimum of $8 \pi^2 / g^2$ at the weak instanton.

When the Higgs fields are included, their contribution to the action can be made smaller by  scaling to smaller distances, till the action reaches the no-Higgs value of $8 \pi^2 / g^2$ for which the configuration is singular.  (This can be easily understood on dimensional grounds.)  This brings us to a crucial caveat of the topological prescription.  The configuration space is a non-compact manifold, and so the non-contractible loops may run off to infinity without giving a solution.  The method is useful to the extent that it suggests the existence of a solution, determines its stability (or lack thereof) and points to the region in configuration space where we can search for the possible solution.  But it does not guarantee the existence of the solution.
\subsubsection{The Weak Sphaleron}
Now we consider time-independent configurations.  We identify an $S_2$ subspace of the $S_3$ domain of the winding 1 map (in \eq{eq:S3_1}) with the boundary of space:
\begin{equation}
U_{\xi} (\theta, \azAngle) = e^{- i \xi \tau_3} U^{(1)}(\xi, \theta, \azAngle) \, , 
\end{equation}
where $\theta, \azAngle$ are the polar and azimuthal angles respectively that span the spatial boundary.  The remaining coordinate, $\xi$, is the interpolation parameter.  The multiplication by $e^{-i \xi \tau_3}$ fixes $U_0=U_\pi=\ID$, without changing the winding of the map.  So $U_{\xi}$ defines a loop of asymptotic configurations, which can be smoothly continued into the bulk of space for each $\xi$, to obtain a loop of configurations.  For example,
\bea
\Phi(\xi) & = & v \left( f_H(r) U_\xi + (1-f_H(r)) \cos^2 \xi \ID \right) \, , \nonumber \\
W_\mu(\xi) & = & f_W(r) \frac{i}{g} U_\xi \partial_\mu U_\xi^\dag \, , 
\eea
where $r$ is the spatial radius and the radial functions $f_H, f_W$ go from 0 at $r=0$ to 1 as $r \rightarrow \infty$.  Suppose we could continuously deform the loop so that for every $\xi$ the configuration is totally-trivial.  Then $U^{(1)}(\xi, \theta, \azAngle)$ could be continuously deformed into the trivial map, which is impossible.  So the loop is non-contractible.  The classical energy along the loop starts at 0 for $\xi=0$, then increases to some maximum value and finally goes down again to 0 at $\xi=\pi$, where the configuration is totally-trivial.  The loop can be made tighter, but cannot be shrunk to a point.  The configuration at which the energy has a minimax (the top of the tightest loop) would be an unstable, static solution.  We find that the minimax of our loop is at $\xi=\pi/2$ and corresponds to the solution to the following equations:
\bea
 \frac{d}{dr} \left(r^2 \frac{d f_H}{dr} \right)  & = & 2 f_H (1-f_W)^2 + 2 \lambda v^2 r^2 (f_H^2-1) f_H \, , \nonumber \\
 r^2 \frac{d^2 f_W}{d r^2} & = & 2 f_W (1-f_W)(1-2 f_W) - \half g^2 v^2 r^2 f_H^2 (1-f_W) \, . 
\eea
Near $r=0$, $f_H \propto r, f_W \propto r^2$.  As $r \rightarrow \infty$, $(1 - f_H) \propto 1/(v r) \mbox{exp}(- 2 \sqrt{\lambda} v r)$ and $(1-f_W) \propto \mbox{exp } ( - g v r / \sqrt{2} )$.  This solution in fact solves the full equations of motion in \eq{eq:EOM} and is the well-known {\it weak sphaleron} \cite{Manton:1983nd,KlinkhamerManton}.  It is the lowest barrier between topologically inequivalent vacua (see Sec.~\ref{The Fermion Number of a Configuration} for this interpretation) and its energy determines the rate of fermion number violating processes at temperatures comparable to the electroweak phase transition scale.  The radial functions $f_H, f_W$ may be found numerically.  For $\lambda=0$ the energy of the weak sphaleron is about $ 8.60 \pi v /g$, which corresponds to about 7.6 TeV when we choose the experimental values v = 177 GeV, g = 0.63.  (For non-zero $\lambda$ the energy is higher.)

We should point out that had we ignored the Higgs fields, the minimax configuration could be driven to lower energies by expanding the size of the configuration.  This is because the pure Yang-Mills theory has no scale (classically) and so the energy must be inversely proportional to the width of the configuration.  The Higgs sector provides the vev scale, which prevents the non-contractible loop from running off to infinity.

\subsubsection{Weak Strings}
Finally we consider static configurations with one trivial dimension (say $z$) and identify an $S_1$ subspace of the $S_3$ domain with the boundary of the $x-y$ plane.  For example,
\be
U_{\xi_1, \xi_2} (\azAngle) = \left[ U^{(1)}(\xi_1, \xi_2, 0) \right]^\dag U^{(1)}(\xi_1, \xi_2, \azAngle) \, , 
\ee
where $\azAngle$ is the azimuthal angle parameterizing the planar boundary, and the remaining domain coordinates, $\xi_1, \xi_2$, are interpolation parameters.  They parametrize a 2-sphere (or 2-loop) of pure-gauge configurations at planar infinity.  Note that the multiplication by $\left[ U^{(1)}(\xi_1, \xi_2, 0) \right]^\dag$ ensures that the boundary of the square spanned by $\xi_1, \xi_2$ is mapped to the identity, without affecting the winding of the map, thereby making the 2-parameter sequence of maps into a 2-sphere of maps.  The continuation into the bulk may be chosen to be
\bea
\Phi(\xi_1, \xi_2) & = & v \left[ f_H(r) U_{\xi_1, \xi_2} + \{ 1-f_H(r) \} (1-\sin \xi_1 \sin \xi_2 )\ID \right] \, , \nonumber \\
W_\mu(\xi_1, \xi_2) & = & f_W(r) \frac{i}{g} U_{\xi_1, \xi_2} \partial_\mu U_{\xi_1, \xi_2}^\dag \, , 
\eea
where $r$ is the planar radius $\sqrt{x^2+y^2}$.  The radial functions $f_H, f_W$ go from 0 to 1 as $r$ goes from 0 to infinity.  Now we have a non-contractible 2-sphere of configurations, indicating the existence of an unstable solution at the top of the tightest sphere.  We find that the configuration at $\xi_1 = \xi_2 = \pi/2$ has the following non-zero components:
\be
\phi_0 = v f_H(r) e^{-i \azAngle} \, , \vec{W}^3 = - f_W(r) \frac{2}{g r} \hat{\azAngle} \, .
\ee
If the radial functions are chosen to satisfy
\bea
\frac{d^2 f_H}{d r^2} & = & - \frac{1}{r}\frac{d f_H}{dr} + \frac{f_H}{r^2}(1 - f_W)^2 + 2 \lambda v^2 f_H ( f_H^2 - 1 ) \, , \nonumber \\ 
\frac{d^2 f_W}{d r^2} & = & \frac{1}{r} \frac{d f_W}{dr} + \half g^2 v^2 f_H^2 (f_W - 1) \, ,
\eea
then we have one of the well-known {\em W-string} solutions \cite{Vachaspati:1992fi,Klinkhamer:1994uy}.  This is in fact a Nielsen-Olesen vortex of the Abelian Higgs model \cite{Nielsen:1973cs}, embedded in a $U(1)$ subgroup of the $SU(2)$ theory.  The above construction uses the $U(1)$ subgroup generated by $\tau_3$, and other choices of the map $U_{\xi_1, \xi_2} (\azAngle)$ result in the solution embedded in other $U(1)$ subgroups, leading to a family of string solutions.  As is the case for the weak sphaleron, the Higgs vacuum expectation value provides a scale that prevents the energy from approaching zero as the configuration width is increased to infinity.  The instability of the weak string solutions detracts from their significance, especially with regard to electroweak baryogenesis where they could have played a crucial role.   In Chap.~\ref{chap:ElectroweakStrings} we discuss this in greater detail and explain how quantum effects could stabilize these configurations.
\subsection{Winding n Map from $S_3$ to $SU(2)$}
The above procedure can be repeated straightforwardly for a winding $n ( \neq 1)$ map from $S_3$ to $SU(2)$:
\be
U^{(n)}(\beta_1, \beta_2, \alpha) = \left[ U^{(1)}(\beta_1, \beta_2, \alpha) \right]^n \, .
\ee
If the entire domain is identified with the boundary of spacetime, then in analogy with the weak instanton, we should obtain a weak {\em multi-instanton} which is topologically stable and carries a topological charge of $n$.  For time-independent configurations, we identify an $S_2$ subspace of the domain with the boundary of space, with one remaining interpolation parameter.  This indicates the existence of a weak {\em multi-sphaleron} \cite{Kleihaus:1994yj} solution with one direction of instability.  This would be the lowest energy barrier between two vacua that differ by winding number $n$.  Finally, for static, planar configurations, we identify an $S_1$ subspace of the domain with the planar boundary and obtain {\em vorticity $n$ W-strings} with 2 directions of instability.

We do not pursue these solutions further because they do not produce qualitatively different physical effects from those of the solutions obtained from the winding 1 map.  For example, the n-instanton would describe the production of $n$ fermions via quantum tunneling through the sphaleron barrier.  However, ($n$ times) repeated instances of the 1-instanton already allows for this process, albeit with a possibly different probability amplitude.  
\subsection{Non-trivial Map from $S_4$ to $SU(2)$}
\label{sec:windingNT}
Now we come to the relatively unexplored realm of classical solutions indicated by the non-trivial topology of maps from $S_4$ to $SU(2)$.  This topology is unique to $SU(2)$, and doesn't exist for in the QCD gauge group $SU(3)$.  Let $\beta_1, \beta_2, \beta_3, \alpha$ denote the angular coordinates that describe a four-dimensional sphere, $S_4$, with $0 \le \beta_i \le \pi$ and $0 \le \alpha < 2\pi$.  The map
\be
U^{(NT)}({\beta_i}, \alpha) = e^{i \beta_1 \tau_3} U^{(1)}(\beta_2, \beta_3, \alpha) e^{-i \beta_1 \tau_3}  \left[ U^{(1)}(\beta_2, \beta_3, \alpha) \right]^\dag
\ee
cannot be continuously deformed to the trivial map $U = \ID$, and it belongs to the non-trivial homotopy class (as denoted by the superscript `NT').  (Recall that $U^{(1)}$ is a winding 1 map from $S_3$ to $SU(2)$, an example of which is in \eq{eq:S3_1}.)
\subsubsection{The $I^*$}
Proceeding as before, when an $S_3$ subspace of the $S_4$ domain is identified with the boundary of spacetime, we obtain a non-contractible loop of configurations in Euclidean spacetime.  We ignore Higgs fields because on dimensional grounds, their contribution to the action approaches 0 as the configuration is driven to a 0 width singularity.  In the pure $SU(2)$ theory, one example of the loop is  
\bea
U_\xi (\hat{x}) & = & e^{i \xi \tau_3} (\hat{x}_0 \ID + i \hat{x}_i \tau_i) e^{-i \xi \tau_3} (\hat{x}_0 \ID - i \hat{x}_i \tau_i) \, , \nonumber \\
W_\mu (\xi) & = & f_W(r) \frac{i}{g} U_{\xi} \partial_\mu U^\dag_{\xi} \, , 
\eea
where $r$ denotes the radius in Euclidean spacetime.  The action along the loop starts at 0 at the totally-trivial configuration $W_\mu(0)$, rises to a maximum and then returns to 0 back at the totally-trivial configuration $W_\mu(\pi)$.  The top of the tightest loop should correspond to a topologically trivial, unstable solution, the $I^*$ \cite{Klinkhamer:1996ad}.  It could contribute significantly to the path integral, since it extremizes the action.  However, within the various ansatze that we consider, we find that the minimax corresponds to a widely separated instanton and anti-instanton pair.  We have not succeeded in finding a new, localized solution.

Another perspective on the possible $I^*$ solution comes from restricting fields to be totally-trivial at $r \rightarrow \infty$.  Then the spacetime manifold ${\mathcal R}_4$ becomes compactified to $S_4$ for the purpose of maps into configurations.  \ClassicalVacuum{} configurations are pure-gauge and fall into two distinct homotopy classes, because $\Pi_4(SU(2)) = {\mathcal Z}_2$.  Now, any interpolation from a \classicalVacuum{} configuration in the trivial class to a \classicalVacuum{} configuration in the non-trivial class must leave the vacuum manifold (otherwise the two configurations would be deformable into each other).  The configuration with the smallest maximum action along all such possible paths would be a saddle point of the action.  It would be the lowest action barrier between topologically inequivalent \classicalVacuum{} configurations in Euclidean spacetime.  This is analogous to the weak sphaleron, but in one higher dimension.

\subsubsection{The $S^*$}
For static solutions we identify an $S_2$ subspace of the $S_4$ domain with the boundary of space and obtain a non-contractible 2-sphere of configurations.  For example,
\bea
U_{\xi_1, \xi_2} (\theta, \azAngle) & = & \left[ U^{(NT)}(\xi_1, \xi2, 0, 0) \right]^\dag U^{(NT)} ( \xi_1, \xi_2, \theta, \azAngle) \, , \nonumber \\
\Phi (\xi_1, \xi_2 ) & = & v \left[ f_H(r) U_{\xi_1, \xi_2} + \{ 1-f_H(r) \} (1-\sin \xi_1 \sin \xi_2) \ID \right] \, , \nonumber \\
W_\mu (\xi_1, \xi_2) & = &  f_W(r) \frac{i}{g} U_{\xi_1, \xi_2} \partial_\mu U^\dag_{\xi_1, \xi_2} \, , 
\eea
where $r, \theta, \azAngle$ denote spatial spherical coordinates.  The radial functions $f_H, f_W$ vanish at the origin and approach 1 as $r \rightarrow \infty$.  The top of the tightest 2-sphere should give a static solution with 2 directions of instability, the $S^*$ \cite{KlinkhamerSstar}.

However, we have not succeeded in finding a solution.  As we tighten our 2-sphere of configurations within different ansatze, we find that the maximum action configuration approaches a separated sphaleron anti-sphaleron pair.

The $S^*$, if it exists, is a promising candidate to form a fermionic soliton that allows a heavy fermion doublet to decouple from the Standard Model (see Sec.~\ref{sec:QuantumSolitonsConclusion}).  
\subsubsection{The $W-string^*$}
Finally, to obtain non-trivial solutions on the $x-y$ plane, we can identify an $S_1$ subspace of the $S_4$ domain with the planar boundary.  This would give a non-contractible 3-sphere of configurations, which indicates the existence of a static, finite energy per unit length solution, with 3 directions of instability.  Daunted by our inability to find the $I^*$ and the $S^*$, we have not pursued this possibility.
\subsection{Caveats and Conclusions}
The topological prescription to find non-\classicalVacuum{} solutions to the classical equations of motion generalizes to any gauge theory.  Once we know the gauge group, for each topologically non-trivial map into the gauge group, we construct a non-contractible $n$-sphere of configurations.  One point on the sphere is chosen to be the totally-trivial \classicalVacuum{} configuration.  Since the sphere cannot be shrunk to a point, the ``tightest'' sphere obtained by minimizing the action for each configuration on the sphere, could give solutions (corresponding to saddle points and local minima of the action).  In the case of the simplified version of the electroweak theory, we have seen that this method allows us to find the well-known weak instanton, weak sphaleron, W-strings, and their higher winding generalizations.  It also suggests the existence of novel solutions, which are analogous to the well-known solutions, except constructed using a different topology.  However, we have been unable to find these solutions.

These topological arguments do not guarantee the existence of the solutions described above, because the configuration space is a non-compact manifold and the non-contractible loops may run off to infinity.  For example, in the absence of the Higgs fields, the weak sphaleron's energy can be lowered by scaling to larger distances and approaches 0 as the solution approaches an infinite width singularity.   Also, it not clear that two non-contractible loops obtained using different maps into the gauge group, give two distinct solutions.  For example, we find that in our search for an $I^*$ using $\Pi_4(SU(2)) = {\mathcal Z_2}$, the minimax configuration tends to break up into two instantons that were already discovered using $\Pi_3(SU(2)) = {\mathcal Z}$.  Furthermore, the method is ignorant of solutions that have trivial gauge fields (such as the kink domain wall).  Nevertheless, the topology points to possible solutions in the vast configuration space and once we know where to look, we can verify whether a solution exists.  Furthermore, the method encodes whether the suggested solution is stable or not, and in the latter case it gives us the directions of instability.

We find that all known and hinted time-independent solutions are unstable.  The classical \HiggsGauge sector seems to have only sphalerons  and no solitons.  Of course, the existence of non-topological solitons (corresponding to local minima of the energy functional) cannot be excluded.  However, having exhausted the topological properties of the theory, we are left with no guiding principle to enable a search for such objects.  But if we consider quantum effects on the classical bosonic sector, then there are compelling reasons to expect the existence of quantum solitons and well-understood mechanisms to guide the search for them (see Chap.~\ref{chap:QuantumSolitons}).  
\section{Solutions in Minkowski Spacetime}
In this section we argue for the existence of approximate breathers in the electroweak theory.  These are unnaturally long-lived, spatially localized configurations that are periodic in real time.

\subsection{$\phi^4$ in 1+1 Dimensions} 
As a lead up to the electroweak breathers, consider the real $\phi^4$ theory in 1+1 dimensions.  This is arguably the simplest of all field theories, and yet it has enough structure to shed light on the mechanism that gives rise to breather configurations in classical field theories.  The action is
\be
S [ \phi ] = \int d^2 x \left[ \half \partial^\mu \phi \partial_\mu \phi - \frac{\lambda}{4} \left( \phi^2 - v^2 \right)^2 \right] \, .
\ee
The equation of motion obtained by extremizing the action is
\be
\ddot{\phi} - \phi'' = -\lambda \left( \phi^2 - v^2 \right) \phi \, .
\label{eq:eom2}
\ee
In addition to the vacuum solutions
\be
\phi (x, t) = \pm v \, , 
\ee
this theory has the well-known solitonic kink (anti-kink) solutions
\be
\phi (x, t) = \pm v \tanh \frac{ v \sqrt{\lambda} ( x - x_0 - ut)}{\sqrt{2 (1-u^2)}} \, , 
\ee
centered initially at $x_0$ and moving with velocity $u$.  Now we shall see that there also exist approximate breathers in the theory.

Consider small oscillations around the vacuum $v$,
\be
\phi = v + \delta \phi \, .
\ee
The linearized equation of motion is
\be
\delta \ddot{\phi} - \delta \phi'' + 2 \lambda v^2 \delta \phi = 0 \, .
\ee
Fourier transforming to momentum space, we get the linear dispersion relation
\be
\omega_{\rm lin} = \sqrt{m^2+k^2} \, , 
\ee
where $m = \sqrt{2 \lambda} v$ is the mass of the scalar particles in the theory.  The crucial observation is that there is a {\em mass gap} and the linear spectrum starts at $m$.  

Now we come to the second critical ingredient in the theory that allows for approximate breathers: the scalar potential is {\em non-linear}.  More specifically, for increasing deviations of $\phi$ from $v$ toward zero (the peak of the so called double-well potential), the curvature of the potential decreases.  If this were a single particle potential, then this would imply that large amplitude oscillations around the vev will have a frequency lower than the linear frequency associated with the quadratic potential.  This generalizes in the case of the field theory, and there are several spatially localized configurations which oscillate in time (or more precisely, the field value at the center oscillates in time) with a fundamental frequency below $m$.  So, these oscillations should be stable against decay by linear modes radiation (i.e. boson radiation).  However, non-linearity also provides a flip side to this argument: the higher harmonics of the fundamental frequency must be present and they will fall in the band of the linear spectrum and cause the configuration to eventually decay.  In discrete systems, the linear spectrum has an upper bound and the harmonics could be constructed to lie beyond the linear spectrum giving rise to stable discrete breathers.  See \cite{CampbellFlachKivshar} for an elementary review of discrete breathers, wherein these ideas are explained in the context of discrete systems.  Also see \cite{Gleiser:1994pt} in which Gleiser has demonstrated the existence of approximate breathers in continuum scalar field theories in 3+1 dimensions (for both symmetric and asymmetric double-well potentials).   

We now give an example of an approximate breather in the 1+1 dimensional $\phi^4$ theory.  We set the dimensionless vev $v$ to 1.  We also choose $\lambda = 1$ and all dimensionful quantities are measured in units of $\sqrt{\lambda}$.  Consider the initial configuration
\be
\phi(x, 0) = \tanh^2 ( x/w ) \, ,
\label{eq:initBreather}
\ee
with the width chosen to be $w=2$.  We start the above configuration at rest and then numerically evolve according to \eq{eq:eom2}.  Since our initial configuration and its time derivative are even functions of $x$, they remain even throughout the time evolution according to the equation of motion.  So we consider only the positive half-line with vanishing spatial derivative boundary condition at the origin (in higher dimensions this condition is required for regular configurations).  The numerical method is the following.  We consider the x-interval between 0 and $L_0$, where $L_0$ is much larger than the initial width of the configuration.  We discretize time and space with lattice spacings $h_t, h_x$ respectively.  Then the discretized equation of motion in \eq{eq:eom2}, accurate to second order in time and space, is
\be
\phi_{i,j+1} = \frac{h_t^2}{h_x^2} \left( \phi_{i+1,j} + \phi_{i-1,j} - 2\phi_{i,j} \right) + 2\phi_{i,j} - \phi_{i,j-1} - h_t^2 \lambda \phi_{i,j}(\phi_{i,j}^2 - v^2) \, ,
\label{discretizedEOM}
\ee  
where the first subscript labels the spatial lattice points and the second labels the temporal lattice points (so that $\phi_{i,j} = \phi(i h_x, j h_t)$).  Note that the spatial lattice points within the interval are labeled by $i = 0, \ldots , N=L_0/h_x$.  The above equation is explicitly in the form that allows $\phi(x)$ at the next time step to be determined if it is known at the current and previous time steps.  For the spatial lattice point $i=0$, the equation requires the value at the point $i=-1$ and we impose the condition that $\phi_{-1, j} = \phi_{1, j}$ which corresponds to the function being even.  At the wall (spatial lattice point $N$), we impose the similar condition, $\phi_{N+1, j} = \phi_{N-1, j}$.  This condition ensures that no energy flows past the wall (since it corresponds to vanishing derivative).  To eliminate reflections from the wall returning to our local region of interest, we move the wall out at the speed of light (i.e. $L = L_0 + t$) when we evolve an initial configuration.  

\begin{figure}[htb]
\begin{center}
\includegraphics{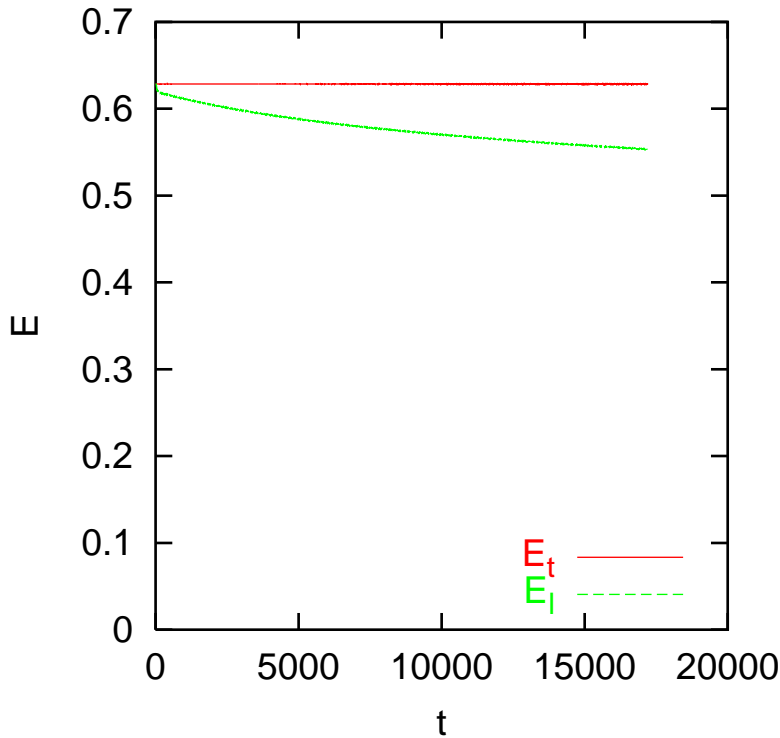}
\end{center}
\caption{\small The total energy, $E_t$, on the positive half-line and the local energy, $E_l$, between $x=0$ and $x=20$, as a function of time for the initial configuration in \eq{eq:initBreather} in the $\phi^4$ theory.  All dimensionful quantities are in units of $\sqrt{\lambda}$ and the vev $v$ is 1.}
\label{fig:breather_energies}
\end{figure}

\begin{figure}[htb]
\begin{center}
\includegraphics{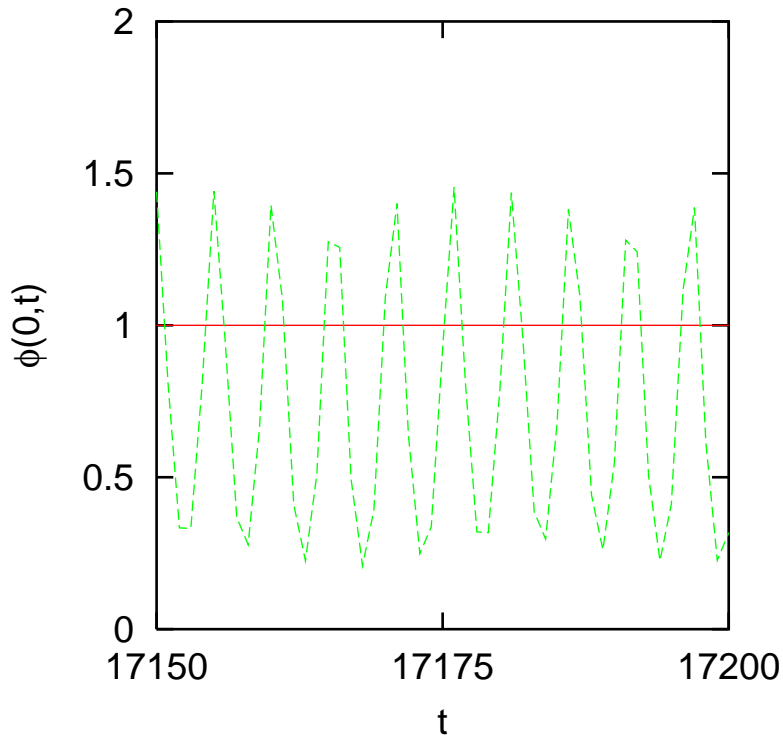}
\end{center}
\caption{\small The field value at the center of the configuration as a function of time, for the initial configuration in \eq{eq:initBreather} in the $\phi^4$ theory.  All dimensionful quantities are in units of $\sqrt{\lambda}$ and the vev $v$ is 1.}
\label{fig:breather_amplitude}
\end{figure}

In Fig.~\ref{fig:breather_energies} we plot the total energy, $E_t$, on the half-line as a function of time.  This is conserved through the evolution.  We also plot the energy, $E_l$, localized between $x=0$ and $x=20$.  This falls very slowly and after a time of about 17,000, only $12 \%$ of the total energy has dissipated out of the local region.  We do not yet understand what sets the scale for this decay rate.  But it is clear that naturalness arguments are violated in this example and the energy is localized for time periods much longer than all dimensionfull scales in the problem.  In Fig.~\ref{fig:breather_amplitude} we plot the value of the field at the origin as a function of time, towards the end of the time of evolution considered.  This oscillates about $\phi=v$ (asymmetrically because the potential is asymmetric about $\phi=v$).  The fundamental period of oscillation is obtained by measuring the interval between peaks and the corresponding fundamental frequency turns out to be about 1.2.  This is lower than the mass of the scalar particle, $m = \sqrt{2} \approx 1.414$, as expected.  The existence of such approximate breathers seems to be a generic feature of the theory, and we have successfully constructed several such configurations with different frequencies and amplitudes.  

We extend the $\phi^4$ theory by considering complex $\phi$ and gauging the global $U(1)$ symmetry.  The gauge boson becomes massive after eating the Goldstone degree of freedom.  So the linear spectrum has mass gaps corresponding to the masses of the scalar and vector bosons.  We find that the breathers in the real $\phi^4$ theory are stable against perturbations in the complex gauged theory, as long as the $U(1)$ charge is large enough so that the fundamental breather frequencies are lower than the mass of the gauge boson (which is proportional to the charge).  However, when we reduce the charge so that the fundamental breather frequency falls within the spectrum of small gauge field oscillations, the breathers dissipate their energies rapidly.  This is further evidence that the existence of approximate breathers is due to the presence of a mass gap and a non-linear potential that allows oscillations with a frequency smaller than the mass of the lightest particle in the theory.

\subsection{Breathers in the Electroweak Theory}
Now that we understand the mechanism that allows long-lived, spatially localized, temporally periodic configurations to exist in a classical field theory, we can ask if there is any possibility for such objects to exist in the Standard Model.  The answer is yes.  When we ignore the hypercharge gauge fields in the electroweak theory and consider the $SU(2)$ Higgs theory (as described in Sec.~\ref{sec:HiggsGauge}), there is no unbroken subgroup, and all three $W$ gauge bosons are massive.  So, the theory fulfills the first requirement of having a mass gap in the linear spectrum.  Also, the Higgs potential is the usual sombrero potential and so it seems likely that we could set up large-amplitude configurations with oscillation frequencies in the mass gap. So the theory appears to  have all the ingredients required for the existence of approximate breathers.  When we include the $U(1)$ sector, the photons remain massless after symmetry breaking.  However, it is conceivable that the breathers can be made electrically neutral so that they don't dissipate by electromagnetic radiation.

We are currently investigating whether such objects exist in the electroweak theory.  They could have many significant implications.  Firstly, since the lifetimes of the breathers is expected to be many orders of magnitude larger than all natural scales in the problem, they could shed light on naturalness in field theories.  Secondly, if these breathers could exist just after the electroweak phase transition, they would set up out-of-equilibrium regions in space, with implications for electroweak baryogenesis.

\chapter{Quantum Solitons}
\label{chap:QuantumSolitons}
We explain how certain \HiggsGauge configurations may be stabilized in the Standard Model, by carrying heavy quark quantum numbers.  These fermionic solitons would maintain anomaly cancellation in the low-energy effective theory and allow heavy fermions to decouple from the chiral gauge theory.  We describe the methodology of looking for such objects in the configuration space.  We present a technique based on scattering theory that enables an efficient, exact and properly renormalized calculation of the one-loop effective energy, which is one crucial ingredient in the stability analysis of quantum solitons.  Finally, we describe our search for a fermionic soliton within a spherical ansatz.  Much of this investigation originally appeared in \cite{KhemaniDecoupling}.

\section{The Idea}
\label{sec:Idea}
A topological soliton is a non-\classicalVacuum, static configuration that is topologically stable.  It carries a conserved topological charge which prevents it from decaying into a \classicalVacuum{} configuration with no topological charge.  Analogously, it may be possible to stabilize a configuration by making it carry a conserved quantum number.  We use the term quantum soliton to refer to any such quantum-stabilized object.  In this section we briefly describe the mechanism and motivation for these solitons, and elaborate on these ideas in the following sections.

There is a natural mechanism in the electroweak theory for the existence of quantum solitons stabilized by carrying fermion number (say top quark number), which stems from the polarization of the fermionic vacuum in a background of gauge and Higgs fields.  Several configurations are known to tightly bind fermions in their vicinity.  These configurations consist of classical solutions (the sphalerons discussed in Sec.~\ref{sec:ClassicalSolutions}) as well as non-solutions.  The existence of tightly-bound levels suggests that it may be energetically favorable for a certain number of fermions (say $N_f$) to be trapped by such backgrounds, with a small associated occupation energy, $E_{\rm occ}^{(N_f)}$.  The binding energy could outweigh the cost in classical energy, $E_{\rm cl}$, to set up the configuration, i.e. $E_{\rm cl}+E_{\rm occ}^{(N_f)} < m_f N_f$, where $m_f$ is the mass of the fermion.  However, to be consistent to order $\hbar$, we must also include the fermion Casimir energy: the renormalized energy shifts of all the other fermion modes.  For static configurations this is the renormalized one-loop fermion vacuum energy $E_{\rm vac}$.  Thus, the minimum total energy of an arbitrary configuration $C=\{\Phi, W_\mu \}$,  which carries fermion number $N_f$, is
\begin{equation}
E^{(N_f)}_{\rm eff}[C] = E_{\rm cl}[C] + E_{\rm occ}^{(N_f)}[C] + E_{\rm vac}[C] \, .
\label{ENf}
\end{equation}
The subscript `eff' denotes that the energy is an effective energy obtained by integrating out the fermions from the theory, as explained in Sec.~\ref{sec:EffectiveEnergy}.  If $C$ decays into a \classicalVacuum{} configuration then it must create $N_f$ perturbative fermions (ignoring anomalous violation of fermion number, which is exponentially suppressed).  However, the strong fermion binding suggests that there exist configurations such that $E_{\rm eff}^{(N_f)}[C]<m_f N_f$.   Then, $C$ can decay only into a quantum soliton at which $E^{(N_f)}$ has a local minimum.

The existence of such fermionic quantum solitons would provide an attractive resolution to the decoupling puzzle in the Standard Model (and other chiral gauge theories) \cite{Appelquist}.  A fermion obtains its mass through Yukawa coupling to the scalar Higgs via the well-known Higgs mechanism.  Explicit mass terms are prohibited by gauge-invariance.  So, as we increase the mass of a fermion (thereby making the denominator in the propagator suppress loop corrections) we also increase the Yukawa coupling which gives a corresponding enhancement from the vertices.  Moreover, the heavy fermion cannot simply disappear from the spectrum because then anomaly cancellation would be ruined.  However, it is plausible that the large Yukawa coupling gives rise to a quantum soliton in the low energy theory.  This carries the quantum numbers of the decoupled fermion and maintains anomaly cancellation using the mechanism described by D'Hoker and Farhi in the $SU(2)_L$ Skyrme model \cite{D'Hoker:1984kr}.
\section{The \HiggsGauge Sector Effective Energy}
\label{sec:EffectiveEnergy}
In addition to the bosonic sector described in Sec.~\ref{sec:HiggsGauge}, the Standard Model has a rich fermion sector of three generations of quarks and leptons.  We will be primarily interested in one-loop quantum effects of heavy fermions on the \HiggsGauge sector.  So we ignore inter-generation mixing and set the Cabibbo-Kobayashi-Maskawa (CKM) matrix to the identity.  Let $\Psi$ denote an isospin doublet, say the top-bottom quark doublet.  Only the left-handed fermions are charged under a gauge transformation, $U(x) \in SU(2)$:
\be
\Psi_L \rightarrow U \Psi_L \, , \Psi_R \rightarrow \Psi_R \, , 
\ee
where the subscript $L, R$ denote left and right handed projections respectively.  The action for the doublet is
\be
\mathcal{S}_F [ \Psi, \Phi, W_\mu] = \int d^4 x \left[ \overline{\Psi}_L i\gamma^\mu D_\mu \Psi_L + 
\overline{\Psi}_R i\gamma^\mu \partial_\mu \Psi_R - f\left( 
\overline{\Psi}_L \Phi\Psi_R + \mbox{h.c.} \right) \right] \, .
\end{equation}
The covariant derivative is 
\be
D_\mu \Psi_L = \left( \partial_\mu - i g W_\mu \right) \Psi_L \, , 
\ee
and $f$ is the Yukawa coupling constant (which may be different for each doublet).  For simplicity we have used the same Yukawa coupling for both components of the doublet, which results in the degenerate mass $m_f = f v$, when the Higgs develops the vev $v$ after spontaneous symmetry breaking.  In the absence of inter-generation mixing, the fermion sector Lagrangian is given by the sum of the Lagrangians for each fermion doublet.  We introduce the potential
\be
V(\Phi, W_\mu) = -g \gamma^\mu W_\mu(x)\frac{1-\gamma_5}{2} + f ( h(x) + ivp^a(x) \tau^a \gamma_5 ) \, , 
\ee
where $h(x), p^a(x)$ are real functions that characterize deviations of the scalar fields from the vev:
\be
\Phi(x) - v = h(x) + ivp^a(x) \tau^a \, . 
\ee
We write the Lagrangian density as the sum of a free part and an interacting part:
\be
S_F [ \Psi, V(\Phi, W_\mu)] = \int d^4 x \left[ \overline{\Psi} \left( i \gamma^\mu \partial_\mu - m_f \right) \Psi - \overline{\Psi} V \Psi \right] \, .
\label{potential}
\ee

Since we have ignored the U(1) hypercharge sector, there are no perturbative anomalies associated with triangle diagrams, and we only have Witten's global anomaly \cite{Witten} to contend with.  This requires an even number of $SU(2)_L$ fermions in the theory, otherwise the theory is not defined.  (In the path integral formulation, the partition function becomes zero.  Alternatively, in a Hamiltonian formulation, the theory has no gauge-invariant states \cite{D'Hoker:1984kr}.)  For now we restrict our attention to a single, heavy fermion doublet and ignore the lighter spectator doublets present for anomaly cancellation.  We investigate if the heavy doublet can be decoupled from the theory by becoming solitonic at a lower energy scale.  The full theory we consider is then given by the \HiggsGauge sector, $S_H$, described in Sec.~\ref{sec:HiggsGauge} and the heavy fermion sector, $S_F$:
\be
S[\Phi, W_\mu, \Psi] = S_H[\Phi, W_\mu] + S_F[\Psi, V(\Phi, W_\mu)] \, .
\ee

We integrate out the fermion doublet from the theory to obtain the 
effective action for the \HiggsGauge sector:
\begin{equation}
e^{i S_{\rm eff}[\Phi, W_\mu]} = e^{i S_H} \frac{
\int[d\Psi][d\overline{\Psi}] e^{iS_F}}{ \int[d\Psi][d\overline{\Psi}] e^{i S_F |_{\rm V=0}}} \, .
\end{equation}
The normalization has been chosen so that the effective action is equal to the classical action for vanishing interaction potential, $V$, defined in eq.~(\ref{potential}).
If $i S_{\rm FD}^{(n)}$ denotes the Feynman diagram with one fermion loop and 
$n$ external insertions of $\left(-i V(\Phi, W_\mu)\right)$, then
\begin{equation}
S_{\rm eff}[\Phi, W_\mu] = S_H + \sum_{n=1}^{\infty} 
S_{\rm FD}^{(n)}\, .
\label{SeffFD}
\end{equation}
The Feynman diagrams can be computed in a prescribed 
regularization scheme. The divergences that emerge as the
regulator is removed are canceled by counterterms. We introduce
the renormalized and counterterm actions, $S_H^{\rm (ren)}$ and $S_H^{\rm (ct)}$ respectively,
by expressing the bare parameters in $S_H$ in terms of 
renormalized parameters and counterterm coefficients to obtain $S_H = 
S_H^{\rm (ren)} + S_H^{\rm (ct)}$.   The renormalized action is  the original \HiggsGauge sector action, eq.~(\ref{eq:HiggsAction}), with
renormalized parameters substituted.  It can be written as the sum of all possible renormalizable, gauge-invariant terms:
\begin{eqnarray}
S_H^{\rm (ct)} & = & \int d^4 x \left[ 
c_1\tr\left(W^{\mu\nu}W_{\mu\nu}\right) + 
c_2\tr\left(\left[D^{\mu}\Phi\right]^{\dag}D_{\mu}\Phi\right ) \right.
\nonumber \\ 
& & \left. + c_3\left[\tr\left(\Phi^{\dag}\Phi\right)-2v^2 \right] + 
c_4 \tr\left[ \Phi^{\dag}\Phi - v^2\right]^2 \right] \, .
\label{Lct}
\end{eqnarray}
The coefficients $c_i$ depend on the regulator.   
 For notational simplicity we do not introduce a different notation for the renormalized parameters and fields ($\Phi, W_\mu, g, \lambda$). 

If we consider static \HiggsGauge fields (in the $W_0 = 0$ gauge) and restrict time to the 
interval $T$, then the effective energy functional is
\begin{equation}
E_{\rm eff}[\Phi, W_i] =  
- \lim_{T \rightarrow \infty} \frac{1}{T}\, S_{\rm eff}[\Phi, W_i]  
\,\equiv\,  E_{\rm cl}[\Phi, W_i] + E_{\rm vac}[\Phi, W_i] \, ,
\label{Eeff}
\end{equation}
where $E_{\rm cl}$ refers to the classical energy of the \HiggsGauge
sector:
\begin{equation}
E_{\rm cl}[\Phi, W_i] =  
\int d^3x \left\{\frac{1}{2} \tr\left(W_{i j}W_{i j}\right) + 
\frac{1}{2}\tr\left(\left[D_{i}\Phi\right]^{\dag} D_{i}\Phi\right) + 
\frac{\lambda}{4}\left[\tr\left(\Phi^{\dag}\Phi\right)-2v^2\right]^2  
\right\} \, . 
\label{Eclassical}
\end{equation}
The fermionic vacuum energy is
\begin{equation} 
E_{\rm vac}[\Phi, W_i] =  
\sum_{n=1}^{\infty} E_{\rm FD}^{(n)}[\Phi, W_i] + E_{\rm ct}[\Phi, W_i]\, ,
\label{Evac1}
\end{equation}
where each regulated Feynman diagram contribution is
\begin{equation}
E_{\rm FD}^{(n)}[\Phi, W_i]  =  
- \lim_{T \rightarrow \infty} \frac{1}{T}S_{\rm FD}^{(n)}\, , 
\end{equation}
and the counterterm contribution is
\begin{eqnarray}
E_{\rm ct}[\Phi, W_i]  & = & \int d^3x \Biggl\{ 
-c_1\tr\left(W_{i j}W_{i j}\right) 
+c_2\tr\left(\left[D_{i}\Phi\right]^{\dag} D_{i}\Phi\right)
\nonumber \\ & & \hspace{1.5cm}
-c_3\left[\tr\left(\Phi^{\dag}\Phi\right)-2v^2\right] 
-c_4 \tr\left[ \Phi^{\dag}\Phi - v^2\right]^2 \Biggr\} \, .
\end{eqnarray}
The entire one-loop effective energy receives contributions also from gauge and Higgs loops.  We ignore these contributions because we believe the fermion loops are fundamentally responsible for the phenomena associated with fermion decoupling.  If we imagine that the fermions have $N_C$ internal degrees of freedom (e.g. color), then this approximation becomes exact for large $N_C$.  Nevertheless we set $N_C=1$.
\subsection{The Counterterms}
The counterterms render $E_{\rm vac}$ finite by canceling the divergences in $E_{\rm FD}^{(n)}$, for $n=1$
through $n=4$, that arise when the regulator is removed. To unambiguously determine the 
counterterm coefficients, $c_i$, we impose
physical on-shell renormalization conditions:
\begin{itemize}
\item[a.]
We choose the vacuum expectation value (vev) 
of $h(x)$ to be 0, which
ensures that the vev $\langle \Phi \rangle = v\ID$ stays fixed at its classical value and
the perturbative fermion mass does not get 
renormalized.  This is equivalent to a ``no-tadpole''
renormalization condition and determines $c_3$.

\item[b.]
We fix the pole of the Higgs propagator to be at
the tree level mass, $m_h=m_h^{(0)}$, 
with residue 1. These conditions determine
$c_2$ and $c_4$.

\item[c.]
We have various choices to fix the remaining undetermined 
counterterm coefficient $c_1$. We choose to set the
residue of the pole of the gauge field propagator to 1 
in unitary gauge. Then the position of that pole, {\it i.e.} the mass
of the gauge field, $m_w$, is a prediction.
\end{itemize}
The resulting counterterm coefficients, $c_i$, are listed
in Appendix \ref{app:Spherical}.
As explained under item c., the mass of gauge fields 
is constrained by the other model parameters when fermion 
loops are included. With our choice of renormalization conditions,
it is the solution to the implicit equation
\begin{eqnarray}
m_w^2 & = & \left(m_w^{(0)}\right)^2
\Biggl[1+\frac{f^2}{8\pi^2} 
\Biggl\{\frac{2}{3}-\frac{m_w^2}{m_f^2}
\left(\frac{1}{6}-\int_0^1dx x^2(1-x)^2\frac{m_w^2}
{\Delta(x,m_w^2)}\right) 
\nonumber \\ && \hspace{1cm}
+6\int_0^1 dx x(1-x)\ln\frac{\Delta(x,m_h^2)}{m_f^2} 
-\int_0^1dx\ln\frac{\Delta(x,m_w^2)}{m_f^2} \Biggr\} \Biggr] \, ,
\label{ModelParamsConstraint}
\end{eqnarray}
with $\Delta (x, q^2) \equiv m_f^2 - x(1-x)q^2$.  Recall that $m_w^{(0)} = gv/\sqrt{2}$ is the tree level perturbative mass of the gauge fields.
\subsection{The Fermion Vacuum Energy}
\label{sec:PhaseshiftsMethod}
We briefly summarize methods introduced in refs.
\cite{PhaseshiftsGeneral,PhaseshiftsFermions}
(see ref.~\cite{Leipzig} for a review and a list of 
   additional references) that enable an {\em
exact} calculation of one-loop effective energies in quantum field theories.  We use these techniques to compute the fermion quantum corrections ($E_{\rm vac}$) to the energies of \HiggsGauge configurations.  As an aside, we want to point out that we have fruitfully used these methods in the study of vacuum energies of quantum fields in the presence of boundary conditions, and the resulting Casimir forces and stresses on the boundaries \cite{KhemaniCasimirLetter,KhemaniCasimirPaper,KhemaniDirichlet}.  The boundary conditions give rise to additional divergences in such calculations, and renormalization is a contentious issue.  We replace the boundaries with background fields that couple to the quantum fields in such a way that in an appropriate limit, they constrain the quantum fields to obey the boundary conditions.  Then we calculate in this renormalizable field theory to obtain properly renormalized, finite results in the presence of the background.  Finally, we take the boundary condition limit on the background to obtain the Casimir energy in the presence of boundaries.  We find that the Casimir energy diverges in the boundary condition limit and this divergence cannot be absorbed into a renormalization of the parameters of the theory.  This means that the Casimir energy and other quantities such as the stress on an isolated surface are sensitive to the physical ultra-violet cutoff.  On the other hand, the Casimir force between objects and the energy density away from the surfaces are finite and well-defined.  Since these topics lie outside the realm of the electroweak theory, we have omitted a more detailed discussion of this here.

In Eq.~(\ref{Evac1}), the fermion vacuum energy in a \HiggsGauge background is formally expressed as an infinite sum of Feynman
diagrams.  We are interested in heavy fermions with large Yukawa couplings, which forbids a perturbative calculation in the couplings.  Also, we will usually consider configurations with widths of the order of the fermion Compton wavelength, which does not allow an approximate calculation using a derivative expansion.  In order to compute non-perturbatively, we  make use of the fact that $E_{\rm vac}$ is also given by a sum over the shift in the zero-point
energies of the fermion modes due to the background fields.  We write this
formal quantity as a sum over bound state energies, $\epsilon_j$,
(times their degeneracies, $D_j$)
and a momentum integral of the energy 
of the continuum states weighted by the change in the density 
of continuum states, $\Delta\rho(k)$, that is
induced by the background fields,
\begin{equation}
E_{\rm vac} = -\frac{1}{2}\sum_j D_j|\epsilon_j| - 
\frac{1}{2}\int_0^\infty dk \sqrt{k^2+m_f^2}\, \Delta\rho(k) + E_{\rm ct} \, .
\label{Evac2}
\end{equation} 
The momentum integral and $E_{\rm ct}$ are both divergent, but 
their sum is finite because the theory is renormalizable.  
We render the integral finite by subtracting the first $N$ 
terms in the Born series expansion of the density of states and adding back in 
exactly the same quantity in the form of Feynman diagrams:
\begin{eqnarray}
E_{\rm vac} & = & -\frac{1}{2}\sum_j D_j|\epsilon_j| 
- \frac{1}{2}\int_0^\infty dk \sqrt{k^2+m_f^2} 
\left( \Delta\rho(k) - \sum_{i=1}^{N}\Delta\rho^{(i)}(k)\right) 
\nonumber\\ & &  
+ \sum_{i=1}^N E_{\rm FD}^{(i)} + E_{\rm ct} \, .
\end{eqnarray}
When we combine $E_{\rm ct}$ with $\sum_{i=1}^N E_{\rm FD}^{(i)}$, 
we cancel the divergences as well as implement the renormalization prescription.
As a result, the above expression is manifestly finite.
The minimal number of required Born subtractions, $N$, 
is easily determined by an analysis of the superficial degree of 
divergence of the one-loop Feynman diagrams.  For our theory, $N=4$.

We will work with background fields in the spherical ansatz
\cite{RatraYaffe}, as described in Sec.~\ref{sec:SphericalAnsatz}.  Then we can express $\Delta\rho(k)$ (and its Born
series) in terms of the momentum derivative of the phase shifts \cite{Schwinger}, induced by the background fields, of the Dirac
wave-functions,
\begin{equation}
\Delta\rho(k) = \frac{1}{2\pi i}\frac{d}{dk}\Tr \ln S(k) 
= \frac{1}{\pi}\frac{d}{dk}\sum_{\sigma\in\{+,-\}}
\sum_G D_G\delta_{G,\sigma}(k) \, .
\end{equation}
Here we have expanded the $S$-matrix in grand spin channels labeled
by $G$.  The grand spin is the vector sum of total angular momentum and isospin. It is conserved by the potential, eq.~(\ref{potential}), for spherically symmetric field configurations.  Note that this method requires us to consider sufficiently symmetric configurations so that the scattering matrix can be expanded in partial waves, and we have chosen spherical symmetry.  We will later consider cylindrical symmetry when we investigate electroweak strings in Chap.~\ref{chap:ElectroweakStrings}.
The asymptotic scattering states are labeled 
by parity $(-1)^G$ and total spin $G\pm1/2$ and we obtain a 
$4\times4\,S$-matrix in general (except in the $G=0$ channel, where it is $2\times2$). We let $\delta_{G,\sigma}(k)$ denote the 
sum of the eigenphase shifts 
at momentum $k$ in channel $G$ and $\sigma=\pm$ specify the sign of the energy 
eigenvalue, $\omega=\pm\sqrt{k^2+m_f^2}$.   Note that the single particle spectrum is not symmetric because the Dirac Hamiltonian is not charge conjugation invariant.  The degeneracy is given by $D_G=2G+1$.  In Appendix \ref{app:Spherical} we show in detail
how to use the Dirac equation to calculate the bound state energies and
scattering phase shifts needed in the computation of $E_{\rm vac}$.  

To simplify the calculation we only subtract the first two 
Born approximants and eliminate the remaining log-divergence in 
the momentum integral by using a limiting function approach 
developed in ref. \cite{DecouplingNoGauge}.  The final expression for the fermion vacuum 
energy is (after integrating by parts and using Levinson's theorem which relates the phase shifts at threshold to the number of bound states)
\bea
E_{\rm vac} & =  & -\frac{1}{2}\sum_j(2G_j+1) \left( |\epsilon_j| - m_f \right) +  E^{(1,2)} + E^{(3,4)} \nonumber \\
 & & + \int_0^\infty \frac{dk}{2\pi} \frac{k}{\sqrt{k^2+m_f^2}} \overline{\delta}(k) + \frac{m_f}{2 \pi} \delta_{\rm lim}(0) \, , 
\label{Evac}
\eea
where $G_j$ is the grand spin associated
with the bound state $j$ and
\begin{equation}
\overline{\delta}(k) =  
\sum_{\sigma\in\{ +,- \}}\sum_{G=0}^\infty (2G+1)
\left(\delta_{G,\sigma}(k) - \delta^{(1)}_{G,\sigma}(k) - 
\delta^{(2)}_{G,\sigma}(k) \right) + \delta_{\rm lim}(k) \, .
\end{equation}
Here $\delta^{(i)}_{G,\sigma}(k)$ denotes the 
$i^{\rm th}$-term in the Born series of $\delta_{G,\sigma}(k)$.
After subtracting  $\delta^{(1)}$ and $\delta^{(2)}$ from
$\delta$, the momentum integral does not contain any
contributions that are linear or quadratic in $V$. 
The limiting function for the sum over all 
eigenphase shifts, $\delta_{\rm lim}(k)$,  eliminates 
the logarithmically divergent pieces
that are third and fourth order in $V$ from the momentum integral. Its 
analytic expression can be extracted from the divergent pieces
of the corresponding Feynman diagrams and
is given in Appendix \ref{app:Spherical} , eq.~(\ref{LimitingPhaseshift}).  Furthermore, $E^{(1,2)}$ denotes the 
contribution up to second order in $V$ from the renormalized Feynman
diagrams. Its explicit expression is displayed in Appendix \ref{app:Spherical}.
Finally, $E^{(3,4)}$ contains the third and fourth order
counterterm contribution combined with the divergences in the third
and fourth order Feynman diagrams that have been subtracted
from the momentum integral via $\delta_{\rm lim}$. Again, its 
explicit form can be found in Appendix \ref{app:Spherical}.

Thus we compute an exact, finite, renormalized, gauge-invariant 
effective energy functional, $E_{\rm eff}[\Phi, W_i]$, defined in eq.~(\ref{Eeff}), for the \HiggsGauge sector, with 
the fermion fields integrated out.   
\section{The Energy of a Fermionic Configuration}
We are interested in exploring the possibility of the emergence of a
stable, fermionic soliton in the \HiggsGauge sector of the theory as 
we increase the Yukawa coupling (thereby making the perturbative fermion heavier).  In the previous section we outlined
the procedure that allows us to calculate the effective energy when
the fermions are integrated out.  Now we analyze the minimum
additional energy required to associate unit fermion number with a particular
\HiggsGauge configuration $C$, where $C\equiv \{ \Phi, W_i \}$.  
First in
section \ref{The Fermion Number of a Configuration} we determine
the integer fermion number $F[C]$ of a configuration.  (This is subtle because we have to contend with the anomalous violation of fermion number).    Then we
occupy or empty levels of the single-particle Dirac Hamiltonian
to obtain the lowest energy state with net fermion number 1.
If $F[C]=0$, then the lightest 
positive bound state needs
to be filled and the occupation energy $E_{\rm occ}^{(1)} =
\epsilon_{\rm lowest}$, where the superscript `(1)' denotes that levels have been occupied/emptied to obtain fermion number 1.  If $F[C]=1$ then $E_{\rm occ}^{(1)}=0$ because $C$
is already fermionic, and so on.  Thus, the minimum total energy of a
single fermion associated with a configuration is
\begin{equation}
E\oneF[C] = E_{\rm cl}[C] + E_{\rm vac}[C]+ E_{\rm occ}^{(1)}[C] \, .
\label{allTheEs}
\end{equation}

In previous works, such as \cite{NolteKunz}, the fermion vacuum contribution was omitted from the above equation and stable
non-topological solitons were found.  We consider such calculations
inconsistent, because $E_{\rm occ}^{(1)}$ and $E_{\rm vac}$ are both order $\hbar$.
We will see explicitly in section \ref{Twisted Higgs} that $E_{\rm vac}$
makes a significant positive contribution when the Yukawa coupling is large enough that the perturbative fermion mass is comparable to the classical energy.

Since we refer to all these different energies frequently in the rest of the paper,  we summarize our notation in Table \ref{EnergiesTable}.\\
\begin{table}[hbt]
	\begin{tabular}{|c|l|} \hline
	\, $E_{\rm cl}$ \, & \, Classical Higgs and gauge energy, eq.~(\ref{Eclassical}) \\ \hline
	\, $E_{\rm vac}$ \, & \, Renormalized fermion vacuum energy, eqs.~(\ref{Evac1}, \ref{Evac2}) \\ \hline
	\, $E_{\rm eff}$ \, & \, One-fermion-loop effective energy, $E_{\rm cl} + E_{\rm vac}$ \\ \hline
	\, $E_{\rm occ}^{(m)}$ \, & \, Smallest occupation energy for fermion number $m$, eq.~(\ref{allTheEs}) \\ \hline  
	\, $E_{\rm eff}^{(m)}$ \, & \, Smallest effective energy in the fermion number $m$ sector, $E_{\rm eff} + E_{\rm occ}^{(m)}$ \\ \hline
	\end{tabular}
\caption{\label{EnergiesTable} Definitions of some of the energies which appear in our analysis.}
\end{table}
\subsection{The Fermion Number of a Configuration}
\label{The Fermion Number of a Configuration}
First we review properties of the \HiggsGauge configuration space 
and the classical energy functional defined on it. From the 
expression for $E_{\rm cl}$ in eq.~(\ref{Eclassical}), it follows 
that configurations
\bea
\Phi & = & vU^{(n)} \, , \nonumber \\
W_j  & = & \frac{i}{g}U^{(n)}\partial_j{U^{(n)}}^\dag \, , 
\label{Vacua3D}
\eea
have $E_{\rm cl}=0$ and we refer to them as {\it \classicalVacuum{} configurations}. Here $U^{(n)}$ is any map from $S^3$ to $SU(2)$ with winding 
number $n$.  We have compactified position space from ${\mathcal R}_3$ to $S_3$ by restricting the fields to be totally-trivial at spatial infinity.  \ClassicalVacuum{} configurations with the same winding number are equivalent 
under small (winding number 0) gauge transformations.
We use ${\mathcal C}^{(n)}$ to 
denote the homotopic class of \classicalVacuum{} configurations with 
winding number $n$.
Topologically inequivalent \classicalVacuum{} 
configurations are related by large (nonzero winding number)
gauge transformations. Along any continuous interpolation between two configurations, one in ${\mathcal C}^{(n)}$ and
 the other in ${\mathcal C}^{(m)}$ (with $n\ne m$), there exists a configuration, $C$, with maximum classical energy.  Since no $U^{(n)}$ can be
continuously deformed into any $U^{(m)}$, $E_{\rm cl}[C] > 0$.  The configuration corresponding to the minimax of these energies, when all interpolations are considered, is the {\em classical sphaleron}.  This is the weak sphaleron discussed in Sec.~\ref{sec:winding1}, where it was constructed using the topological prescription of non-contractible loops.  When the fermion vacuum energy is added to the classical energy to obtain the effective energy ($E_{\rm eff}=E_{\rm cl}+E_{\rm vac}$), not only does the magnitude of the minimax energy change,
but its location in configuration space shifts as well.  We therefore define the {\em quantum-corrected sphaleron} to
be the configuration that has the lowest of the maximum effective energies 
along all interpolations between topologically inequivalent \classicalVacuum{}
configurations. 

We associate any configuration $C$ with a unique class of
\classicalVacuum{} configurations by continuously deforming $C$ in the direction of the
negative gradient of the classical energy until we get a configuration in ${\mathcal C}^{(n)}$ for some $n = n(C)$.  We call ${\mathcal C}^{(n(C))}$ the 
{\it connected ${\mathcal C}$-class} of $C$ and we say that $C$ is 
in the winding number $n$ basin.  Note that the classical sphaleron and all configurations that descend to it are on the boundary between different basins.  Also note that we do not fix the gauge during the gradient descent, otherwise all configurations would descend to the same topological \classicalVacuum.

For any two configurations $C_1$ and $C_2$, we define the 
{\it spectral flow} $S[C_1, C_2]$ to be the number of 
eigenvalues (levels) of the single particle Dirac 
Hamiltonian that cross zero from above 
minus the number that cross zero from below 
along any interpolation from $C_1$ to $C_2$.  See \cite{Christ} for an early discussion on the relation between level crossings and the anomalous violation of fermion number.  The fermion number of a configuration $C$ is defined as
\begin{equation}
F[C] = S[C_1   , C]  
\label{fermionNumber}
\end{equation}
with $C_1 \in {\mathcal C}^{(n(C))}$.  Since the Dirac spectrum is gauge-invariant, $F[C]$ does not depend on which particular $C_1$ is chosen from the connected ${\mathcal C}$-class.  Moreover, $F[C]$ is gauge-invariant even under large gauge transformations.  Also, $F[C]$ does not depend on the chosen interpolation because the spectral flow is the same for all interpolations.  This definition of the fermion number can be readily understood for a C 
that has classical energy less than the energy 
of the classical sphaleron. A continuous interpolation from any configuration in ${\mathcal C}^{(n(C))}$ to $C$ preserves net fermion number 
because anomalous fermion number violations require the \HiggsGauge fields to 
cross the sphaleron barrier.  However, this does not preclude level crossings corresponding to pair production during the interpolation.  Thus, defining 
\classicalVacuum{} configurations to have zero
fermion number leads to eq.~(\ref{fermionNumber}).  Configurations
that have classical energies larger 
than the classical sphaleron are not separated by an
energy barrier from topologically inequivalent basins, so it is not
clear what their fermion number should be, although our  definition
does assign a unique $F$ to them.  

Having determined the fermion number of a configuration, we can use
eq.~(\ref{allTheEs}) to find $E\oneF $, the minimum effective energy of a configuration in the fermion number 1 sector.
\subsection{Stability of the Soliton}
We would like to know if there exists a configuration at
which the one-fermion effective energy functional, $E\oneF $, has a local
minimum.  This configuration would be a fermionic soliton.  We carry out a variational search, looking for a
configuration $C$ such that $E\oneF [C]<m_f$ and
$E\oneF [C]<E_{\rm q.s.}$, where $E_{\rm q.s.}$ is the effective energy of the quantum-corrected sphaleron.  The first condition ensures that
$C$ cannot simply decay into a configuration with
zero classical energy plus a perturbative fermion.  The second condition prevents $C$ from rolling over the  quantum-corrected sphaleron, giving up its
fermion number and then rolling down the $E\zeroF $ surface to a
\classicalVacuum{} configuration.  Finding a configuration with these
properties would guarantee the existence of a nontrivial
local minimum of $E\oneF $.
\section{The Search for a Spherical Soliton}
In this section we describe our search for the soliton.  
We first review the spherical ansatz for the gauge and Higgs fields.  
We then outline the restrictions imposed on the variational ansatze 
used to search for a soliton.  Finally, we report on our search within 
two physically motivated sets of ansatze: the ``twisted Higgs'' and ``paths 
over the sphaleron''.  Note that throughout this section the 
perturbative fermion mass is set to 1 so that energies 
and lengths are measured in units of 
$m_f$ and $1/m_f$, respectively. 
\subsection{The Spherical Ansatz}
\label{sec:SphericalAnsatz}
We only consider static gauge and Higgs fields in the spherical ansatz.   This enables us to expand the fermion S matrix in terms of partial waves labeled by the grand spin $G$, as described in Appendix \ref{app:Spherical}.  Our method for calculating the fermion vacuum energy requires
such an expansion. Under these restrictions (and in the $W_0 = 0$ gauge, which for smooth fields is obtained by a non-singular gauge transformation), the fields
can be expressed in terms of five real functions of $r$:
\begin{eqnarray}
W_i (\vec{x}) & = & -W^i(\vec{x})=\frac{1}{2g}\left[ a_1(r)\tau_j\hat{x}_j\hat{x}_i + 
\frac{\alpha(r)}{r}(\tau_i - \tau_j\hat{x}_j\hat{x}_i ) + 
\frac{\gamma(r)}{r}\epsilon_{ijk}\hat{x}_j\tau_k \right] \, , 
\nonumber \\
\Phi (\vec{x}) & = & v \left[ s(r) + ip(r)\tau_j\hat{x}_j \right] \, ,
\label{SphericalAnsatz}
\end{eqnarray}
where $\hat{x}$ is the unit three-vector
in the radial direction.  

The ansatz transforms under a $U(1)$ subgroup of
the full $SU(2)$ gauge symmetry, corresponding to elements of the 
form
\begin{equation}
g(\vec{x}) = e^{if(r)\tau_j\hat{x}_j/2} \, ,
\label{sphGauge}
\end{equation}
with $a_1$ transforming as a 1 dimensional vector field, $s+ip$
as a scalar with charge $1/2$, and $\alpha + i(\gamma-1)$ as a
scalar with charge $1$.  It is convenient to introduce the moduli $\rho, \Sigma$ and phases $\theta, \eta$ for the charged scalars:
\begin{equation}
-i\rho e^{i\theta} \equiv \alpha + i(\gamma-1)
\quad{\rm and}\quad
\Sigma e^{i\eta} \equiv s + ip   \, .
\label{polarfields}
\end{equation}
From now on we will specify a configuration using the five functions $a_1(r), \rho(r), \theta(r), \Sigma(r)$ and $\eta(r)$.  Regularity of $W_i(\vec{x})$ and $\Phi(\vec{x})$ at $\vec{x}=0$ requires that
\begin{eqnarray}
\rho(0) & = & 1 \, , \nonumber \\
\rho'(0) & = & 0 \, , \nonumber \\
\theta(0) & = & -2n_{\theta}\pi \, , \nonumber \\ 
a_1(0) & = & \theta '(0) \, ,  \nonumber \\
\mbox{either } \Sigma(0) & = & 0 \mbox{ or } \eta(0) = -n_{\eta}\pi \, .
\label{BoundCond0}
\end{eqnarray} 
Here $n_{\theta}, n_{\eta}$ are integers and primes denote derivatives with respect to the radial coordinate.

For the gauge transformation $g(\vec{x})$ in eq.~(\ref{sphGauge}) to be non-singular, we require $f(0)=-2n\pi$, where $n$ is an integer and we denote this boundary condition as a superscript: $f(r) \equiv f^{(n)}(r)$.  If $f^{(n)}(r)$ is restricted to be $0$ as $r\rightarrow\infty$ (which is equivalent to
$g(r\rightarrow \infty)=\ID$) then $n$ is the winding of the map
$g(\vec{x}):S^3\rightarrow SU(2)$.  So the topology of the \classicalVacuum{} configurations persists in the spherical ansatz. The classical energy of eq.~(\ref{Eclassical}) takes the form 
\begin{eqnarray}
E_{\rm cl} & = & 4\pi \int_0^\infty dr 
\left\{ \frac{1}{g^2}\left[\rho^{\prime^2}
+ \rho^2(\theta^\prime  - a_1)^2 
+ \frac{(\rho^2-1)^2}{2r^2}  \right]  \right. 
\nonumber \\ &&\hspace{2cm} 
\left. +\frac{1}{f^2}\left[ r^2 \Sigma^{\prime 2 }
+ r^2\Sigma^2(\eta^\prime  - \frac{1}{2}a_1)^2 
+\frac{r^2}{4}m_h^2(\Sigma^2-1)^2  
 \right. \right.  \nonumber \\ && \hspace{3cm} 
\left. \left. + \frac{1}{2}\Sigma^2\left( (\rho - 1)^2 
+ 4\rho^2\sin^2\frac{\theta-2\eta}{2} \right) \right]\right\} \, ,
\label{eclchiral}
\end{eqnarray} 
and winding number $n$ \classicalVacuum{} configurations of eq.~(\ref{Vacua3D}) now become 
\begin{eqnarray}
\rho(r)&=&1\, , \quad \Sigma(r)=1 \, , \nonumber \\
\theta(r)&=&f^{(n)}(r)\, , \quad
\eta(r)=\frac{f^{(n)}(r)}{2}\, , \quad
a_1(r)=f^{(n)\prime}(r)\, .
\label{Vacua1D}
\end{eqnarray}

We want the \HiggsGauge fields to have finite classical energy.   So we require a field configuration of the form eq.~(\ref{Vacua1D}) as $r\rightarrow\infty$, and the restriction that $f^{(n)}(\infty)=0$ uniquely specifies the boundary conditions on the fields at infinity.  At $\vec{x}=0$, the boundary conditions on $\rho$ (specified in eq.~(\ref{BoundCond0})) make the energy density finite and we do not require any additional constraints. 

The anomalous violation of fermion number is given by the anomaly equation
\begin{equation}
\partial_\mu \left(\overline{\Psi}\gamma^\mu\Psi\right) 
= \partial_\mu K^\mu \, ,
\end{equation}   
where $K^\mu$ is the Chern-Simons current.  It is useful to consider the charge associated with it:
\begin{eqnarray}
N_{\rm CS} & = & -\frac{g^2}{8\pi^2}\epsilon_{ijk}\int d^3x 
\tr \left(W_i\partial_j W_k - \frac{2}{3} i g W_i W_j W_k  \right) 
\nonumber \\
& = & \frac{1}{2\pi}\int_0^\infty dr 
\left[a_1+\rho^2(\theta^\prime - a_1)\right] \, .
\end{eqnarray}
This is a non-integer in general, and is equal to the integer winding
number of $f^{(n)}$ for the 
configurations of eq.~(\ref{Vacua1D}), and a half-integer for the
sphaleron \cite{KlinkhamerManton}.  Under a winding $n$ gauge transformation, $N_{\rm CS}
\rightarrow N_{\rm CS}+n$.  For background fields that interpolate between
topologically distinct \classicalVacuum{} configurations, the net
fermion number produced is given by the change in $N_{\rm CS}$.
\subsection{Restrictions on the Variational Ansatze}
\label{Restrictions}   
Our methods allow us to consider any static, spherically symmetric
configuration, $C$, in the \HiggsGauge sector, specified by 
the five real
functions $a_1(r)$, $\rho(r)$, $\theta(r)$, $\Sigma(r)$ and
$\eta(r)$.  In principle, we could numerically minimize the fermionic
energy, $E\oneF [C]$, in terms of the five functions and determine if
a soliton exists.  In practice however, that is not feasible.
So instead we limit ourselves to the variation of a few parameters in
ansatze motivated by physical considerations.

We restrict our variational ansatze to those that obey the 
above described boundary conditions at the origin 
and at infinity.
In addition, we restrict the
Higgs fields to lie within the chiral circle, $\Sigma(r)<1$, because
otherwise the effective potential is
unbounded from below.  (The leading terms in the derivative expansion of \eq{SeffFD} can be found in \cite{Ball:xg}).  Finally, the effective theory (obtained by integrating out the fermions) has a Landau pole in the ultraviolet, reflecting new dynamics at some
cutoff energy scale or equivalently at a minimum distance scale.
Configurations that are large compared to this distance scale 
are relatively
insensitive to the new dynamics at the cutoff, but smaller
configurations are sensitive.  For small widths and large couplings,
the Landau pole becomes significant and leads to unphysical negative
effective energies in eq.~(\ref{Eeff}) \cite{Ripka:ne,Hartmann:ai}.  We have to be wary of this in
estimating the reliability of our results.  See
\cite{DecouplingNoGauge} for more detailed discussions on the
effective potential and the Landau pole.
\subsection{Twisted Higgs}
\label{Twisted Higgs}
We first consider twisted Higgs configurations, with $n_{\eta}=1$ so that
$\eta(r)$ goes from $-\pi$ at $r=0$ to $0$ as
$r\rightarrow\infty$.  The other functions are trivial: $a_1(r)=0, \rho(r)=1, \theta(r)=0$ and $\Sigma(r)=1$.  If we smoothly interpolate from a
\classicalVacuum{} configuration (in the connected ${\mathcal C}$-class)
to such a configuration, we observe that one fermion bound state, that originates in the positive continuum, has its energy decrease sharply.  The wider the final twisted
Higgs configuration, the closer this level ends up to the negative
continuum.  At a width of order $1/m_f$, it has energy zero,
eliminating  any occupation energy contribution, $E_{\rm occ}^{(1)}$, to the
fermionic energy, $E\oneF $, associated with it.  The existence of
this level makes such twisted Higgs configurations attractive
candidates for the variational search.

We consider one such twisted Higgs configuration with a width
characterized by a variational parameter $w$,
\begin{equation}
\eta =  -\pi e^{-r/w}
\label{twistedHiggs}
\end{equation}
and add various perturbations to it.  For instance, one among many of our variational ansatze (in the $\theta=0$ gauge) is a four parameter ansatz with parameters $p_0,\ldots,p_3$: 
\begin{eqnarray}
\eta & = & -\pi e^{-r/w} + p_0 \frac{r/w}{1+(r/w)^2}e^{-r/w} \, , \nonumber \\
\Sigma & = & 1 + p_1 \frac{1}{1+(r/w)}e^{-r/w} \, , \nonumber \\
a_1 & = & p_2 \frac{r/w}{1+(r/w)^2}e^{-r/w} \, , \nonumber \\
\rho & = & 1 + p_3 \frac{(r/w)^2}{1+(r/w)^3}e^{-r/w} \, , 
\end{eqnarray}
where $-1<p_1<0$ (to keep the Higgs field within the chiral circle and
its magnitude positive) and $p_3>-5.23$ (to keep $\rho$ positive).  
For a prescribed set of theory parameters ($m_w, m_h$ and $f$) we determine the gauge coupling $g=\sqrt{2}m_w^{(0)}/v$ from the renormalization constraint eq.~(\ref{ModelParamsConstraint}).  We then vary the ansatz parameters ($w, p_0, \ldots , p_3$) 
to lower the fermionic energy $E\oneF $.  We find
that the gain in binding energy is insufficient to compensate for the
increase in the effective energy $E_{\rm eff}$.   Through all our
variations, $E\oneF $ is strictly greater
than $m_f$ and we find no evidence for a soliton.  The same result was
obtained in \cite{DecouplingNoGauge} without gauge fields, and the
extra gauge degrees of freedom do not seem to help in the twisted
Higgs ansatz.

\begin{figure}[htb]
\begin{center}
\includegraphics{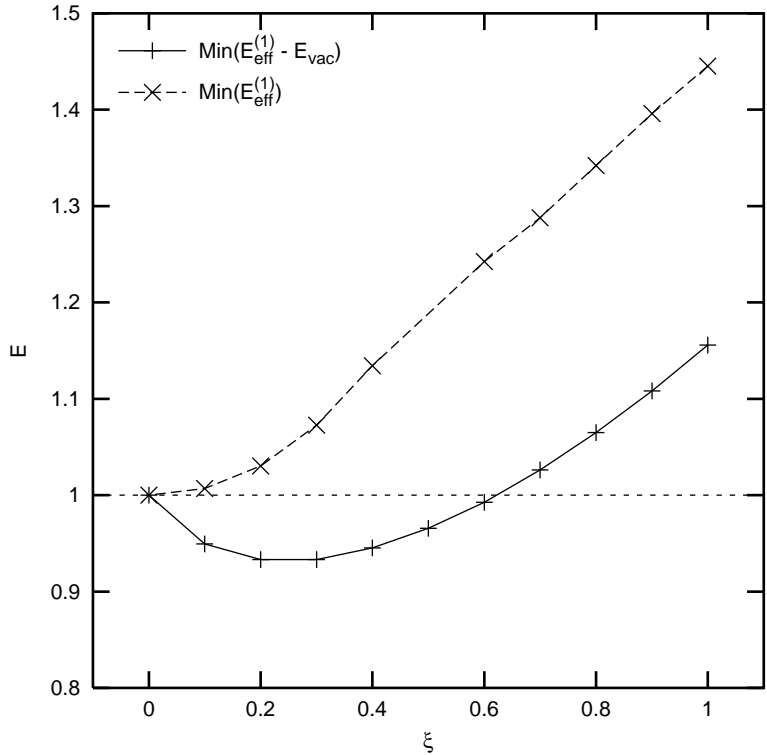}
\end{center}
\caption{\small Minimum fermionic effective energies (in units of $m_f$), with as well as without $E_{\rm vac}$ contributions, along the interpolation in eq.~(\ref{twistedHiggsInterpolation}). }
\label{soliton}
\end{figure}

We also find that the fermion vacuum contribution, $E_{\rm vac}$, to the fermionic energy, $E\oneF $ destabilizes would-be solitons.  Consider a linear interpolation from the trivial \classicalVacuum{} configuration to the twisted Higgs configuration in eq.~(\ref{twistedHiggs}) with gauge fields set to zero.  We introduce the interpolating parameter $\xi$ which goes from 0 to 1:
\begin{equation}
\Sigma e^{i\eta} = 1 - \xi + \xi \exp\left( -i\pi e^{-r/w} \right) \, .
\label{twistedHiggsInterpolation}
\end{equation}  
We choose the Yukawa coupling to be $f=10$ and the Higgs mass to be $v/\sqrt{2}$.  Since the gauge fields are trivial, the classical energy as well as the Dirac spectrum are independent of the gauge coupling $g$ and the gauge bosons mass $m_w$.  For each value of $\xi$, we compute $E\oneF $ and $E\oneF  - E_{\rm vac}$ for different values of the width parameter $w$.  In Fig. \ref{soliton} we plot $E\oneF $ and $E\oneF  - E_{\rm vac}$ as functions of $\xi$, choosing the width at every point to minimize the energy (we do not allow $w$ to be less than 1 so that we remain relatively insensitive to the Landau pole).  For all points on the plot there is no spectral flow and so $E\oneF  = E_{\rm cl} + \epsilon_{\rm lowest} + E_{\rm vac}$ in accordance with eq.~(\ref{allTheEs}), where $\epsilon_{\rm lowest}$ is the smallest positive bound-state energy in the Dirac spectrum.  If $E_{\rm vac}$ is omitted, for $0<p<0.6$ we have configurations that have fermionic energies lower than the mass of the perturbative fermion ($m_f = 1$ in our units).  These configurations indicate the existence of a local minimum on the $E\oneF -E_{\rm vac}$ surface which would be a soliton.  The $E_{\rm vac}$ contribution, however, raises the energies of the configurations to above $m_f$ as shown in the figure, and the would-be solitons are destabilized.
\subsection{Paths over the Sphaleron}
\label{Paths over the Sphaleron}
The gauge fields introduce another mechanism for strongly binding a
fermion state: there is a zero mode in the background of the
sphaleron \cite{Boguta,Ringwald}.  Along an interpolation of the background fields from a
configuration in ${\mathcal C}^{(n)}$ to a configuration in ${\mathcal C}^{(n+1)}$, a
fermion level leaves the positive continuum, crosses 
zero from above and
finally enters the negative continuum.  The lowering of the occupation
energy, $E_{\rm occ}^{(1)}$, as we approach zero from above is offset by
the rising effective energy $(E_{\rm eff}=E_{\rm cl}+E_{\rm vac})$, so
we must investigate whether the former can dominate the latter.  We also use
such interpolations to study the effects of a large Yukawa coupling on the sphaleron.  We approximate the
quantum-corrected sphaleron by minimizing the effective energy
barrier between topologically inequivalent \classicalVacuum{}
configurations, with respect to 
the variational parameters of our interpolations.  We also 
investigate the possible emergence of new barriers in the 
one-fermion energy surface when the perturbative fermion becomes heavier than the 
quantum-corrected sphaleron.  
These last two phenomena affect the stability of the heavy 
fermion and in some models may be significant for baryogenesis.
  
We make the following choices for the theory parameters: we fix the
Yukawa coupling at $f=10$, which is large enough that fermion effects
are significant, but small enough to prevent our configurations
from being affected by the Landau pole.  Indeed we encounter no
negative energy instabilities in our computations of 
$E_{\rm vac}$ for this coupling.  
We keep the Higgs mass fixed at $v/\sqrt{2}$,
which corresponds to $m_h = \frac{m_f}{\sqrt{2}f} \approx 0.07 m_f$.
We choose $g$ to keep the quantum-corrected sphaleron energy
comparable to $m_f$, since it is plausible that the mass of the
sphaleron sets the scale for decoupling.  If the sphaleron is
too heavy compared to the fermion, the binding energy gained by
the level crossing would be washed out by the effective energy of the
sphaleron.  For $g=6.5$ our best approximation to the
quantum-corrected sphaleron is approximately degenerate with the
fermion. When the gauge coupling is given, the mass $m_w$ of the gauge boson is determined from the
renormalization constraint eq.~(\ref{ModelParamsConstraint}):
$m_w \approx 0.63 m_f$ for $g=6.5$ and $m_w\approx0.98 m_f$ for $g=10$.
These theory parameters are of course large deviations from the
Standard Model parameters. We exaggerate them to 
see the effects of the heavy perturbative fermion.  Another concern is
that for large $g$, we should consider quantum fluctuations of the \HiggsGauge fields. However, we believe that anomaly cancellation drives the creation of a soliton, which would suggest that the fermion vacuum contains the essential physics, and our methods allow us to exactly
compute this contribution to the energy for any Yukawa coupling.

First we consider a linear interpolation between a 
winding-0 and a winding-1 \classicalVacuum{} 
configuration. The interpolation parameter $\xi$ goes from 0 to 1:
\begin{eqnarray}
\Phi & = & v(1-\xi)\ID + \xi v U^{(1)} \, , \nonumber \\
W_j & = & \xi \frac{i}{g}U^{(1)}\partial_j{U^{(1)}}^{\dag} \, .
\label{linearpath1}
\end{eqnarray}
In the spherical ansatz, $U^{(1)}(\vec{x})=g(\vec{x})$ is specified by a single 
function, as in eq.~(\ref{sphGauge}), that we choose to be 
\begin{equation}
f^{(1)}(r) = -2\pi e^{-r/w} \, ,
\label{linearpath2}
\end{equation}
where $w$ characterizes the width of the configuration.  
We interpolate from $\xi=0$ (the trivial configuration with $N_{\rm CS}=0$) to $\xi = 1/2$ (a configuration with
$N_{\rm CS}=1/2$ in the presence of which the fermion has a zero mode) and
vary $w$ along the interpolation to minimize $E\oneF $.  This gives
an upper bound on the minimum $E\oneF $ as a function of $N_{\rm CS}$.
For $N_{\rm CS}=1/2$, this is an upper bound on the quantum-corrected
sphaleron energy as well, because $E\oneF  = E_{\rm eff}$ in the presence of
a fermion zero mode (the occupation energy is then 0).  (Our numerical methods do not allow us to consider 
$N_{\rm CS}$ exactly equal to $1/2$, because the Higgs
magnitude vanishes at $r=0$ and the second order Dirac equations
develop a singularity as discussed in Appendix \ref{app:Spherical}.  But we can
compute $E\oneF $ very close to $N_{\rm CS}=1/2$.)  An exploration of
$N_{\rm CS}$ between 0 and 1/2 is sufficient to map to all values of
$N_{\rm CS}$, since configurations with $N_{\rm CS}$ between 1/2 and 1 are 
obtained by
charge conjugation, and configurations with $N_{\rm CS}<0$ or $N_{\rm CS}>1$ are
large-gauge-equivalent to configurations with $N_{\rm CS}$ between 0 and 1.

We also consider an instanton-like configuration where the Euclidean
time $\xi=x_4$ is the interpolation 
parameter (which varies from $-\infty$ to $\infty$) between two topologically
inequivalent \classicalVacuum{} configurations:
\begin{eqnarray}
W_\mu & = & h(r,\xi)\frac{i}{g} 
U_{\rm inst}(\vec{x},\xi)\partial_\mu 
U^{\dag}_{\rm inst}(\vec{x},\xi) \, , \nonumber \\
\Phi & = & v \sqrt{h(r,\xi)}\,U_{\rm inst}(\vec{x},\xi) \, ,
\label{definst1}
\end{eqnarray}
where 
\begin{equation}
U_{\rm inst}(\vec{x},\xi)=
\frac{\xi + i \tau_{j}x_{j}}{\sqrt{r^2+\xi^2}}
\label{definst2}
\end{equation}
is the canonical winding-1 map from $S^3$ (space-time infinity) to
$SU(2)$.  Furthermore $h$ is a function of the 
Euclidean space-time radius ($\sqrt{r^2+\xi^2}$)
and goes from 0 to 1 as this radius goes from 0 to $\infty$.
't Hooft's electroweak instanton \cite{'tHooft} is
constructed as a self-dual gauge field configuration
in the topological charge one sector,
and a Higgs field configuration that minimize the covariant kinetic
term in the Lagrangian density.  (In Sec.~\ref{sec:winding1}, this was constructed for the SU(2) theory with no Higgs fields.)  This gives
\begin{equation}
h(r,\xi)=\frac{r^2+\xi^2}{r^2+\xi^2+w^2}
\end{equation}
for any width $w$ (the classical theory with no Higgs field is
scale-invariant).  
We modify this radial function to exponentially
approach its asymptotic value of 1, so that the potential in our Dirac
equation falls off fast enough to have a 
well-defined scattering problem (as described in Appendix \ref{app:Spherical}).   We choose
\begin{equation}
h(r,\xi) = 1-e^{-(r^2+\xi^2)/w^2} \, .
\label{definst3}
\end{equation}
This choice does not minimize any part of the classical Euclidean
action (in the topological charge one sector). Since we are interested  in minimizing $E\oneF $,
which has fermion vacuum energy and occupation energy contributions in
addition to the \HiggsGauge sector classical energy, we do not need our configurations to minimize the classical energy.  In fact, as we
describe later, the configurations that minimize $E\oneF $ are
rather different from those that minimize $E_{\rm cl}$.

In order to compute the Dirac spectrum for this background 
using the methods described in Appendix \ref{app:Spherical}, 
we gauge transform 
to $W_0=0$ and 
$\lim_{r\rightarrow\infty}\eta=\lim_{r\rightarrow\infty}\theta=0$ 
using the transformation function
\begin{equation}
f(r,\xi) = \int_{-\infty}^{\xi}d\xi^\prime \frac{2r}{r^2+(\xi^\prime)^2}
h(r,\xi^\prime) - 2\pi
\end{equation}
in eq.~(\ref{sphGauge}).
Finally, we go from $\xi = -\infty$ (the trivial configuration with 
$N_{\rm CS}=0$) to $\xi=0$ (a configuration with $N_{\rm CS}=1/2$ in the 
presence of which the fermion has a zero mode) and for each $\xi$ 
we choose the $w$ that minimizes $E\oneF $.

\begin{figure}[t]
\begin{center}
\includegraphics{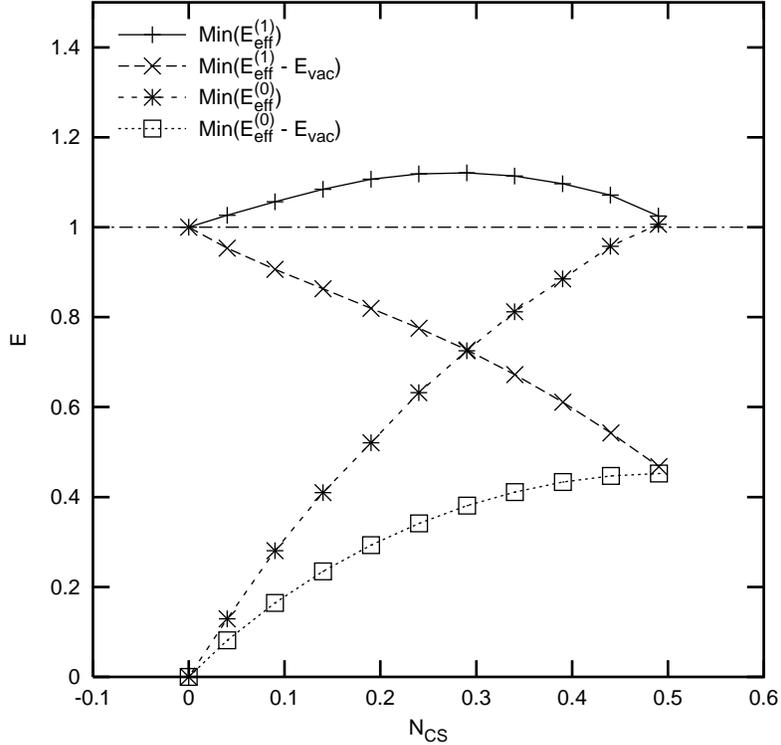}
\end{center}
\caption{\small Minimum effective energies (in units of $m_f$) along the linear path in eqs.~(\ref{linearpath1}, \ref{linearpath2}), in both the zero-fermion and one-fermion sectors (with as well as without the $E_{\rm vac}$ contributions).}
\label{quantum}
\end{figure}
In Fig.~\ref{quantum} we plot the minimum effective energies in both
the zero-fermion sector ($E\zeroF $) 
and the one-fermion sector
($E\oneF $), minimized within our variational ansatz for the linear
interpolation, as
functions of the Chern-Simons number, $N_{\rm CS}$ (see eqs.~(\ref{linearpath1},\ref{linearpath2})).  
As mentioned before, we fix the theory
parameters at a Yukawa coupling of $f=10$, a gauge coupling of $g=6.5$
and a Higgs mass of $m_h\approx 0.07 m_f$.
These parameters determine the mass of the gauge bosons to be $m_w
\approx 0.63 m_f$ from the renormalization constraint,
eq.~(\ref{ModelParamsConstraint}).  To isolate and highlight the
contribution of the fermion vacuum energy, $E_{\rm vac}$,
we also plot the energies minimized with $E_{\rm vac}$ subtracted.

First consider the zero-fermion sector with the two curves $E\zeroF $
and $E\zeroF  - E_{\rm vac}$.  For all points on the plot there is
no spectral flow and so $E\zeroF  - E_{\rm vac}= E_{\rm cl}$.  At
$N_{\rm CS}=0$, both $E\zeroF $ and $E_{\rm cl}$ are minimized
at the trivial \classicalVacuum{} configuration, eq.~(\ref{Vacua1D}) with $f^{(0)}(r)=0$.  At $N_{\rm CS}=1/2$, $E\zeroF $
is minimized at the quantum-corrected sphaleron while $E_{\rm cl}$
is minimized at the classical sphaleron.  Within our variational
ansatz, we find the parameters that minimize $E\zeroF $ at
$N_{\rm CS}=1/2$ are different from those that minimize $E_{\rm cl}$.  So
our approximation to the quantum-corrected sphaleron is distinct from
our approximation to the classical sphaleron.  Moreover, the fermion
vacuum energy correction to the sphaleron turns out to be
rather large.  
Our classical sphaleron has an energy of $0.45 m_f$ (which
agrees well with the numerical estimate of $E=1.52\frac{4\pi
v\sqrt{2}}{g}$ in \cite{KlinkhamerManton}), while our
quantum-corrected sphaleron has an energy of $1.02 m_f$. 

Next consider the $E\oneF $ and $E\oneF  - E_{\rm vac}$ plots in
the one-fermion sector in Fig.~\ref{quantum}.  
Again, for all points on the plot there is no
spectral flow and so $E\oneF  = E_{\rm cl}+ \epsilon_{\rm lowest} + E_{\rm vac}$
in accordance with eq.~(\ref{allTheEs}), where $\epsilon_{\rm lowest}$ is the
smallest positive bound-state energy in the Dirac spectrum.  Since the
classical sphaleron has an energy much smaller than the 
perturbative fermion mass, one would 
expect that the perturbative
fermion would have an unsuppressed decay mode over the
sphaleron, as first pointed out by Rubakov in \cite{Rubakov}. 
The
$E\oneF -E_{\rm vac}$ curve indeed displays this decay path.  The
fermion vacuum energy modifies things in two crucial
ways.  First, the fermion quantum corrections to the sphaleron raise
its energy.  (The theory parameters have been chosen so that the quantum-corrected sphaleron is approximately degenerate with the fermion, as mentioned before.)  So
the threshold mass is significantly increased.  Second, in the plot of
$E\oneF $ we observe that there is an energy barrier 
between the fundamental fermion
and the quantum-corrected sphaleron.  This indicates that even when
the fermion becomes heavier than the sphaleron, there might exist a
range of masses for which the decay continues to be exponentially
suppressed (since it can proceed only via tunneling).

\begin{figure}[htb]
\begin{center}
\includegraphics{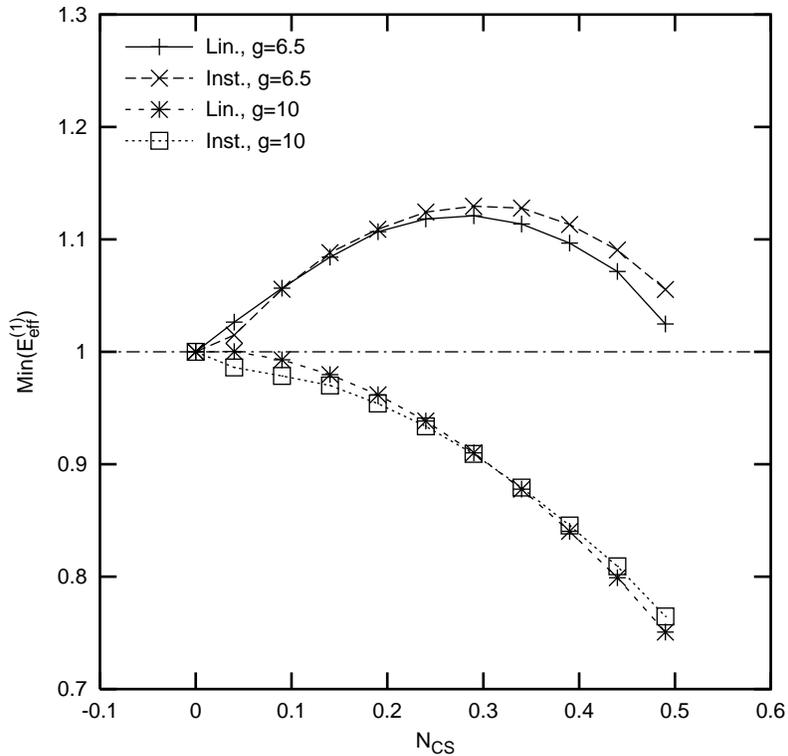}
\end{center}
\caption{\small Minimum $E\oneF $ (in units of $m_f$) for paths from 
a \classicalVacuum{} configuration to the sphaleron. Curves 
denoted 'Lin.' refer to the linear path in eqs.~(\ref{linearpath1}, \ref{linearpath2})
while those labeled 'Inst.' are associated with the instanton 
path, eqs.~(\ref{definst1},\ref{definst2},\ref{definst3}).}
\label{pathOverSph}
\end{figure}
In Fig.~\ref{pathOverSph} we restrict our attention to the 
one-fermion sector and consider the linear interpolation and the
instanton-like interpolation.  In addition to $g=6.5$ we consider $g=10$, which
corresponds to $m_w \approx 0.98 m_f$.  For each of these two gauge
couplings, we plot the effective energy minimized within our ansatze
as a function of $N_{\rm CS}$ for the two interpolations.

The two seemingly different interpolating configurations produce very
similar minimum $E\oneF $ curves.  Furthermore,
when we enlarge the variational ansatze in our interpolations, we
are unable to reduce the energies by any significant amount.  Thus we
speculate that the plotted curves may be close to tight upper bounds
on the true minimum $E\oneF $ curve.  This is the justification for
considering only the linear interpolation in Fig.~\ref{quantum} and
taking the evidence for the emergence of a new barrier and the
significant energy change of the sphaleron seriously.

Note that as the gauge coupling increases, lowering the energy
of the quantum-corrected sphaleron, the barrier between the
fundamental fermion and the sphaleron does not persist indefinitely.
We observe that for $g=10$, when $m_f$ is approximately 1.3 times our
quantum-corrected sphaleron, there is no barrier and the decay mode
is finally unsuppressed.
 
Just as in the case of the twisted-Higgs variational ansatz, in our
ansatze of paths from a \classicalVacuum{} configuration 
to the sphaleron,
we have not found a configuration with the associated fermion energy
lower than both the perturbative fermion and the quantum-corrected
sphaleron.  Thus, we find no evidence for the existence of fermionic
solitons in the low energy spectrum of the Standard Model within our
 ansatze.
\section{Conclusions and Discussion}
\label{sec:QuantumSolitonsConclusion}
We have explored quantum effects of a (hypothetical) heavy fermion on a chiral gauge theory (the electroweak theory, ignoring the hypercharge gauge fields) for certain parameters.  The quantum-corrected sphaleron is heavier than the classical sphaleron by an energy of the order of the perturbative fermion mass.  This higher barrier suppresses fermion number violating processes.  We also observe the creation of an energy barrier between the perturbative fermion and the sphaleron, so that a fermion with energy
slightly above the sphaleron can still only decay through tunneling.  This new barrier exists only for an intermediate range of perturbative fermion masses, and a heavy enough perturbative fermion is indeed unstable.  We do not, however, find evidence for the existence of a soliton in the spectrum of the theory.  The fermion vacuum contribution seems to destabilize any would-be solitons.  

In hindsight it is not too surprising that we do not find a quantum soliton within the spherical ansatz.  As mentioned in Sec.~\ref{sec:Idea}, the existence of such solitons would maintain anomaly cancellation when fermions are decoupled from the electroweak theory.  Without the hypercharge gauge field, the only anomaly is Witten's global anomaly \cite{Witten} due to topologically non-trivial maps from $S_4$ to $SU(2)$.  However, in the spherical ansatz, the theory reduces to a $U(1)$ theory in which $\Pi_3(SU(2))$ persists as $\Pi_1(U(1))$ (and so we have topologically inequivalent vacua and the weak sphaleron) but there is no remnant of $\Pi_4((SU(2))$.  So, the quantum soliton that could resolve the decoupling puzzle probably lies outside the ansatz.

One route beyond the spherical ansatz is the use of non-trivial $\Pi_4(SU(2))$ to construct non-contractible loops and corresponding novel classical sphalerons (see Sec.~\ref{sec:windingNT}).  We expect fermion zero-modes in these sphaleron backgrounds and thus the required tight-binding-mechanism exists.  Moreover, since the non-trivial topology that gives rise to Witten's anomaly is built into the construction of such configurations, they are promising candidates for objects that maintain anomaly cancellation in the decoupled theory.  However, an evaluation of the stability of such a state would require a
calculation of the one-loop effective energy, which is more difficult in the
absence of spherical symmetry.  We have not pursued this direction any further.

Another interesting non-spherical configuration is the W-string solution discussed in Sec.~\ref{sec:winding1}.  Our motivation goes beyond fermion decoupling.  Stable electroweak strings are a crucial ingredient in a scenario for electroweak baryogenesis even without a first order electroweak phase transition\cite{Brandenberger}.  It is likely that the classically unstable W-string could be quantum-stabilized by having it carry fermion number.  The string configuration gives rise to fermion zero modes and it is possible that that several massless quarks and leptons are trapped along the string in a stable manner.  We discuss this in greater detail in the next chapter.
\chapter{Electroweak Strings}
\label{chap:ElectroweakStrings}
In this chapter we begin by reviewing the \HiggsGauge sector of the electroweak theory, after reintroducing the hypercharge gauge fields that we have ignored so far.  We then describe the electroweak string solutions in the theory by constructing them as embedded Nielsen-Olesen vortices.  We explain the significance of these objects and how they may be stabilized by trapping heavy quarks.  An investigation of this stabilizing effect of quarks requires us to compute fermion vacuum energies in the backgrounds of strings.  Our computational method involves scattering data of fermions.  However, the string backgrounds generate long-range potentials leading to difficulties with scattering theory.  In order to isolate and examine these issues, we study the simpler problem of magnetic flux tubes in QED, as a stepping stone to the electroweak strings analysis.

\section{The \HiggsGauge Sector}
In Sec.~\ref{sec:HiggsGauge}, we described a simplified version of the bosonic sector of the electroweak theory in which we ignored the hypercharge gauge fields, $B_\mu$.  Now we re-introduce them and consider the full $SU(2) \times U(1)$ gauged Higgs theory.  It is defined by the action
\bea
S_H [ \phi, W_\mu, B_\mu ] & = & \int d^4 x \left[ - \frac{1}{4} B_{\mu\nu}B^{\mu\nu} -\frac{1}{2} \tr \left(W^{\mu\nu}W_{\mu\nu}\right) \right. \nonumber \\*
 & & + \left. \left(D^{\mu}\phi \right)^{\dag} D_{\mu}\phi - \lambda \left( \phi^{\dag} \phi - v^2 \right)^2 \right] \, .   
\eea
The non-Abelian field strength tensor, $W_{\mu \nu}$, is defined in Sec.~\ref{sec:HiggsGauge}.  The Abelian field strength tensor is
\be
B_{\mu\nu} = \partial_\mu B_\nu - \partial_\nu B_\mu \, .
\ee
The Higgs doublet is
\be
\phi =  \Biggl( \begin{array}{c} \phi_+ \\[-0.1in] \phi_0 \end{array} \Biggr) \, ,
\ee
and the covariant derivative is
\be
D_\mu \phi = \left( \partial_\mu  - i g W_\mu \Phi  - i g' Y_\phi B_\mu \right) \phi \, , 
\ee
where the Higgs hypercharge is $Y_\phi = \half$.  Under a gauge transformation $V = V_1 \times V_2$, where $V_1 = \exp ( i \beta(x) Y_\phi ) \in U(1)$ and $V_2 = \exp ( i \alpha^a(x) \tau^a/2 ) \in SU(2)$,
\bea
\phi & \rightarrow & V_1 V_2 \phi \, , \nonumber \\
B_\mu & \rightarrow & B_\mu + \frac{1}{g'} \partial_\mu \beta \, , \nonumber \\
W_\mu & \rightarrow & V_2 \left( W_\mu + \frac{i}{g} \partial_\mu \right) V_2^\dag \, .
\label{eq:GaugeTransform} 
\eea  

After spontaneous symmetry breaking, when the Higgs acquires a vev ($\langle \Phi \rangle = v \ID$), the gauge fields eat the three Goldstone bosons and acquire masses.  The mass eigenstates of the gauge fields (in unitary gauge) are
\bea
A_\mu & = & \sin\theta_w W_\mu^3 + \cos\theta_w B_\mu \, , \nonumber \\
Z_\mu & = &  \cos \theta_w W_\mu^3 - \sin\theta_w B_\mu \, , \nonumber \\
W_\mu^{\pm} & = & \frac{1}{\sqrt{2}} \left( W_\mu^1 \mp W_\mu^2  \right) \, , 
\eea
where the mixing angle is defined as
\begin{equation}
\tan \theta_w = \frac{g'}{g} \, .
\end{equation}
The photon $A_\mu$ is massless (due to the unbroken $U(1)$ subgroup of electromagnetism), the neutral $Z_\mu$ boson has mass $m_Z = \frac{1}{\sqrt{2}} \sqrt{(g')^2+g^2} v$, and the two charged W bosons $W_\mu^{\pm}$ are degenerate with mass $m_W = \frac{1}{\sqrt{2}} g v$.  The neutral Higgs field $\phi_0$ has mass $m_H = 2 v \sqrt{\lambda}$.  In terms of the mass eigenstates, the action becomes
\bea
S_H [ \Phi, Z_\mu, W_\mu^\pm, A_\mu ] & = & \int d^4 x \left[ - \frac{1}{4} A_{\mu\nu}A^{\mu\nu} - \frac{1}{4} Z_{\mu\nu}Z^{\mu\nu} - \frac{1}{4} (W^+)^{\mu\nu}W^-_{\mu\nu} \right. \nonumber \\
 & & \left. + \left(D^{\mu} \phi \right)^{\dag} D_{\mu}\phi - \lambda \left( \phi^{\dag}\phi - v^2 \right)^2 \right] \, .   
\eea
The field strength tensors are Abelian, the covariant derivative is
\be
D_\mu \phi = \left( \partial_\mu - i \sqrt{g^2 + g'^2} Q_Z Z_\mu - i e Q_A A_\mu - i \frac{g}{\sqrt{2}} \left( W_\mu^+ T^+ + W_\mu^- T^- \right) \right) \phi \, ,
\ee
where
\bea
e & = & \frac{g g'}{\sqrt{g^2+g'^2}} \, , \nonumber \\
Q_A & = & T^3 + Y \, , \nonumber \\
Q_Z & = & \cos^2(\theta_w) T^3 - \sin^2 (\theta_w) Y = T^3 - \sin^2(\theta_w) Q_A \, , \nonumber \\
T^\pm & = & \half (\tau^1 \pm i \tau^2) \, , \nonumber \\
T^3 & = & \half \tau^3 \, .
\label{eq:charges}
\eea

We use the experimental values for the mixing angle and the fine structure constant ($\sin^2\theta_w = 0.23, \alpha = 1/137$) to determine
\begin{eqnarray}
g & = & \frac{e}{\sin\theta_w} \approx 0.63 \, , \nonumber \\
g' & = & \frac{e}{\cos\theta_w} \approx 0.31 \, .
\end{eqnarray}
The observed Z mass of 90 GeV fixes the vev at $v \approx$ 177 GeV.  Finally, choosing the Higgs mass to be 115 GeV we get the self-coupling $\lambda \approx 0.11$.

\section{The String Solutions}
Now we consider static configurations that are trivial in one direction, say $z$, and have finite energy per unit length in a localized region of the $x-y$ plane.  Since the region in space in which the energy is localized extends in the $z$ direction, such configurations are called {\em strings}.  They are also called {\em cosmic strings} to distinguish them from fundamental strings in string theory.  Finite energy requires the configuration to be pure gauge at planar infinity.  Let $(\rho, \azAngle)$ denote the radial and angular coordinates on the plane.  If $V_1, V_2$ are elements of $U(1), SU(2)$ respectively, then the asymptotic configuration at $\rho \rightarrow \infty$ is defined by a map $V_1(\azAngle) V_2(\azAngle)$ from $S_1$ (the planar boundary spanned by $\azAngle$) to the gauge group:
\bea
\phi^{(\infty)} & = & V_1 \times V_2 \;  \biggl( \begin{array}{c} 0 \\[-0.15in] v \end{array} \biggr) \, , \nonumber \\
B_\mu^{(\infty)} & = & \frac{i}{g' Y} V_1 \partial_\mu V_1^\dag \, , \nonumber \\
W_\mu^{(\infty)} & = & \frac{i}{g} V_2 \partial_\mu V_2^\dag \, . 
\eea  
If the gauge group had been $U(1)$, then these maps would be characterized by winding numbers ($\Pi_1 (U(1)) = {\mathcal Z}$), and we would obtain topologically stable Nielsen-Olesen vortex solutions \cite{Nielsen:1973cs}.  (They are called vortices because the gauge fields are azimuthally directed, as we shall see later.)  However, the gauge group is $SU(2) \times U(1)$.  But we can still embed Nielsen-Olesen solutions in this theory by considering maps from the planar boundary to various $U(1)$ subgroups of the gauge group that break completely when the electroweak symmetry breaks.  These solutions are unstable.  This is readily seen if the string solutions are constructed using non-contractible loops of configurations, as was done in the $SU(2)$ theory in Sec.~\ref{sec:winding1}.  In that case it was shown that the strings are sphalerons with two directions of instability.  For a comprehensive review of electroweak strings, see \cite{VachaspatiReview}.  

\subsection{The Z-string}
First consider the $U(1)$ subgroup generated by the charge $Q_Z$ associated with the $Z_\mu$ gauge bosons, as defined in \eq{eq:charges}.  The maps that specify the asymptotic pure-gauge configurations are 
\be
V(\azAngle) = e^{-2 i n \azAngle Q_Z} \, , 
\ee  
where $n$ is an integer (otherwise the Higgs fields are not single-valued under $\azAngle \rightarrow \azAngle + 2 \pi$).  Then, the winding $n$ Z-string solution is 
\bea
\phi_0 & = & f_H(\rho) v e^{i n \azAngle} \, , \nonumber \\
\vec{Z} & = & f_G(\rho) \frac{2 n}{\sqrt{g^2+g'^2} \rho} \hat{\azAngle} \, ,  
\eea  
with all the other fields vanishing.  The unit vector in the azimuthal direction is
\be
\hat{\azAngle} = \left( \begin{array}{c} - \sin \azAngle \\ \cos \azAngle \end{array} \right) \, . 
\ee
The radial functions $f_H, f_G$ vanish at the origin (for smooth fields) and approach 1 at infinity (for finite energy per unit length).  They satisfy the Nielsen-Olesen differential equations
\bea
0 & = & f_H'' + \frac{1}{\rho} f_H' - \frac{n^2}{\rho^2}f_H ( 1 - f_G )^2 + 2 \lambda v^2 f_H (1 - f_G^2) \, , \nonumber \\
0 & = & f_G'' - \frac{1}{\rho} f_G' + \half (g^2+g'^2) v^2 f_H^2 (1-f_G) \, , 
\eea  
and can be determined numerically.

The Z-string is a vortex configuration carrying Z magnetic flux of $4 \pi n / \sqrt{g^2+g'^2}$.  The magnetic pressure tends to spread out the flux.  However, the Higgs condensate is suppressed in the region of the Z magnetic field, and this tends to compress the flux.  The two competing effects result in an equilibrium size for the Z-string.  However, had we considered the unbroken electromagnetic $U(1)$ subgroup generated by $Q_A$, the Higgs condensate would be unaffected in regions of magnetic field, and the configuration would be driven to a singularity.  (The energy could be made closer to zero by increasing the size of the area of the flux and independently decreasing the area in which the Higgs condensate is suppressed.)  

\subsection{The W-strings}
Consider the 1-parameter family of $U(1)$ subgroups generated by
\be
T_\xi = \cos(\xi) T^1 + \sin(\xi) T^2 \, ,
\ee
where $\xi$ is between 0 and $\pi$.  The maps that specify the asymptotic pure-gauge configurations are
\be
V_\xi (\azAngle) = e^{ i n \azAngle T_\xi } \, 
\ee
where $n$ is an integer.  This gives a 1-parameter family of winding $n$ W-string solutions:
\bea
\phi & = & f_H(\rho) v \left( \begin{array}{c} i e^{-i \xi} \sin(n\azAngle) \\ \cos(n \azAngle) \end{array} \right) \, , \nonumber \\
\vec{W}^1 & = & - \cos(\xi) f_G(\rho) \frac{2 n}{g} \hat{\azAngle} \, , \nonumber \\
\vec{W}^2 & = & - \sin(\xi) f_G(\rho) \frac{2 n}{g} \hat{\azAngle} \, .
\eea
The radial functions satisfy the Nielsen-Olesen differential equations.

In analogy with the Z-string, W-strings are vortex configurations carrying $W^{1,2}$ magnetic flux of $4 \pi n /g$.  Their size is stabilized by the competition between the magnetic pressure and the potential energy.

\section{Fermions on Strings}
Although the electroweak strings are classically unstable, they may be stabilized by their interactions with quarks and leptons.  Since the Higgs condensate is suppressed in the core of the string, the fermions become massless on the string, and could resist its decay.  Of course, as discussed in Chap.~\ref{chap:QuantumSolitons}, to be consistent to order $\hbar$, the fermion vacuum energy contribution must also be included in the stability analysis.  The minimum total energy associated with $N_f$ fermions (with perturbative mass $m_f$)  trapped on the string is
\be
E_{\rm eff}^{(N_f)} = E_{\rm cl} + E_{\rm occ}^{(N_f)} + E_{\rm vac} \, , 
\ee
with the different energies above defined in Table \ref{EnergiesTable}.  It is conceivable that for some value of $N_f$,
\be
E_{\rm eff}^{(N_f)} < m_f N_f \, , 
\ee
leading to stable multi-quark objects with possibly rich phenomenology.

Electroweak strings, if stabilized by the fermion binding mechanism, could have profound cosmological consequences.  A gas of strings would have negative pressure associated with it (the strings tend to contract in length) and could contribute to the dark energy that is required to explain the recently observed cosmic acceleration \cite{Perlmutter:1998np,Riess:1998cb}.  Also, as pointed out by Nambu \cite{Nambu}, Z-strings are expected to terminate in monopole-antimonopole pairs, which could give rise to a primordial magnetic field.  Furthermore, a network of stable strings at the electroweak phase transition would provide a scenario for electroweak baryogenesis without requiring a first-order phase transition \cite{Brandenberger}.  As the strings contract, they provide out-of-equilibrium regions, and the core of the string has copious baryon number violation due to the suppressed Higgs condensate.  This is an alternative to the usual idea of bubble-nucleation baryogenesis which requires a first order phase transition to go out of thermal equilibrium.

Our calculational method for the fermion vacuum energy uses scattering data of fermions.  This raises some technical issues because the string configurations give long-range potentials with subtleties associated with scattering theory.  So, as a stepping stone to coupling fermions to electroweak strings, we have considered the simplest example of a vortex configurations: magnetic flux tubes in quantum electrodynamics (QED).  This allows us to isolate and examine the problem of scattering fermions off vortices.  We discuss this in detail next.

\section{Flux Tubes in QED}
\label{sec:QED}
We investigate quantum corrections to the energy of a magnetic flux
tube in QED, up to one-loop order.  We consider space-time
dimensions of D=3 and D=4.  In D=3 the configuration is a vortex (rather than a tube).  In D=4 we consider the energy per unit length of the flux tube.  Our primary motivation is that this
analysis will shed light on vortices (strings) in more complicated
field theories, especially the Z-string in the standard electroweak
theory. 

We compare the energies in three and four space-time dimensions to
see if the simpler case of D=3 (with no divergences) bears any
resemblance to the D=4 calculation.  The classical energy is of course
the same in the two cases.  The quantum correction to the energy could
possibly be very different, especially because of the different
divergence structure in the two cases.  In D=4,
the bare one-loop energy is divergent and only after renormalization
conditions are imposed on the calculation do we get a finite result.
In D=3, in contrast, the bare energy is finite.  But, the comparison
between the two cases is sensible only when we use the same
renormalization conditions.  This requires a finite counterterm in the
D=3 case to keep the renormalized photon field and electric charge
fixed.  Without this finite renormalization, the D=3 and D=4 energies
are qualitatively different.  But, after proper renormalization, we
find that the energies are closely related.  This suggests that if
quantum corrections stabilize the Z-string in two spatial dimensions,
then that might be indicative of a stable Z-string in three spatial
dimensions. 

We also use this problem to get a handle on several technical issues
associated with the computation of the one-loop energy of a flux
tube.  As described in Sec.~\ref{sec:FunctionalDeterminant}, an
efficient way to compute the energy is to use scattering data of
fermions in the background of the flux tube.  However, vortex
configurations give long range potentials, which violate standard
assumptions in scattering theory \cite{Newton:1982qc}.  This leads to
inconsistent and difficult to interpret results, some of which are
discussed in Sec.~\ref{sec:subtleties}.  We show that these puzzles
are due to the fact that an isolated flux tube is unphysical, and once
a region of return flux is included to make the net flux 0, the
scattering is well-defined and the puzzles disappear.  In the limit
of the return flux being infinitely spread out, the energy has a well
defined limit, which is associated with the localized flux tube.    

\subsection{The Theory} 
We consider QED in space-time dimensions D=3 and D=4, with a
four-component fermion field, $\psi$, in both cases.  The Lagrangian
density is 
\be
\mathcal{L}^{(D)} = - \frac{1}{4} F_{\mu \nu} F^{\mu \nu} + \bar{\psi} (i
\dslash + e \Aslash - m ) \psi + \mathcal{L}^{(D)}_\EctSub \, ,
\ee  
where the indices run from 0 to D-1 and $\mathcal{L}^{(D)}_\EctSub$ is
the counterterm Lagrangian in D space-time dimensions.  In D=4, this
describes electromagnetism with a single fermion flavor of mass m and
charge e.  In D=3, we have parity-invariant electromagnetism with two
flavors of fermions (of equal mass m and equal charge e).  The
four-component fermion contains two two-component fermions.  This can
be thought of as a planar version of the D=4 theory in which all
particles (including virtual  ones) are constrained to lie on the x-y
plane. 

We are interested in static magnetic flux tubes.  These are localized,
cylindrically symmetric magnetic fields (pointing in the z direction
in D=4), with a net flux \flux through the x-y plane.  For example, a
Gaussian flux tube of width $w$ is 
\be
B_\gaussSub ( \rho ) = B_\gaussSub( 0 ) e^{-\rho^2/w^2} \, , 
\label{eq:gauss}
\ee
where $\rho^2 = x^2 + y^2$.  Its flux is
\be
\flux_\gaussSub = \pi w^2 B_\gaussSub( 0 ) \, .
\ee
We often find it convenient to specify the flux in units of $2 \pi /e$
and define 
\be
\flux = \res{\flux} \frac{2 \pi}{e} \, .
\ee 
In the radial gauge, vortex configurations of gauge fields give flux tubes:
\bea
A_0 & = & 0 \, , \nonumber \\
\vec{A} & = & \frac{\flux}{2 \pi \rho} f(\rho)\hat{\azAngle} 
\label{eq:vortex}
\eea
where $f(\rho)$ goes from 0 to 1, and  
\be
B ( \rho ) = \frac{F}{2 \pi \rho} \frac{d f(\rho)}{d \rho} \, .
\label{eq:vortexB}
\ee
For small $\rho, f(\rho) \propto \rho^2$, so that $B$ is not singular.  

\subsection{The Energy}
We want to compute the energy, $E^{(D)}$, (per unit z-length in D=4)
up to one-loop order of magnetic flux tubes.  For static
configurations, this is minus the one-loop effective action per unit
time (per unit z-length in D=4).  Since the theory is Abelian, the
photons do not self-interact and the one-loop effective action is
obtained by integrating out the fermion field.  The photon
fluctuations start contributing only at two loops and higher, and we
ignore their contributions.  Then, 
\be
E^{(D)} = E_\EclSub + E^{(D)}_\EvacSub
\ee
where the classical energy (in both D=3 and D=4) is
\be
E_\EclSub = \frac{1}{2} \int d^2 x B^2 \, ,
\ee
and the renormalized fermion vacuum energy (the renormalized one-loop
energy) is 
\be
E_{\rm vac}^{(D)} = \lim_{T\{,L\} \rightarrow \infty} \frac{i}{T\{L\}}
\left[ \ln \Det (i\dslash + e\Aslash - m) - \ln \Det (i\dslash - m)
  \right] + E^{(D)}_\EctSub \, . 
\label{eq:det}
\ee
The terms in $\{\}$ above are present only in D=4 where the energy per
unit z-length is required.  The counterterm energy,
$E^{(D)}_\EctSub$, and the functional determinant contribution to
$E^{(D)}_\EvacSub$ are discussed in the following subsubsections.  Note
that the energy is defined relative to the absence of electromagnetic
fields. 

\subsubsection{Renormalization}
The fermion fluctuations renormalize the bare photon field and fermion charge,
\bea
A^\mu & = & (1+C^{(D)})^{-1/2} A_{\rm bare}^\mu \, , \nonumber \\
e & = & (1+C^{(D)})^{1/2} e_{\rm bare} \, ,   
\eea
giving the counterterm energy
\be
E^{(D)}_\EctSub = C^{(D)} \half \int d^2 x B^2 \, .
\ee
In the absence of photon fluctuations, the bare fermion field and mass
do not get renormalized.  In $D=4$ the counterterm is divergent and
combines with the divergent functional determinant to give a finite,
renormalized one-loop energy.  In $D=3$ however, the renormalization
of the bare photon field and charge is finite resulting in a finite
counterterm.  Nevertheless, the counterterm must be taken into
account, otherwise the photon field is not properly normalized and the
charge is the (unobservable) bare charge.  Renormalization is the
process of fixing the unobservable bare parameters in the theory by
imposing physical conditions.  In D=4 this process also removes the
ultraviolet divergences (by allowing the bare parameters to be
divergent), but the lack of divergences in D=3 does not mean that
renormalization is not required (as is often stated in the
literature).  In fact, one of our aims is to compare the energies in
D=3 and D=4, and this is sensible only when we impose the same
renormalization conditions in the two cases.   

We choose the on-shell renormalization condition (residue of the
$q^2=0$ pole of the photon propagator is 1) to fix the counterterm
coefficient: 
\be
C^{(D)} = - \frac{4 e^2}{3 (4\pi)^{D/2}} \frac{\Gamma \left(2 -
  \frac{D}{2} \right)}{m^{4-D}} \, . 
\ee
In $D=3$, it is the finite quantity
\be
C^{(3)} = - \frac{e^2}{6 \pi m} \, ,
\ee
and in $D=4$ it is the divergent quantity
\be
C^{(4)} = - \frac{e^2}{12 \pi^2} \left(
\frac{2}{\epsilon} - \gamma + \ln \frac{4 \pi}{m^2} \right) \, .  
\ee 
Note that we have dimensionally regulated the divergent counterterm coefficient in D=4 by introducing the regulator $\epsilon = 4 - D$.
\subsubsection{Functional Determinant}
\label{sec:FunctionalDeterminant}
We use the phase shift approach, described in Sec.~\ref{sec:PhaseshiftsMethod}, to exactly compute the functional
determinant that contributes to the one-loop energy as in
\eq{eq:det}.  Recall that the
basic idea is that the fermion vacuum energy is given by the sum over
the shift in the zero-point energies of the fermion modes due to the
presence of the magnetic background.  In the continuum this is a sum
over bound state energies and an integral over the continuum energies
weighted by the change in the density of states.  Since the vortex
configuration that characterizes the flux tube is cylindrically
symmetric, the scattering matrix of the fermion may be decomposed into
partial waves labeled by the z-component of the total angular
momentum, $M$.  Then, in D=3,  the change in the density of states is
given by the phase shifts of the fermion scattering wave functions in
the flux tube background: 
\be
\Delta \rho(k) = \sum_M \frac{1}{\pi} \frac{d \delta_M (k)}{dk} \, , 
\ee
where $k$ is the magnitude of the momentum.  The phase shifts above
are defined as the sum over both signs of the energy.  In D=4, there
is an additional factor in the change in the density of states
corresponding to the trivial $z$ dimension, and we use the interface
formalism \cite{GrahamInterface} to straightforwardly account for this.
We subtract the first two terms in the Born series expansion of the
change in the density of states and add the corresponding energy back
in the form of the two-point Feynman diagram (a fermion loop with two
insertions of the background potential).  In D=4, the Born subtraction
renders the integral over the continuum energies convergent and
isolates the ultraviolet divergences in the Feynman diagram, which
when combined with the counterterm gives a properly renormalized
finite result.  In D=3 we did not have to Born subtract, but we do so
anyway to maintain a close similarity with D=4.  In Appendix
\ref{app:QED} we provide the details on how we compute
the phase shifts and the born series. 

Our final expression for the renormalized fermion vacuum energy is
\be
E^{(D)}_\EvacSub = E^{(D)}_\EctSub + E^{(D)}_\EfdSub + E^{(D)}_\EdeltaSub \, .
\ee
The unrenormalized Feynman diagram energy (unrenormalized two-point energy) is
\be
E^{(D)}_\EfdSub = \frac{8 \pi \res{\flux}^2}{(4 \pi)^{D/2}}
\int_0^\infty dp \left( \int_0^\infty d\rho \frac{d f(\rho)}{d \rho}
J_0(p \rho)  \right)^2 \int_0^1 dx \frac{x(1-x) p \Gamma( 2-
  D/2)}{(m^2+p^2 x(1-x))^{2-D/2}} \, , 
\ee
which combines with the counterterm energy to give the renormalized
Feynman diagram energy (in D=4 this cancels the divergence).  The
phase shifts energy, which corresponds to the contributions from the
three-point and higher functions, is 
\bea
E^{(3)}_\EdeltaSub & = & + \frac{1}{2\pi} \int_0^\infty dk
\frac{k}{\sqrt{k^2+m^2}} \sum_M \bar{\delta}_M(k) \, , \\ 
E^{(4)}_\EdeltaSub & = &  - \frac{1}{2\pi^2} \int_0^\infty dk k \ln
\frac{\sqrt{k^2+m^2}}{m} \sum_M \bar{\delta}_M (k) \, , 
\eea
where
\be
\bar{\delta}_M(k) = \delta_M (k) - \delta_M^{(1)} (k) -
\delta_M^{(2)}(k) \, ,   
\ee
and $\delta^{(i)}_M$ denotes the $i^{\rm th}$ term in the Born series
expansion of the phase shift. 

\subsection{A Few Puzzles}
\label{sec:subtleties}
In the previous subsections we have described how to compute properly
renormalized energies of magnetic flux tubes, to one-loop.  However,
before proceeding with the calculation, there are several subtleties
that need to be addressed.  The main issue is that even for
exponentially localized magnetic fields, the photon field falls off
too slowly (like $1/\rho$ in the vortex configuration) for the
standard theorems of scattering theory to hold.  Since our
computational method relies on scattering phase shifts, we need to
resolve this.  In addition to scattering theory problems, there seem
to be rather unexpected and physically senseless results associated
with flux tubes.  In this subsection we enumerate some of these. 

As described in Appendix \ref{app:QED}, we use the
second order Dirac equations in two radial functions, $g_1(\rho),
g_2(\rho)$, to calculate phase shifts in the background of a flux
tube. The asymptotic form of these equations is 
\bea
0 &=& g''_1 + \frac{g_1'}{\rho} + \left( k^2 -
\frac{(M-\smallhalf+\res{\flux})^2}{\rho^2} \right) g_1 \, , \nonumber \\  
0 &=& g''_2 + \frac{g_2'}{\rho} + \left( k^2 -
\frac{(M+\smallhalf+\res{\flux})^2}{\rho^2} \right) g_2 \, . 
\eea
So, instead of the free equations, we have the equations in the
presence of an ideal flux tube (with profile function $f(\rho)=1$ and
singularities at $\rho=0$).   The centrifugal barrier is shifted by
the flux and the regular solutions are Bessel functions: 
\be
g_1(\rho) = J_{|M- \smallhalf +\res{F}|} (k \rho) \, , g_2(\rho) =
J_{|M+\smallhalf+\res{F}|} (k \rho) \, . 
\ee
From the asymptotic forms of these ($k \rho >> 1$), we find a phase
shift relative to the trivial configuration ($A_\mu=0$) by the amount 
\be
\delta_{M, \pm, \rm free} (k) = \frac{\pi}{2} \left( |M- \smallhalf| +
|M+\smallhalf| - |M-\smallhalf+\res{\flux}|  - |M+\smallhalf+\res{\flux}| \right) \, ,      
\ee 
where the second index labels the sign of the energy $\omega$
\cite{Ruijsenaars:1983fp}.  The second order equations allow us to
compute phase shifts relative to the ideal flux tube, $\delta_{M, \pm,
  \rm ideal}$, to which $\delta_{M, \pm, \rm free}$ must be added to
obtain the phase shift relative to the trivial configuration: 
\be
\delta_{M, \pm} = \delta_{M, \pm, \rm ideal} + \delta_{M, \pm, \rm free} \, .
\ee  
However, for $k=0$, there is no $\rho$ large enough where the
asymptotic form of the Bessel functions holds, and it is unclear what
$\delta_{M, \pm, \rm free}(0)$ (and hence $\delta_{M, \pm}(0)$) should
be. 

Now let's restrict our attention to momenta strictly greater than 0.
In a delta-shell background of magnetic flux at $\rho=\rho_0$, the
phase shifts can be calculated analytically \cite{Jaroszewicz:1986ss}.
We consider more general localized flux tubes and compute phase shifts
numerically.  We agree with the analytical calculation on the
expressions for the phase shifts at large and small momenta.  These
expressions depend only on the net flux through the plane and not on
the particular profile of the magnetic field.  In the $J_z = M$
channel, at large momentum, 
\be
\lim_{k \rightarrow \infty} \delta_{M, \pm} (k) = 0 \, .
\ee  
For small momentum, 
\be
\lim_{k \rightarrow 0^+} \delta_{M, \pm} (k) =  \left\{ \begin{array}{lll} 
  \pi \left( |M+\smallhalf| - |M+\smallhalf+\res{F}| \right) &:& M \ge \smallhalf \\ 
  \pi \left( |M-\smallhalf| - |M-\smallhalf+\res{F}| \right) &:& M \le -\smallhalf \\ 
\end{array} \right.
\ee

Armed with the limiting values of the phase shifts at large and small
momenta, and using the fact that the momentum derivative of the phase
shifts gives the density of states, we can ask how many states leave
the continuum as the flux tube is turned on.  This should correspond
to the number of bound states.  For concreteness, consider
$\res{\flux}=1.2$.  In this background there are two bound states: one
with $J_z= 1/2$ and $\omega=-m$, and another with $J_z= -1/2$ and
$\omega=m$.  However, the number of states that have left the
continuum in channel $J_z = M$ is 
\be
\Delta_M = \frac{1}{\pi} \left( \delta_{M, +}(0^+) + \delta_{M, -} (0^+) \right) = \left\{ \begin{array}{ccc}
	1.6 &:& M = -\half \\
	2.4 &:& M < -\half \\
	-2.4 &:& M \ge \half
	\end{array} \right.
\ee
So, in no channel is there an agreement between the number of states
that leave the continuum and the number of bound states and we have a
puzzling discrepancy in the number of states. 

Apart from the formal scattering theory problems discussed above,
there seem to be physical puzzles involving a flux tube.  These show
up in D=3 with a single two-component fermion.  In this case, the net
fermion charge in the background of the flux tube is $\res{\flux}/2$,
as is well-known (see \cite{Niemi:1983rq} for example).  This implies
that as the flux tube is turned on, charge conservation is violated.
Even if we ignore this problem of how to generate the flux tube
without violating charge conservation and assume that the flux tube is
given as an initial condition, we run into a problem.  As the flux
tube is spread out so that the magnetic field approaches 0 everywhere
(but the flux is held fixed), the total energy of the flux tube
approaches 0 (as shown in Sec.~\ref{sec:fixedFlux}).  This means that
we have charged, massless objects in the theory and the fundamental
fermion cannot be stable. 

\subsection{Embedding}
\label{sec:embedding}
All the subtleties and puzzles discussed in Sec.~\ref{sec:subtleties}
could perhaps have resolutions.  However, we now argue that it is
impossible to create a configuration carrying non-zero net flux
starting from nothing.  We demonstrate how the flux tube can be
embedded in a physical no-net-flux configuration in which the
localized flux tube has a spread out region of return flux.  Now the
scattering potential falls into the class of potentials well
understood in scattering theory and all the associated puzzles
disappear.  As the return flux is spread out, we show that the energy
of the embedded configuration approaches a well-defined limit which
we interpret to be the energy of the flux tube.  This is exactly the
energy obtained by ignoring all the problems with the phase shifts in
the background of a flux tube and using them to compute the energy. 

In \Qf we have the Bianchi identity
\be
\epsilon^{\alpha \beta \mu \nu} \partial_\beta F_{\mu \nu} = 0 \, .
\ee
This is a mathematical identity and not an equation of motion from the
action principle.  In terms of electric and magnetic fields it gives
us the two Maxwell equations: 
\bea
\nabla \cdot \vec{B} & = & 0 \, , \nonumber \\
\frac{\partial \vec{B}}{\partial t} & = & - \vec{\nabla} \times \vec{E} \, .
\eea
The fact that the magnetic field is divergence-less requires all
magnetic flux lines to be closed.  So, there can be no net flux
through the $x-y$ plane.  In \Qt, only the second of the two equations
remains, in the form 
\be
\frac{\partial B}{\partial t} =  - \partial_x E_y + \partial_y E_x \, .
\ee
Integrating both sides over the area of the x-y plane, we get:
\be
F(t) = F(t=0) - \int_C \vec{E} \cdot d \vec{l} \, , 
\ee
where C denotes the circle at $\rho \rightarrow \infty$.  So, if we
start with net flux zero, we can create a non-zero net flux only if
we have an electric field at spatial infinity (with non-zero curl),
which is physically impossible.  Thus, both in D=3 and in D=4, every
flux tube that is set up locally must have a compensating return flux
to make the net flux zero.  This return flux is not a crutch to enable
the use of scattering theory to compute energies, but rather a
physical requirement. 

The presence of the return flux not only removes the subtleties
associated with scattering theory, but also eliminates the physical
puzzles discussed in the single fermion D=3 theory at the end of
Sec.~\ref{sec:subtleties}.  If we can't create a net flux, we can't
create a net charge either.  We can only polarize the Dirac sea and
create local regions of charge with net charge zero.  Furthermore, by requiring the net flux to be zero, we have eliminated
the presence of massless charged objects in the theory. 

Now we demonstrate that as the return flux is spread out so that the
associated magnetic field, $B_R$,  locally approaches 0, the energy
approaches a well-defined limit.   This can be understood
analytically by expressing the total energy as a series in inverse
powers of the width of the magnetic field.  As shown in
Sec.~\ref{sec:fixedFlux}, for a fixed flux, as the magnetic field is
spread out, the classical energy goes to 0 like $1/w^2$ and the
renormalized one-loop energy goes like $1/w^4$ (in both D=3 and D=4).
Thus, in the limit of spread-out return flux, the energy should limit
to that associated with the local flux tube alone.  We show that this
value corresponds to computing the energy using the phase shifts in
the unphysical problem with no return flux. 

\begin{figure}
\centerline{
\includegraphics{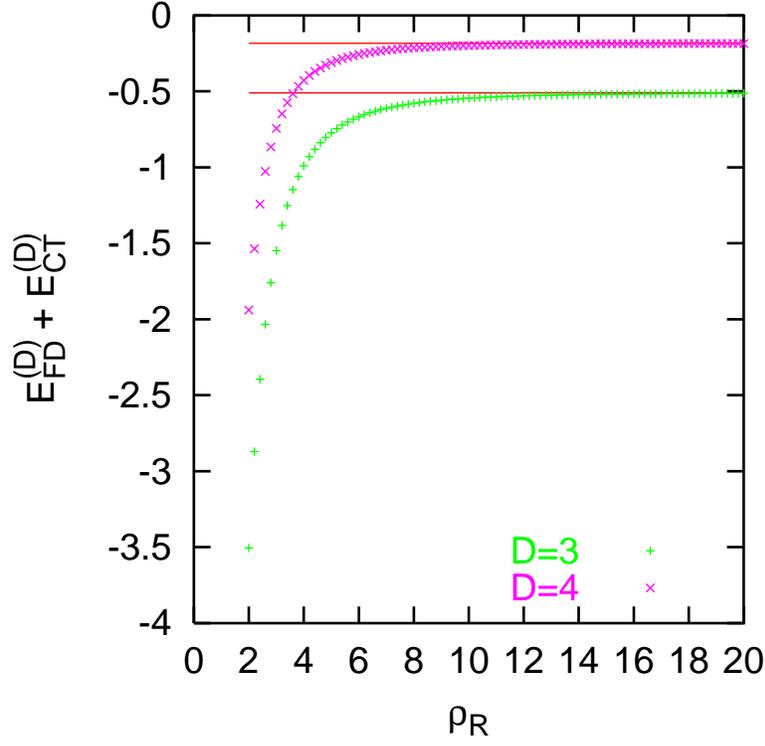}
}
\caption{\label{fig:embedFD} Renormalized Feynman Diagram energies (in
  D=3 and D=4) as a function of the return flux radius.  The solid
  lines correspond to the energies without the return flux.} 
\end{figure}

\begin{figure}
\centerline{
\includegraphics[width=8cm]{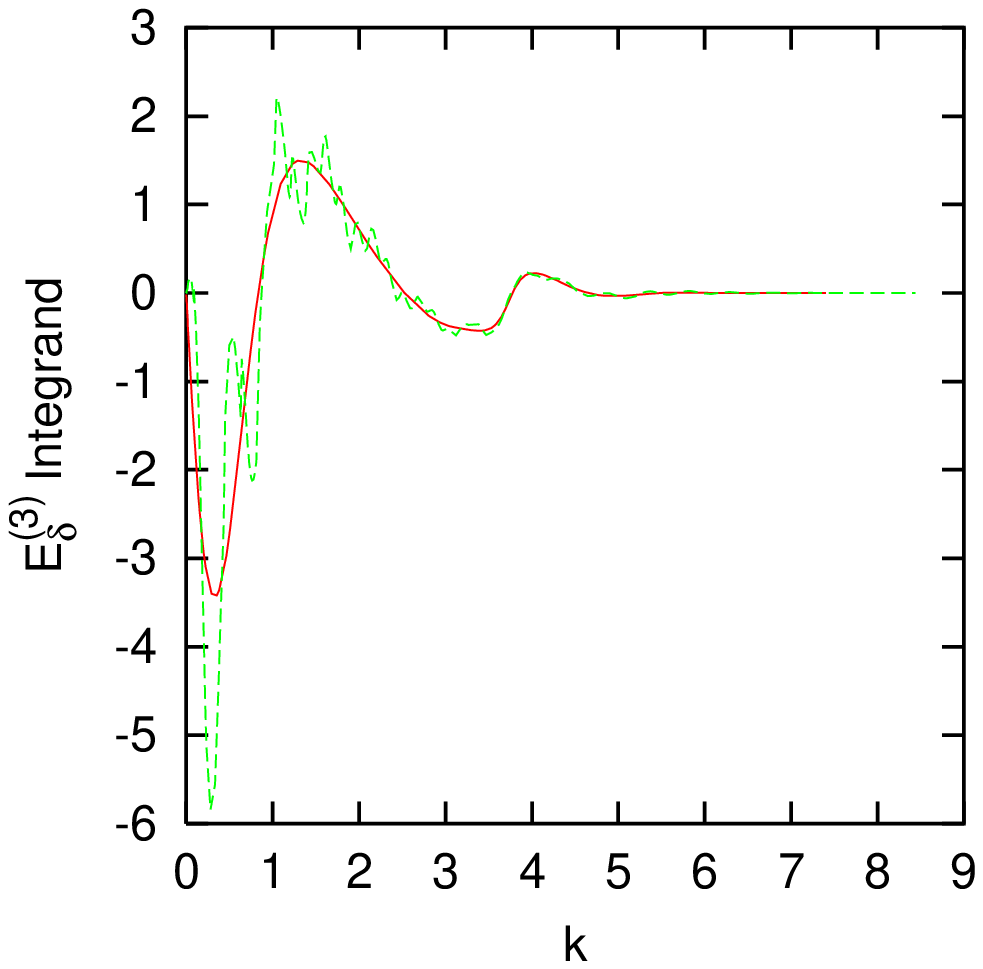}
\hskip 1.0cm
\includegraphics[width=8cm]{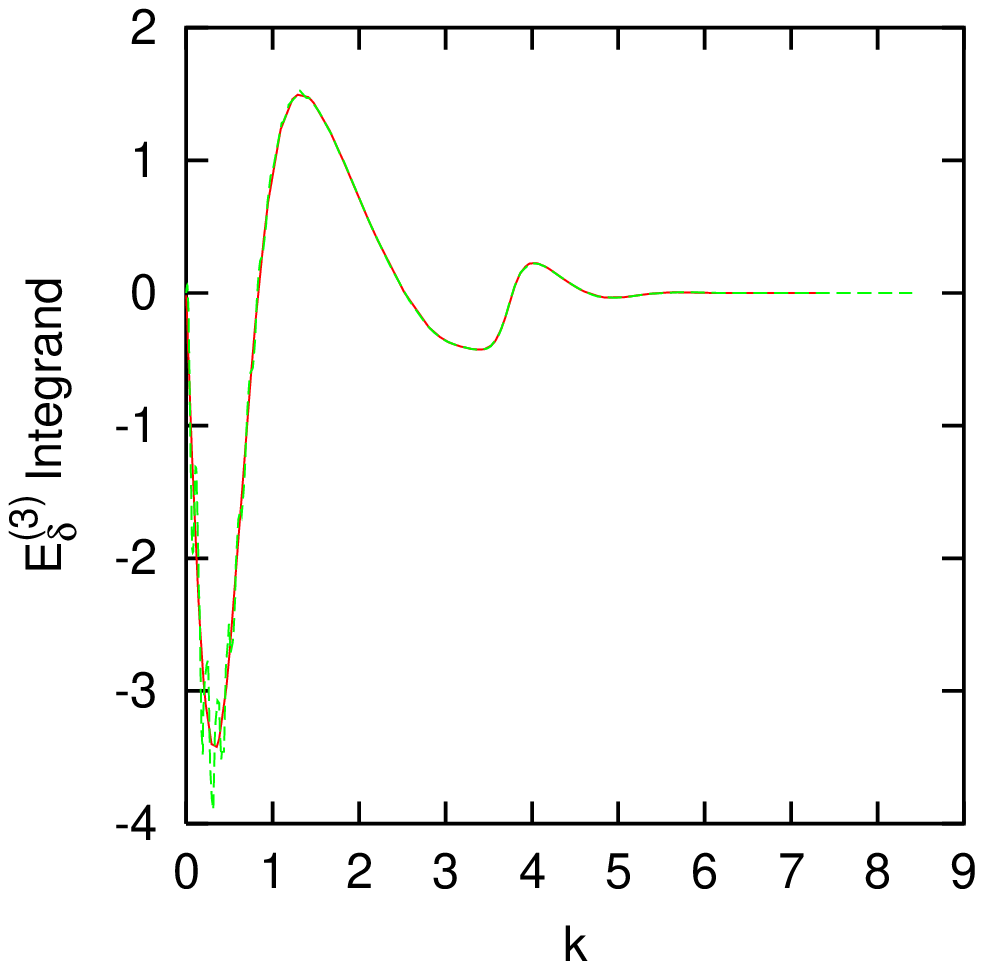}
}
\caption{\label{fig:embedDelta} Integrand of the phase shift
  contribution to the energy in D=3 for $\rho_R=6$ (left panel) and
  $\rho_R=26$ (right panel).  The solid lines correspond to the
  isolated flux tube problem and the dashed lines to the embedded
  problem.} 
\end{figure}

Consider the Gaussian flux tube, $B_\gaussSub(\rho)$, defined in
\eq{eq:gauss} with flux $F$.  Choose the return flux to be centered at
a ring of radius $\rho_R$ (with a width proportional to $\rho_R$),
e.g. 
\be 
B_R(\rho) = - \frac{16 F}{\pi \rho_R^2 \left( 1+ 256 \left(
  \rho^2/\rho_R^2 - 1\right)^2 \right) \left( \pi/2 + \arctan{16}
  \right)} 
\ee
The local flux tube is superimposed with the return flux to get the
no-net-flux embedding 
\be
B_0 (\rho) = B_\gaussSub(\rho) + B_R(\rho) \, . 
\ee
The limit of spread out return flux corresponds to $\rho_R \rightarrow
\infty$.  The total energy consists of three parts: the classical
energy, the renormalized Feynman diagram energy and the phase shifts
energy.  We consider these different energy contributions for $B_0$ as
a function of $\rho_R$.  We scale all dimensionful quantities in units
of the fermion mass, $m$ and choose $\res{\flux} = 4.8$ and $w=1$.
Using the expression for the classical energy, it is trivial to see
that 
\be
\lim_{\rho_R \rightarrow \infty} E_\EclSub [B_0] = E_\EclSub [B_\gaussSub] \, .
\ee
In Fig.~\ref{fig:embedFD}, we show the approach of the renormalized
Feynman diagram energy, $E^{(D)}_\EfdSub + E_\EctSub$, in the
no-net-flux configuration to that in the flux tube configuration, as
the return flux is spread out.  In both D=3 and D=4 there is good
convergence for $\rho_R > 10$. 
In Fig.~\ref{fig:embedDelta}, we show the integrand of the phase shift
part of the energy, in both the embedded configuration (for two fixed
values of $\rho_R$) as well as in the flux tube configuration, for
D=3.  Note that the integrands disagree only for small values of the
momentum (for which the scattering fermion is sensitive to the
presence of the spread out return flux).  The integrand in the
embedded configuration oscillates around the integrand in the flux
tube configuration, with an amplitude that decreases as $k$ increases.
The region of disagreement gets pushed to smaller values of $k$ as
$\rho_R$ increases.  In the spread-out limit, we find that the phase
shift contribution to the energy is identical in the embedded problem
and the flux tube problem.  We have verified that the same results
hold in D=4 and do not plot the corresponding results. 

Thus, when the flux tube is embedded in a physical no-net-flux
configuration and the return flux is spread-out, we find that the
energy in the embedded problem approaches the energy in the isolated
flux tube problem.  This allows us to use the phase shifts in the
background of the isolated flux tube to compute the energy, with the
implicit understanding that there is a spread-out return flux
present, which eliminates all the puzzles and does not contribute to
the energy.    

\subsection{Results}
\subsubsection{Derivative Expansion}
\label{sec:derivativeExpansion}
For a fixed magnetic field at the origin, we vary the width of the
flux tube (thereby varying the flux as well).  As the width increases,
the configuration approaches a non-vanishing constant magnetic field,
and the first few terms in the expansion of the energy in derivatives
of the magnetic field should be a good approximation.  In this subsection
we investigate the convergence of the derivative expansion for the
Gaussian flux tube of \eq{eq:gauss}. 

In D=3, the derivative expansion of the unrenormalized one-loop
energy is given by \cite{Cangemi:1995by,Gusynin} 
\be
E^{(3)}_\EvacSub - E^{(3)}_\EctSub = E^{(3)}_{\EvacSub, 0} +
E^{(3)}_{\EvacSub, 2} \, ,  
\ee
to lowest non-trivial order in the derivative, where
\bea
 E^{(3)}_{\EvacSub, 0} & = & \int d^2 x \frac{|eB|^{3/2}}{4 \pi^{3/2}}
 \int_0^\infty ds e^{-sm^2/|eB|} s^{-3/2} \left( \coth(s) -
 \frac{1}{s} \right) \, , \nonumber \\ 
 E^{(3)}_{\EvacSub, 2} & = & \frac{1}{4} \int d^2 x
 |\vec{\nabla}(eB)|^2 |4 \pi eB|^{-3/2} \int_0^\infty ds
 e^{-sm^2/|eB|} s^{-1/2} \frac{d^3 (s \coth s)}{ds^3} \, . 
\eea
Note that the above expressions are twice those for a two-component
fermion.  Since our Gaussian flux tube is a function of $\rho/w$, we can rescale the x-y coordinates in units of $w$ and obtain
$E^{(3)}_{\EvacSub, 0} \propto w^2$ and $E^{(3)}_{\EvacSub, 2} \propto
w^0$.  On dimensional grounds, it is clear that the counterterm
contribution to the energy is proportional to $w^2$ and for the
Gaussian flux tube,   
\be
E^{(3)}_\EctSub = - \frac{e^2 B_\gaussSub^2(0) w^2}{24 m} \, .
\ee
We add the counterterm to get the renormalized derivative expansion
\be
E^{(3)}_\EvacSub = E^{(3)}_\EctSub + E^{(3)}_{\EvacSub, 0} +
E^{(3)}_{\EvacSub, 2} \, ,  
\ee
where the first two terms are proportional to $w^2$, the last term is
independent of $w$, and all omitted terms go to 0 for large $w$. 

In D=4, the lowest non-trivial order derivative expansion of the
renormalized one-loop energy is \cite{Lee:1989vh} 
\be
E^{(4)}_\EvacSub = E^{(4)}_{\EvacSub, 0} +  E^{(4)}_{\EvacSub, 2} \, , 
\ee
where
\bea
 E^{(4)}_{\EvacSub, 0} & = & \int d^2 x \frac{|eB|^2}{8 \pi^2}
 \int_0^\infty ds e^{-sm^2/|eB|} s^{-2} \left( \coth(s) - \frac{1}{s}
 - \frac{s}{3} \right) \, , \nonumber \\ 
 E^{(4)}_{\EvacSub, 2} & = & - \int d^2 x |\vec{\nabla}(eB)|^2 |32
 \pi^2 eB|^{-1} \int_0^\infty ds e^{-sm^2/|eB|} \left( 1 - 4 \coth^2 s
 + 3 \coth^4 s + \right. \nonumber \\ 
 & & \left. \frac{3 \coth s}{s}(1 - \coth^2 s)  \right) \, .
\eea
Again we rescale the coordinates in units of $w$ and obtain
$E^{(4)}_{\EvacSub, 0} \propto w^2$ and $E^{(4)}_{\EvacSub, 2} \propto
w^0$.  

\begin{figure}
\centerline{
\includegraphics[width=8cm]{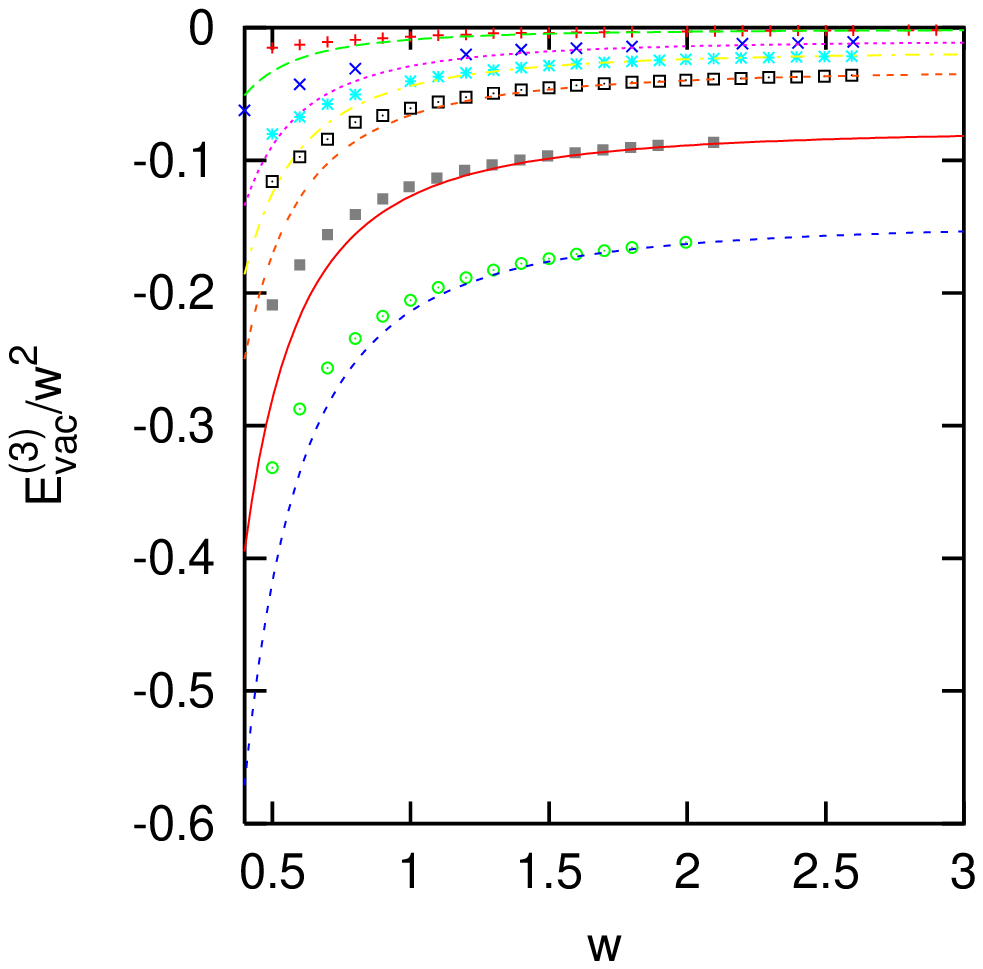}
\hskip 1.0cm
\includegraphics[width=8cm]{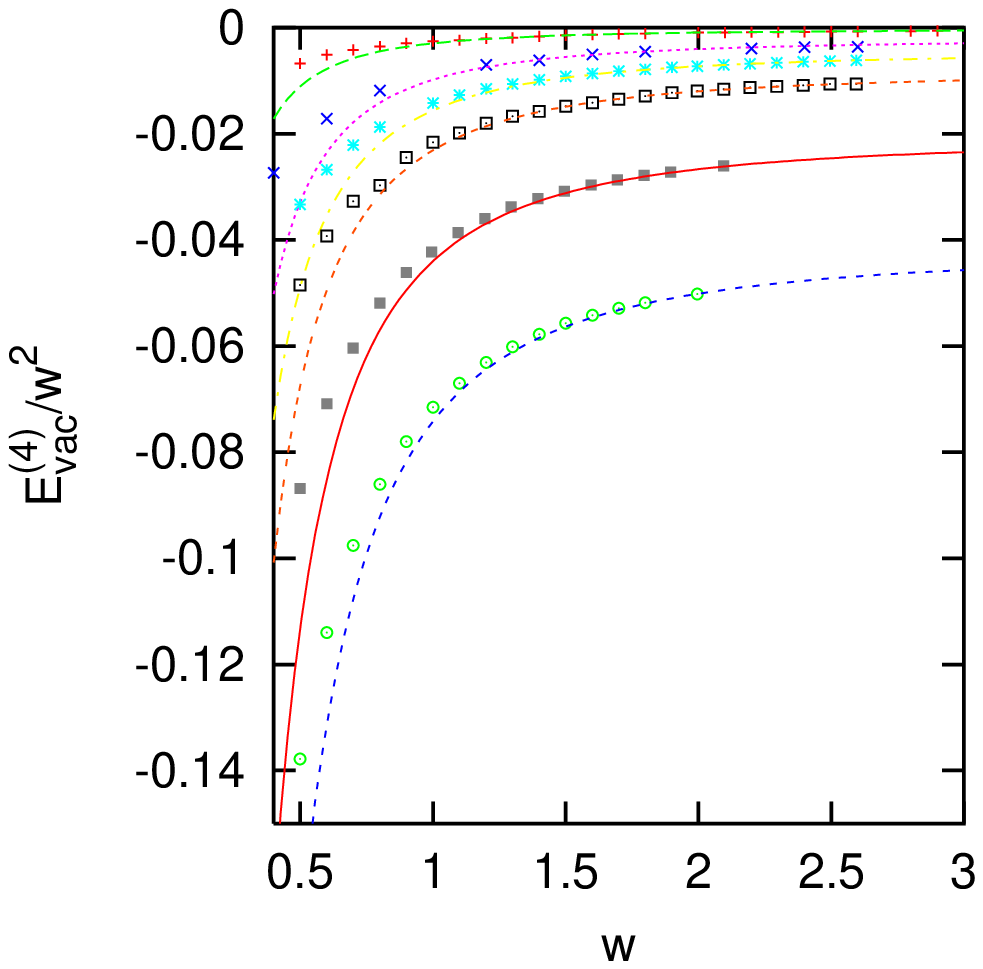}
}
\caption{\label{fig:derivExp} Renormalized one-loop energies in D=3
  (left panel) and D=4 (right panel), for fixed values of the magnetic
  field at the origin, as a function of the width of the Gaussian flux
  tube.  The lines correspond to the derivative expansion to lowest
  non-trivial order in the derivative.  From top to bottom, $e
  B_\gaussSub (0) = 1.1, 2, 2.5, 3, 4, 5$.} 
\end{figure}

In Fig.~\ref{fig:derivExp}, we compare the exact $E^{(D)}_\EvacSub$
with the renormalized derivative expansion approximation, for
different fixed values of $eB_\gaussSub(0)$ and find excellent
agreement for widths larger than 1.  There appears to be no
qualitative difference between the energies in D=3 and D=4.  This is in contrast to the claim in \cite{Langfeld:2002vy} that in D=4, renormalization effects cause the convergence to the derivative expansion result as a function of the width to be slower.  Had we not included the counterterm in D=3 (thereby having a different
renormalization condition from the one in D=4), the corresponding
energies would have been positive. 
\subsubsection{Fixed Flux}
\label{sec:fixedFlux}
Now we consider flux tubes with fixed flux and varying widths.  We
show that in the limit of the width going to infinity, the energy of
the flux tube goes to zero. 

First consider the classical energy:
\bea
E_\EclSub & = & \frac{ \pi \res{\flux}^2}{e^2 w^2} \int_0^\infty dx
\frac{1}{x}\left( \frac{d^2 f(x)}{d x^2}  \right)^2 \nonumber \\ 
 & = & \frac{\pi \res{\flux}^2}{e^2 w^2} \textrm{ for a Gaussian flux
  tube} \, , \\ 
\eea
where $x \equiv \rho/w$.  For large widths, this goes to 0 like
$1/w^2$.  In the large width limit, the magnetic field becomes weak
and so the dominant contribution to the one-loop energy comes from
the two-point function.  In D=3, the unrenormalized two-point energy
can be expressed as a series in $1/w^2$.  The leading term
proportional to $1/w^2$ turns out to be exactly equal to minus the
counterterm energy and only the sub-leading term remains in the
renormalized two-point energy.  For a Gaussian flux tube, 
\bea
E^{(3)}_\EfdSub & = & \frac{\res{\flux}^2}{6 m w^2} \left( 1 -
\frac{1}{5 m^2 w^2} + \mathcal{O}(1/(mw)^4) \right) \nonumber \\ 
E^{(3)}_\EfdSub + E^{(3)}_\EctSub & = & - \frac{\res{\flux}^2}{30 m^3
  w^4}  \, .\\ 
\eea
So, for large widths, renormalization not only changes the sign of the
one-loop energy, but also the rate at which 0 is approached.  The
renormalized results in D=3 are similar to the results in D=4: 
\be
E^{(4)}_\EfdSub + E^{(4)}_\EctSub  \propto - \frac{\res{\flux}^2}{m^2 w^4}
\ee
to leading order in $1/w^2$.  For a Gaussian flux tube the
proportionality constant is $1/(30 \pi)$.  Note that in the $ w
\rightarrow \infty$ limit, the total energy in both dimensionalities
is 0.  This is why in our embedding of the flux tube in a a
no-net-flux configuration, the energy approached a well-defined
limit as the return flux was spread out (see
Sec.~\ref{sec:embedding}). 

\begin{figure}
\centerline{
\includegraphics[width=8cm]{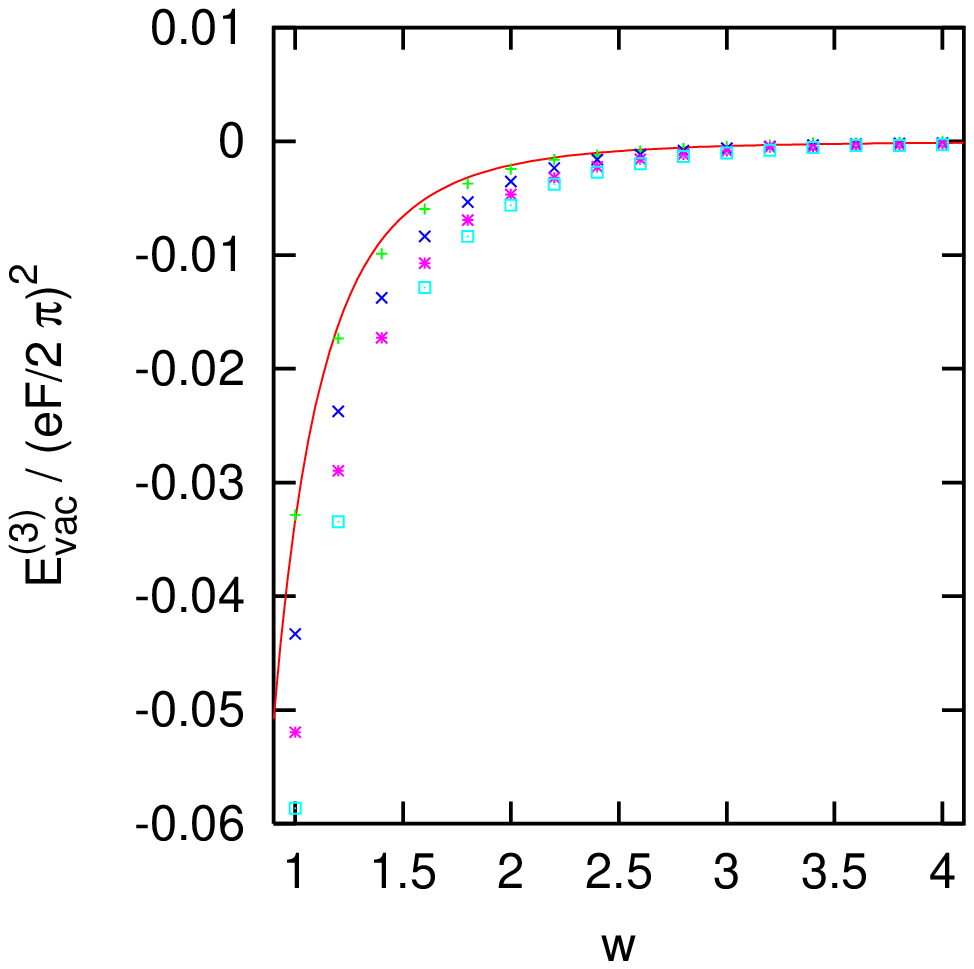}
\hskip 1.0cm
\includegraphics[width=8cm]{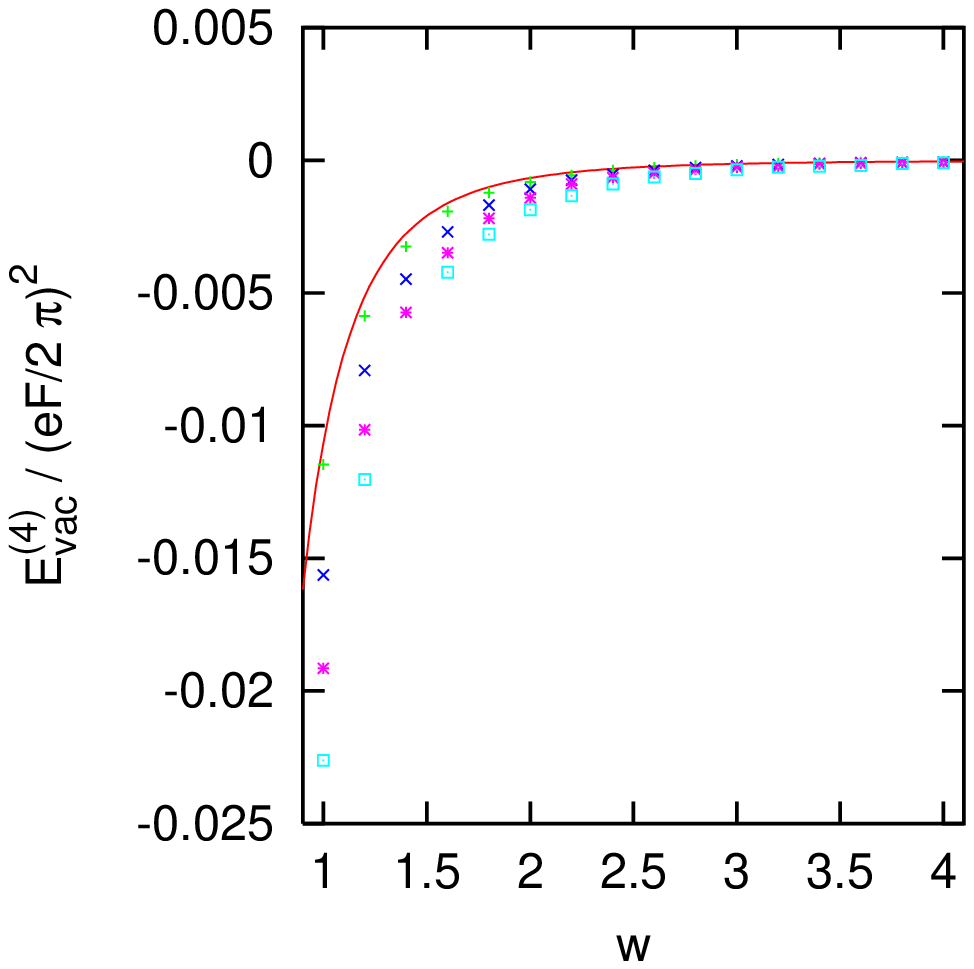}
}
\caption{\label{fig:fixedFlux} Renormalized fermion vacuum energy
  (normalized in units of $\res{\flux}^2$) as a function of the width,
  for various fixed values of the flux $\res{\flux}$ in the Gaussian
  flux tube ($\res{\flux} = 2.5, 4.5, 6.5, 8.5$ from top to bottom).  The left panel is for D=3 and the right panel for D=4.} 
\end{figure}

In Fig.~\ref{fig:fixedFlux}, we plot the exact one-loop energies for
various values of the flux as a function of the width.  We normalize
the energies in units of $\res{\flux}^2$ so that all differences
between the different fluxes are due to three-point and higher
contributions.  We compare the energies with the leading order
two-point energies.  We find good agreement for large widths.  As the
flux increases, we need to go to larger widths to get a weak magnetic
field everywhere, and so the convergence to the leading order
two-point energy occurs at larger widths.  

\section{Conclusions}
We have argued that electroweak strings could trap quarks and give rise to novel objects with rich phenomenology.  These could be very significant for electroweak baryogenesis and could also be dark energy candidates.  In order to gear up for the electroweak strings stability analysis, we studied the simpler problem of fermions in the background of magnetic flux tubes in QED.  We found that all puzzles and problems associated with the long-range nature of the potential disappear when we consider a region of return flux (which is the physical scenario).  This suggests that the electroweak strings calculation could be more tractable if we embed the background into one with no net flux.  Also, we find that when the energies are properly renormalized, there are no qualitative differences between two and three spatial dimensions.  So, it may be sensible to consider electroweak vortices in two spatial dimensions, and look for a stable fermionic soliton (in analogy with the quantum solitons discussed in Chap.~\ref{chap:QuantumSolitons}).  If we find such an object, then it may suggest that a stable object exists in the physical case of three spatial dimensions.  We are currently investigating this.
\chapter{Conclusions and Summary}

We have used the non-trivial topology of maps from $S_3$ and $S_4$ into the gauge group $SU(2)$ of the weak theory to construct non-contractible loops of \HiggsGauge configurations in Euclidean spacetime.  A winding 1 map from $S_3$ to $SU(2)$ allows us to construct the well-known {\em weak instanton}, {\em weak sphaleron} and the family of {\em W-strings}.  When the winding is chosen to be an integer $n$, the method suggests the existence of {\em multi-instantons} (with topological charge $n$), {\em multi-sphalerons} (barriers between winding 0 vacuum and winding n vacuum) and {\em vorticity $n$ W-strings}.  However, these do not produce qualitatively different physical effects from those of the $n=1$ solutions.   For example, the n-instanton would mediate the production of $n$ fermions via quantum tunneling through the sphaleron barrier.  But ($n$ times) repeated instances of the 1-instanton already allows for this process, albeit with a possibly different probability amplitude.  A topologically non-trivial map from $S_4$ to $SU(2)$ indicates the existence of novel unstable solutions.  In Euclidean spacetime, it hints at a barrier between zero-action configurations in the trivial and non-trivial sectors (when spacetime is compactified to $S_4$) -- the {\em $I^*$}.  The Euclidean action would have a saddle point at the $I^*$ and it could significantly contribute to the path integral.  When we consider static configurations, then the map suggests the existence of a solution with two directions of instability -- the {\em $S^*$}.  Quarks and leptons are expected to have zero modes in the background of this configuration, which makes it a promising candidate for a fermionic soliton that allows a heavy fermion to decouple from the theory.  Restricting static configurations to be trivial in one spatial direction, the map suggests the existence of string solutions with three directions of instability -- the {\em $W-string^*$}.  However, we have not succeeded in constructing these solutions.  The topological prescription suggests their existence and points to the region in configuration space where we can look for them.  But it does not guarantee that the solutions exist.  

In $1+1$ dimensional $\phi^4$ theory, we have reviewed the existence of approximate breathers -- long-lived configurations that are localized in space and oscillate in time.  Their existence hinges on two crucial properties of the theory: (1) there should be no massless particles in the theory;  (2) the field potential should be non-linear in such a way that large amplitude oscillations would have a frequency lower than the lowest linear frequency (the lightest particle mass).  When these two conditions are met, a configuration with fundamental frequency below the lowest linear frequency is stable against linear decay.  However, it is not absolutely stable because the non-linearity means that higher harmonics of the fundamental frequency are present, and these couple to the linear modes.  Nevertheless, such approximate breather configurations have lifetimes many orders of magnitude larger than all scales in the theory, which presents a challenge to the notion of naturalness in field theory.  The \HiggsGauge sector of the electroweak theory has both properties (when the hypercharge gauge fields are ignored so that there is no unbroken subgroup and all three W bosons are massive) required for the existence of breathers.  So we expect {\em approximate breathers} in the theory, and we are in the process of looking for them within a spherical ansatz.  Their physical import could lie in the fact that during the electroweak phase transition they would create out-of-equilibrium regions in space and thus contribute to electroweak baryogenesis.  

We have looked for stable solitons in the electroweak theory that carry the quantum numbers of a heavy fermion doublet and could allow it to decouple.  We consider \HiggsGauge configurations within a spherical ansatz and compute fermion vacuum fluctuation contributions to the energy, but ignore those of the bosonic fields.  We find significant quantum corrections to the height of the sphaleron barrier.  As we make the fermion heavier than the quantum-corrected sphaleron, we see the emergence of a new barrier maintaining the exponential suppression of the fermion decay.  For even larger values of the Yukawa coupling, however, the barrier disappears and the fermion's decay is unsuppressed.  We do not see any evidence for a fermionic soliton, and find that the fermion vacuum energy destabilizes would-be solitons.  However, it is still possible, indeed likely, that the fermionic soliton that maintains anomaly cancellation in the low energy theory and allows a heavy fermion to decouple lies outside the spherical ansatz.  A most promising candidate is the {\em $S^*$}, which is expected to exist due to the same non-trivial topology that gives rise to Witten's global anomaly.  However, in the absence of sufficient symmetry, the calculation of the one-loop effective energy required to analyze its stability becomes intractable.

We have argued that it is likely that classically unstable electroweak strings may be stabilized when quantum effects of quarks are considered, especially when several quarks are trapped along the string.  Such stable multi-quark objects would have rich phenomenology associated with them, especially with regard to electroweak baryogenesis and dark energy.  Such backgrounds generate long-range potentials for fermions, thereby presenting difficulties with our method to compute one-loop energies.  So we perform the calculation for a similar, but simpler, problem of magnetic flux tubes in QED, as a stepping stone to the electroweak strings calculation.  We find that when magnetic flux tubes are embedded in a no-net-flux configuration (by including a spread-out region of return flux), the potential becomes short-range, and all subtleties and puzzles disappear.  Moreover, as the return flux is made more diffuse, the energy approaches a well-defined limit which corresponds to the energy of an isolated flux tube.  This suggests that a similar embedding may be useful in the electroweak strings calculation.  We also find that the quantum corrections to the energy in two and three spatial dimensions are very similar when we impose the same renormalization conditions in the two cases (which means a finite renormalization in two spatial dimensions).  This has lead us to currently investigate whether electroweak strings in two spatial dimensions can be stabilized by fermion quantum effects.  This should be relevant for the stability in the physical case of three spatial dimensions.

Undoubtedly, exploring physics beyond the Standard Model is intriguing and critical.  But, as argued in this work, there are several unexplored objects, with wide-ranging significance, within the Standard Model itself.  They address notions of naturalness and decoupling of fermions in chiral gauge theories.  On a phenomenological front, they provide viable scenarios for electroweak baryogenesis and dark energy.  We have explored some of these possibilities and much still remains to be discovered.

\appendix
\chapter{Feynman Diagrams and Phaseshifts in the Spherical Ansatz}
\label{app:Spherical}
\section{Results from Feynman Diagrams}

Here we list the results from the Feynman diagram
calculations mentioned in Sec.~\ref{sec:EffectiveEnergy}.

In dimensional regularization ($d=4-\epsilon$)
the counterterm coefficients, defined in eq.~(\ref{Lct}), read
\begin{eqnarray}
c_1 & = & \frac{1}{6}\frac{g^2}{(4\pi)^2} \Biggl[ {\mathcal D} - \frac{1}{2} 
- 3\int_0^1dx x(1-x) \Biggl( 2\ln\frac{\Delta(x,m_w^2)}{m_f^2} 
- x(1-x)\frac{m_w^2}{\Delta(x,m_w^2)}  
\Biggr)  \Biggr] \, , \nonumber \\
c_2 & = & -\frac{f^2}{(4\pi)^2}\left[ {\mathcal D} - \frac{2}{3} - 
6\int_0^1 dx x(1-x)\ln \frac{\Delta(x,m_h^2)}{m_f^2} \right] \, , 
\nonumber \\
c_3 & = & 2m_f^2\frac{f^2}{(4\pi)^2}\left({\mathcal D}+1\right) \, , 
\nonumber \\
c_4 & = & \frac{f^4}{4(4\pi)^2}\left[ 4{\mathcal D} - \frac{m_h^2}{m_f^2} 
- 6\int_0^1dx\ln\frac{\Delta(x,m_h^2)}{m_f^2} \right] \, .
\end{eqnarray}
We have introduced the abbreviations
\begin{equation}
\Delta(x,q^2) \equiv m_f^2 - x(1-x)q^2 \quad {\rm and}\quad
{\mathcal D} \equiv \frac{2}{\epsilon} - \gamma + \ln 
\frac{4\pi\mu^2}{m_f^2} \, ,
\end{equation}
where $\mu$ is the momentum scale introduced to
maintain the canonical dimensions of the parameters 
when regularizing in fractional dimensions. 

In eq.~(\ref{Evac}) $E^{(1,2)}$ denotes the contribution
to the vacuum polarization energy from first and second order 
renormalized Feynman diagrams. Its explicit expression reads
\begin{eqnarray}
E^{(1,2)} & = & 
\frac{-2}{(4\pi)^2}\int\frac{d^3q}{(2\pi)^3} 
\Biggl\{f^2\fourier{h}(\vec{q})\fourier{h}(-\vec{q}) 
\Biggl[ -(q^2+m_h^2) + 6\int_0^1 dx \Delta(x,-q^2)
\ln\frac{\Delta(x,-q^2)}{\Delta(x,m_h^2)} \Biggr]
\nonumber \\ && \hspace{2cm}
- m_f^2 \fourier{p}^a(\vec{q})\fourier{p}^a(-\vec{q}) 
\Biggl[q^2-6q^2\int_0^1 dx x(1-x) 
\ln\frac{\Delta(x,-q^2)}{\Delta(x,m_h^2)} 
\nonumber \\ && \hspace{7cm}
-2m_f^2\int_0^1 dx \ln\frac{\Delta(x,-q^2)}{m_f^2} \Biggr]  
\nonumber \\ & & \hspace{1cm}
+\frac{g^2}{2}\tr\left(\vec{q}\cdot\vec{\fourier{W}}(\vec{q})\vec{q}
\cdot\vec{\fourier{W}}(-\vec{q})  \right) 
\Biggl[\frac{1}{6} - 
2\int_0^1 dx x(1-x)\ln\frac{\Delta(x,-q^2)}{\Delta(x,m_w^2)}  
\nonumber \\ &&\hspace{7cm}
-\int_0^1 dx x^2(1-x)^2\frac{m_w^2}{\Delta(x,m_w^2)} \Biggr] 
\nonumber \\ & & \hspace{1cm}
+\frac{g^2}{2}\tr\left(\vec{\fourier{W}}(\vec{q})
\cdot\vec{\fourier{W}}(-\vec{q})\right)
\Biggl[-\frac{q^2}{6} - \frac{2}{3}m_f^2 
+ 2q^2\int_0^1 dx x(1-x)\ln\frac{\Delta(x,-q^2)}{\Delta(x,m_w^2)}  
\nonumber \\ & & \hspace{4cm}
+q^2\int_0^1 dx x^2(1-x)^2\frac{m_w^2}{\Delta(x,m_w^2)} 
+ m_f^2\int_0^1 dx \ln\frac{\Delta(x,-q^2)}{\Delta(x,m_h^2)}  
\nonumber \\ & & \hspace{4cm}
-5m_f^2\int_0^1 dx x(1-x)\ln\frac{\Delta(x,m_h^2)}{m_f^2} \Biggr] 
\nonumber \\ & & \hspace{1cm}
-igm_f^2\vec{q}\cdot\vec{\fourier{W}}^a(\vec{q})\fourier{p}^a(-\vec{q}) 
\Biggl[-\frac{2}{3}
+\int_0^1 dx \ln\frac{\Delta(x,-q^2)}{m_f^2}  
\nonumber \\ & & \hspace{5cm}
-6\int_0^1 dx x(1-x)\ln\frac{\Delta(x,m_h^2)}{m_f^2} \Biggr] \Biggr\} \, ,
\end{eqnarray}
with the Fourier transform of a field $\varphi(\vec{x})$ defined in the
usual way as $\fourier{\varphi}(\vec{q}) = \int d^3x
\varphi(\vec{x})e^{i\vec{q}\cdot\vec{x}}$.  The third and fourth order
counterterm contribution combined with the divergences in the third
and fourth order Feynman diagrams is
\begin{eqnarray}
E^{(3,4)} & = & \int \frac{d^3 x}{(4\pi)^2}\, \tr\, \Biggl\{ 
\frac{g^3}{6}\left(4i\partial_iW_j + g[W_i,W_j]\right)[W_j,W_i] 
\Biggl[\frac{1}{2} + 6\int_0^1 dx x(1-x)\ln\frac{\Delta(x,m_w^2)}{m_f^2}  
\nonumber \\ &&\hspace{7cm}
-3\int_0^1 dx x^2(1-x)^2\frac{m_w^2}{\Delta(x,m_w^2)}\Biggr] 
\nonumber\\ & & \hspace{1cm}
-g\vec{W}\cdot\left[g\vec{W}\left(\phi\phi^\dag+2vh\right)  
+ 2i\left(\vec{\partial}\phi\right)\phi^\dag \right]
\Biggl[-\frac{2}{3} 
-6\int_0^1 dx x(1-x)\ln\frac{\Delta(x,m_h^2)}{m_f^2}\Biggr]
\nonumber\\ &&\hspace{3cm}
+ f^4 \left( \phi\phi^\dag+ 4vh \right)
\phi\phi^\dag \Biggl[\frac{m_h^2}{4m_f^2}           
+ \frac{3}{2}\int_0^1 dx \ln\frac{\Delta(x,m_h^2)}{m_f^2}\Biggr] 
\Biggr\} \, ,
\end{eqnarray}
where $\phi=\Phi-v$ parameterizes the deviation of the
Higgs field from its vev.

\section{The Dirac Equation}

In this section, we describe how we obtain the bound state energies of
the Dirac equation and the scattering phase shifts (and their Born
series) in the presence of a background potential.  These quantities
are required to compute the vacuum polarization energy in eq.~(\ref{Evac}).

The fermion field obeys the time-independent Dirac equation
\begin{equation}
H_D \Psi = \omega \Psi \, ,
\end{equation}
where 
\begin{equation}
H_D = -i\gamma^0\gamma^i\partial_i + \gamma^0 
\left[m_f + V(\Phi, W_i)\right] \, ,
\label{DiracEqn}
\end{equation}
and $V$ is given in eq.~(\ref{potential}).  It is most convenient
to use the chiral representation of the Dirac matrices,
\begin{equation}
H_D \equiv \left( \begin{array}{cc} h_{11} & h_{12} \\ h_{21} & h_{22} 
\end{array} \right)  = \left( \begin{array}{cc} i\sigma_j\partial_j + 
g\sigma_jW_j & m_f(s+ip\tau_j\hat{x}_j ) \\ m_f(s-ip\tau_j\hat{x}_j ) & 
-i\sigma_j\partial_j \end{array} \right) \, .
\end{equation}    
The grand spin $\vec{G}$ is defined as the vector sum of isospin, spin and 
orbital angular momentum.  It commutes with $H_D$ as long as the fermion 
doublet is degenerate in mass and the background fields are in the spherical
ansatz.  We satisfy both conditions.  For a given grand spin quantum
number $G$ (we suppress the grand spin projection label $M$ throughout),
the Dirac spinor $\Psi_G$ has eight components and may be 
written in terms of generalized
spherical harmonic functions ${\mathcal Y}_{j,l} ({\hat{x}})$
with $j=G \pm \frac{1}{2}$ and $l=j \pm \frac{1}{2}$ as
\begin{equation}
\Psi_G (\vec{x}) = \left( \begin{array}{c} ig_1{\mathcal
Y}_{G+\frac{1}{2},G+1} + g_2{\mathcal Y}_{G+\frac{1}{2},G} +
g_3{\mathcal Y}_{G-\frac{1}{2},G} + ig_4{\mathcal
Y}_{G-\frac{1}{2},G-1} \\ i f_1{\mathcal Y}_{G+\frac{1}{2},G+1} +
f_2{\mathcal Y}_{G+\frac{1}{2},G} + f_3{\mathcal Y}_{G-\frac{1}{2},G}
+  if_4{\mathcal Y}_{G-\frac{1}{2},G-1} \end{array} \right) \, ,
\end{equation}
where $g_i$ and $f_i$ are radial functions and we have suppressed the
grand spin labels on them.  Note that in this chiral 
theory modes of different parity, {\it e.g.} $g_1$ and $g_2$
mix. The spherical harmonics are two-component spinors in both 
spin and isospin space.  The special case $\Psi_0$ is defined
only in terms of ${\mathcal Y}_{\frac{1}{2},1}$ and ${\mathcal
Y}_{\frac{1}{2},0}$ and does not contain $g_3, g_4, f_3, f_4$.

The matrix elements of operators like $\tau_j \hat{x}_j$ between the
spherical harmonics may be found in the literature \cite{Herbert}. 
We use them to write the Dirac equation (\ref{DiracEqn})
as a a set of eight coupled first-order linear differential equations 
in the radial functions, for fixed $G$. From these equations we
obtain the bound state
solutions ($| \omega |<m_f$) in  each grand spin channel using
shooting algorithms.  From  Levinson's theorem we determine the
number, $N_G^{\rm bound}$, of bound states to 
shoot for, using phase shifts,  $\delta_G (\omega)$, of the
scattering state solutions of the Dirac equation:
\begin{equation}
N^{\rm bound}_G = \frac{1}{\pi} \left( \delta_G(m_f) - 
\delta_G(\infty) + \delta_G(-m_f) - \delta_G(-\infty) \right) \, . 
\end{equation}
To construct these scattering state 
solutions we re-write the Dirac equation as a 
set of second-order differential equations in the radial functions.
Formally they read,
\begin{equation}
\left[ h_{12}h_{21} - h_{12}(h_{22}-\omega)h_{12}^{-1}(h_{11}-\omega) 
\right]\Psi_G^U = 0 \, ,
\label{upperdirac}
\end{equation}
where 
$$
\Psi_G^U=ig_1{\mathcal
Y}_{G+\frac{1}{2},G+1} + g_2{\mathcal Y}_{G+\frac{1}{2},G} +
g_3{\mathcal Y}_{G-\frac{1}{2},G} + ig_4{\mathcal
Y}_{G-\frac{1}{2},G-1}
$$
denotes the upper two-component spinor in $\Psi_G$.  
In the chiral representation of the Dirac matrices, we require 
$s^2(r)+p^2(r)>0$ so that $h_{12}$ is invertible. As 
mentioned in Sec.~\ref{Restrictions} this is a restriction on our 
variational ansatze. Using the known matrix elements 
for the spin-isospin operators like $\tau_i\hat{x}_i$, we 
then project eq.~(\ref{upperdirac}) onto grand spin
channels and obtain the desired second order differential
equations.  These are rather lengthy and are listed a the end of this Appendix.  
They may be written in the form,
\begin{equation}
\sum_{j=1}^{4} \left\{ D_G(r) + N_G(r)\frac{\partial}{\partial r} +
M_G(r) \right\}_{ij}g_j(r) = 0 \, ,
\label{eq:2ndOrderMatrixEqn}
\end{equation}
with
\begin{eqnarray}
D_G(r) & = & \ID \left( \frac{\partial^2}{\partial r^2} +
\frac{2}{r}\frac{\partial}{\partial r} + k^2 \right) - \frac{1}{r^2}O_G \,
, \cr\cr
O_G & = & {\rm diag}\left((G+1)(G+2), G(G+1), G(G+1), (G-1)G\right)\, ,
\end{eqnarray}
and $k^2=\omega^2-m_f^2$. The matrices $N_G(r)$ and
$M_G(r)$ are given in terms of the functions 
$s(r),\ldots,\gamma(r)$ that specify the static background fields in the spherical ansatz, as in eq.~(\ref{SphericalAnsatz}). 
Their elements can be read off from the equations displayed at the end of this Appendix.
As $r\rightarrow\infty$, $N_G(r)\to0$ and $M_G(r)\to0$ and  
the differential equations decouple, as long as the potential goes 
to zero  sufficiently fast. 

We have a four-channel scattering problem.  We express the four wavefunctions 
and four boundary conditions in matrix form, ${\mathcal 
G}_{ij}(r)=g_i^{(j)}(r)$, where the linearly independent boundary 
conditions are labeled by $j=1,2,3,4$.  
We then write ${\mathcal G}(r)$ as a multiplicative modification of 
the matrix solution to the 
free differential equations, ${\mathcal G}(r)\equiv F(r)\cdot H(kr)$, 
where $H(x) = \mbox{diag}(h_{G+1}^{(1)}(x), h_{G}^{(1)}(x), 
h_{G}^{(1)}(x), h_{G-1}^{(1)}(x) )$ with
$h_{\ell}^{(1)}(x)$ denoting spherical 
Hankel functions of the first kind such that $D_G(r)\cdot H(kr)=0$. 
(The $4\times4$ matrices $F$ and $H$ depend on the 
grand spin quantum number $G$. For convenience we omit that
label from now on.)
Imposing the boundary conditions 
$F(r\rightarrow\infty)=0$ and 
$F'(r\rightarrow\infty)=0$, it is clear that the $i^{th}$ row of 
${\mathcal G}$ describes an outgoing spherical wave in the $i^{th}$ 
channel.  Similarly, ${\mathcal G}^{*}$ describes incoming spherical 
waves.  The scattering wavefunction can be written as 
\begin{equation}
{\mathcal G}_{\rm sc}(r) = -{\mathcal G}^{*}(r) + {\mathcal G}(r)S(k) 
\, ,
\end{equation}
and requiring this to be regular at the origin gives the scattering matrix
\begin{equation}
S(k) = \lim_{r\rightarrow 0}H^{-1}(kr)F^{-1}(r)F^{*}(r)H^{*}(kr) 
\, .
\end{equation}
We are interested in the sum of the eigenphase shifts in a given grand spin channel,
\begin{equation}
\delta(k) = \frac{1}{2i}\Tr \ln S(k) = 
\frac{1}{2i}\lim_{r\rightarrow 0}\Tr\ln \left( F^{-1}(r)F^{*}(r) 
\right) \, .
\end{equation}
An efficient way to avoid any ambiguities in 
additive contributions of multiples of $\pi$ in $\delta(k)$ is 
to define
\begin{equation}
\delta(k,r) = \frac{1}{2i}\Tr\ln \left( F^{-1}(r)F^{*}(r) \right) \, ,
\end{equation}
with $\delta(k)=\delta(k,0)$.
We then integrate
\begin{equation}
\frac{\partial \delta (k,r)}{\partial r} = - \Im\Tr\left( F'F^{-1} 
\right) 
\end{equation}
along with $F(k,r)$ from infinity to 0 with the boundary condition 
$\lim_{r\rightarrow\infty}\delta(k,r)=0$ to obtain $\delta(k)$ as a 
smooth function of $k$.  The differential equation for the matrix $F(k,r)$,
\begin{equation}
0 = F'' + \frac{2}{r}F' + 2F'L'+\frac{1}{r^2}[F,O]+N(F'+FL')+MF \, ,
\end{equation}
is obtained from  
\begin{equation}
\left[ \left\{ D(r)+ N(r)\frac{\partial}{\partial r} + M(r) \right\} 
{\mathcal G}(r) \right] H^{-1}(kr) = 0 \, ,
\end{equation}
where $L(kr)\equiv \ln H(kr)$ and primes denote
derivatives with respect to the radial coordinate.  
The components of $L'(kr)$ can be 
expressed as simple rational functions, which avoids any numerical 
instability that would be caused by the oscillating Hankel functions.

To construct the Born series for $\delta(k)$, we introduce 
$F^{(n)}(k,r)$ where $n$ labels the order in the background fields in an expansion around the \classicalVacuum{} configuration with $f^{(0)}(r) = 0$ in eq.~(\ref{Vacua1D}).  We obtain the corresponding differential equations
\begin{eqnarray}
0 & = & {F^{(1)}}'' + \frac{2}{r}{F^{(1)}}' + 
2{F^{(1)}}'L'+\frac{1}{r^2}[F^{(1)},O]+N^{(1)}L'+M^{(1)} \, , \nonumber 
\\
0 & = & {F^{(2)}}'' + \frac{2}{r}{F^{(2)}}' +
2{F^{(2)}}'L'+\frac{1}{r^2}[F^{(2)},O]+N^{(1)}\left({F^{(1)}}' +
F^{(1)} L'\right)  
\nonumber \\ && \hspace{1cm}
+N^{(2)}L' + M^{(1)}F^{(1)} + M^{(2)} \, ,
\end{eqnarray}
where the matrices $N^{(i)}$ and $M^{(i)}$ are obtained from 
$N$ and $M$ by expanding to order $i$ in the deviation of the 
background fields from the above described \classicalVacuum{} configuration.
We integrate these differential equations with the boundary
conditions $F^{(i)}(k,\infty)=0$ and $F^{(i)\prime}(k,\infty)=0$ and
obtain
\begin{eqnarray}
\delta^{(1)}(k)&=&-\Im \tr \left(F^{(1)}(k,0)\right)\,,
\nonumber \\ 
\delta^{(2)}(k)&=&-\Im \tr \left(F^{(2)}(k,0)-\frac{1}{2}
F^{(1)}(k,0)^2\right)\,.
\end{eqnarray}
We eliminate the quadratic divergence 
from the vacuum polarization energy by subtracting these from $\delta(k)$ and adding them back in as 
renormalized first and second order Feynman diagrams.  

There still remains the logarithmic divergence whose 
elimination would require third and fourth order Born subtractions.
These become considerably more complicated, so instead we use the 
limiting function approach as described in \cite{DecouplingNoGauge}.  The 
idea is to subtract only the local contributions to the third and fourth 
Born approximants to the phase shift by identifying them with the divergent 
contributions to the third and fourth order Feynman diagrams.  
To this end we formally manipulate these divergent 
Feynman diagrams. To extract the local contributions we 
set the external momenta to zero and then integrate over
the energy and the two spatial angles of the loop momenta, 
$k^\mu$, such that
a (regularized) integral over $k=|\vec{k}|$ is left. We 
write its integrand in the form as in eq.~(\ref{Evac}),
$$
\frac{1}{2\pi}\sqrt{k^2+m_f^2}\,\frac{d\delta_{\rm lim}(k)}{dk}
$$
where
\begin{eqnarray}
\delta_{\rm lim}(k) & = & \frac{1}{8\pi}\left(\frac{k}{k^2+m_f^2} + 
\frac{1}{m_f}\arctan\frac{m_f}{k} \right)
 \nonumber \\ && \hspace{0.5cm} \times
 \int d^3x\, \tr 
\Biggl\{\frac{g^3}{6}\left(4i\partial_iW_j+g[W_i,W_j]\right)[W_i,W_j] 
-m_f^4 \left(\phi\phi^\dagger + 4vh \right)\phi\phi^\dagger
\nonumber\\ && \hspace{3cm} 
+ m_f^2 g\vec{A}\cdot
\left[g\vec{A}\left(\phi\phi^\dagger+2vh\right)  
+ 2i\vec{\partial}\phi\phi^\dag \right]\Biggr\}
\label{LimitingPhaseshift}
\end{eqnarray}
in the $W_0 = 0$ gauge and where $\phi=\Phi-v$ denotes the 
deviation of the Higgs field from its vev.  

Thus we numerically determine the bound state energies, the
phase shifts, their Born series, and the limiting function.  These are
all ingredients in the expression for the fermion vacuum polarization
energy in eq.~(\ref{Evac}).
\clearpage
\newpage
\subsection{The Second Order Equations}
Here we list the four coupled second order equations in the four radial functions specifying the left-handed spinor for a fixed grand spin G, obtained from the Dirac equation.  The matrix elements of $M_G, N_G$ in \eq{eq:2ndOrderMatrixEqn} can be read off from these, and then the phaseshifts computed, as described above.  The radial functions $\alpha, \gamma, a_1, \Sigma$ and $\eta$ specify static background \HiggsGauge fields in the spherical ansatz, as in eqs.~\ref{SphericalAnsatz}, \ref{polarfields}.  
\begin{eqnarray*}
0 & = & g_1'' + g_1'\Bigl[\frac{2}{r} -
\frac{1}{2\Sigma^2}\frac{d\Sigma^2}{dr} + \frac{1+G}{(1+2G)r}(2\sin^2
(\eta) - \gamma)\Bigr] \\
 & & + \frac{g_2'}{(1+2G)r}\Bigl[-\frac{1}{2}a_1 r - (1+G)\alpha -
r\eta' + (1+G)\sin (2\eta)\Bigr] \\
 & & + \frac{g_3'\sqrt{G(G+1)}}{(1+2G)r}\Bigl[ra_1 + \alpha + 2r\eta'
- \sin (2\eta)\Bigr] + \frac{g_4'\sqrt{G(G+1)}}{(1+2G)r}\Bigl[\gamma -
2\sin^2 (\eta) \Bigr]  \\
 & & + g_1\Bigl[-m^2 \Sigma^2 + \omega^2 +
\frac{1}{(1+2G)r}\Bigl( -
\frac{1}{r}(2+7G(1+G)+2G^3-(1+G)^2\gamma + r(1+G)\gamma') \\
& & + \frac{1}{2}\omega a_1 r - \omega \alpha (1+G) - \eta'(\omega r +
\frac{1}{2}(1+2G)a_1 r - \frac{1}{2}(1+G)\alpha ) \\
& & + \sin (2\eta)(1+G)(\omega + \frac{1}{2}a_1 - \frac{\alpha}{r}) -
\frac{\sin^2 (\eta)}{r}(-4-6G-2G^2 +2(1+G)\gamma) \\ 
& & -\frac{1}{2\Sigma^2}\frac{d\Sigma^2}{dr}((1+2G)(2+G) -
(1+G)\gamma)\Bigr)\Bigr] \\
 & & + \frac{g_2}{(1+2G)r}\Bigl[(1+G)^2\frac{\alpha}{r} + \frac{G}{2}a_1
- \frac{1}{2}a_1'r + (1+G)(\omega\gamma - \alpha') +
\eta'(G-(1+G)\gamma ) \\ 
& & + \frac{\sin (2\eta)}{r}(1+G)(\gamma - G) -
\sin^2 (\eta) (1+G) (a_1 + 2\omega + \frac{2\alpha}{r}) \\ 
& & + \frac{1}{2\Sigma^2}\frac{d\Sigma^2}{dr}((1+2G)\omega r + (1+G)\alpha +
\frac{1}{2}a_1 r) \Bigr] \\
& & + \frac{g_3 \sqrt{G(G+1)}}{(1+2G)r} \Bigl[-(1+G)\frac{\alpha}{r} -
Ga_1 + \alpha' + ra_1' - \omega \gamma + \eta'(2+2G+\gamma) \\
& & - \frac{\sin (2\eta)}{r} (1+G+\gamma ) + \sin^2 (\eta)(2\omega
+ \frac{2\alpha}{r}+a_1) -
\frac{1}{2\Sigma^2}\frac{d\Sigma^2}{dr}(a_1 r + \alpha) \Bigr] \\ 
& & + \frac{g_4 \sqrt{G(G+1)}}{(1+2G)r} \Bigl[-(1+G)\frac{\gamma}{r} -
\omega r a_1 + \omega \alpha + \gamma' + \eta'(-\alpha +
2\omega r) \\
& & + \sin (2\eta)(-\omega - \frac{1}{2}a_1 +
\frac{\alpha}{r}) + \frac{2\sin^2 (\eta)}{r}(\gamma + G - 1) - \frac{1}{2\Sigma^2}\frac{d\Sigma^2}{dr}\gamma \Bigr] \\ 
\end{eqnarray*}
\newpage
\begin{eqnarray*}
0 & = & g_2'' 
	+ \frac{g_1'}{r(1+2G)}\Biggl[- \alpha (1+G) + \half a_{1}r  +
	\eta'r + \sin (2\eta)(1+G) \Biggr] \\
 & & + \frac{g_2'}{r(1+2G)}\Biggl[2(1+2G) + \gamma (1+G) -2\sin^2
	(\eta) (1+G) - \frac{1}{\Sigma}\Sigma'r(1+2G) \Biggr] \\
 & & + \frac{g_3'\sqrt{G(G+1)}}{r(1+2G)}\Biggl[ -\gamma + 2\sin^2
	(\eta) \Biggr] + \frac{g_4'\sqrt{G(G+1)}}{r(1+2G)}\Biggl[
	\alpha - a_{1}r - 2\eta'r - \sin (2\eta) \Biggr] \\
 & & + \frac{g_1}{r(1+2G)}\Biggl[ -\frac{\alpha}{r}(1+G)^2 -
	\alpha'(1+G) + \gamma\omega (1+G) + a_{1}(1+\half G) +
	\half a_{1}'r \\
 & & + \eta'(2+G-\gamma (1+G)) + \frac{1}{\Sigma} \Sigma' (-\omega
	r(1+2G) + \alpha (1+G) - \half a_1 r ) \\
 & & + \sin^2(\eta)(1+G) (-2\omega + \frac{2\alpha}{r} - a_1) +
	\frac{\sin (2\eta)}{r}(1+G)(2 + G - \gamma) \Biggr] \\
 & & + \frac{g_2}{r(1+2G)}\Biggl[ -\Sigma^2 m^2 r(1+2G) + \omega ^2
	r(1+2G) - \frac{G(1+G)(1+2G)}{r} \\
 & & + \alpha \omega (1+G) +
	\frac{\gamma}{r}(1+G)^2 + \gamma'(1+G) + \half a_{1}\omega
	r + \frac{1}{\Sigma}\Sigma'(G(1+2G)-\gamma (1+G)) \\
 & & - \eta'(\omega r + \alpha (1+G) + \half a_{1}r(1+2G)) +
	\frac{2\sin^2(\eta)}{r}(1+G)(G - \gamma) \\
 & & - \sin (2\eta)(1+G)(\omega + \frac{\alpha}{r} + \half a_1 ) \Biggr]\\
 & & + \frac{g_3\sqrt{G(G+1)}}{r(1+2G)}\Biggl[ -\alpha\omega -\frac{\gamma}{r}(1+G)
	- \gamma' -a_{1}\omega r + \frac{\Sigma'}{\Sigma}\gamma
	+ \eta'(2\omega r + \alpha) \\
 & & + 	\frac{2\sin^2(\eta)}{r}(1+G+\gamma) + \sin (2\eta)(\omega +
	\frac{\alpha}{r} + \half a_1 ) \Biggr]\\
 & & +  \frac{g_4\sqrt{G(G+1)}}{r(1+2G)}\Biggl[\frac{\alpha}{r}(1+G) +
	\alpha' - \gamma \omega -a_1 (2+G) - a_1'r +
	\frac{\Sigma'}{\Sigma}(-\alpha + a_1 r) \\
 & & + \eta'(2G-2+\gamma) + \sin^2(\eta)(2\omega
	-\frac{2\alpha}{r}+a_1 ) + \frac{\sin (2\eta)}{r}(G-1+\gamma)
	\Biggr]\\
\end{eqnarray*}
\newpage
\begin{eqnarray*}
0 & = & g_3'' 
	+ \frac{g_1'\sqrt{G(G+1)}}{r(1+2G)}\Biggl[\alpha - a_1 r -
	2\eta'r - \sin (2\eta) \Biggr] +
	\frac{g_2'\sqrt{G(G+1)}}{r(1+2G)}\Biggl[ -\gamma
	+2\sin^2(\eta) \Biggr] \\
 & & + \frac{g_3'}{r(1+2G)}\Biggl[2(1+2G) + \gamma G -
	\frac{\Sigma'}{\Sigma}r(1+2G) - 2\sin^2(\eta)G \Biggr] \\
& & + 	\frac{g_4'}{r(1+2G)}\Biggl[ -\alpha G - \half a_1 r -
	\eta'r + \sin(2\eta)G \Biggr] \\
& & + 	\frac{g_1\sqrt{G(G+1)}}{r(1+2G)}\Biggl[ -\frac{\alpha}{r}G +
	\alpha' - \gamma\omega +a_1 (G-1) - a_1'r +
	\frac{\Sigma'}{\Sigma}(-\alpha + a_1 r) \\
& & + \eta'(-2(2+G)+\gamma) + \frac{\sin (2\eta)}{r}(-2-G+\gamma)
	+ \sin^2(\eta)(2\omega -\frac{2\alpha}{r} +a_1) \Biggr] \\
& & + \frac{g_2\sqrt{G(G+1)}}{r(1+2G)}\Biggl[ -\alpha\omega +
	\frac{\gamma}{r}G - \gamma' - a_1 \omega r +
	\frac{\Sigma'}{\Sigma}\gamma + \eta'(2\omega r +
	\alpha)   \\
& & + \sin (2\eta)(\omega + \frac{\alpha}{r} + \half a_1)
	+ 2\frac{\sin^2(\eta)}{r}(-G+\gamma) \Biggr] \\ 
& & + \frac{g_3}{r(1+2G)}\Biggl[ -\Sigma^2m^2r(1+2G) + \omega^2r(1+2G)
	-\frac{G(G+1)(1+2G)}{r} + \alpha\omega G \\
& & - \frac{\gamma}{r}G^2
	+ \gamma'G - \half a_1 \omega r -
	\frac{\Sigma'}{\Sigma}((1+G)(1+2G)+\gamma G) +
	\eta'(\omega r - \alpha G - \half a_1 r (1+2G)) \\
& & - \sin
	(2\eta)G(\omega + \frac{\alpha}{r} + \half a_1 ) -
	\frac{2\sin^2(\eta)}{r}G(\gamma + 1+G) \Biggr] \\
& & + \frac{g_4}{r(1+2G)}\Biggl[ \frac{\alpha}{r}G^2 - \alpha'G +
	\gamma\omega G + \half a_1 (G-1) - \half a_1' r +
	\frac{\Sigma'}{\Sigma}(-\omega r (1+2G) + \alpha G \\
& & + \half a_1 r ) + \eta'(G-1-\gamma G) + \frac{\sin (2\eta)}{r}G(1-G-\gamma )
	+ \sin^2(\eta)G(-2\omega + \frac{2\alpha}{r} - a_1) \Biggr] \\
\end{eqnarray*}
\newpage
\begin{eqnarray*}
0 & = & g_4'' + \frac{g_1'\sqrt{G(G+1)}}{r(1+2G)}\Biggl[\gamma -
2\sin^2(\eta) \Biggr] + \frac{g_2'\sqrt{G(G+1)}}{r(1+2G)}\Biggl[
\alpha + a_1 r + 2\eta'r - \sin (2\eta) \Biggr] \\
& & \frac{g_3'}{r(1+2G)}\Biggl[ -\alpha G + \half a_1 r + \eta'r +
\sin (2\eta)G \Biggr] + \frac{g_4'}{r(1+2G)}\Biggl[ 2(1+2G) -
\gamma G \\
& & - \frac{\Sigma'}{\Sigma}r(1+2G) + 2\sin^2(\eta)G \Biggr] +
\frac{g_1\sqrt{G(G+1)}}{r(1+2G)}\Biggl[ \alpha\omega +
\frac{\gamma}{r}G + \gamma' - a_1 \omega r -
\frac{\Sigma'}{\Sigma}\gamma \\
& & + \eta'(2\omega r - \alpha ) +
\sin (2\eta)(-\omega + \frac{\alpha}{r} + \half a_1 ) -
2\frac{\sin^2(\eta)}{r}(G+2-\gamma ) \Biggr] \\
& & + \frac{g_2 \sqrt{G(G+1)}}{r(1+2G)}\Biggl[ \frac{\alpha}{r}G +
\alpha' - \gamma\omega + a_1 (1+G) + a_1'r -
\frac{\Sigma'}{\Sigma}(\alpha + a_1 r) \\
& & + \eta'(-2G + \gamma ) + \frac{\sin (2\eta)}{r}(G-\gamma ) +
\sin^2(\eta)(2\omega + \frac{2\alpha}{r} + a_1) \Biggr] \\
& & + \frac{g_3}{r(1+2G)}\Biggl[ -\frac{\alpha}{r}G^2 - \alpha'G +
\gamma\omega G + \half a_1 (1+G) + \half a_1'r +
\frac{\Sigma'}{\Sigma}(\omega r(1+2G) + \alpha G \\
& & - \half a_1 r ) + \eta' (1+G-\gamma G) + \frac{\sin
(2\eta)}{r}G(1+G+\gamma ) - \frac{\sin^2(\eta)}{r}G(2\omega r +
2\alpha - a_1 r ) \Biggr] \\
& & + \frac{g_4}{r(1+2G)}\Biggl[ -\Sigma^2m^2r(1+2G) + \omega^2r(1+2G)
- \frac{(G-1)G(1+2G)}{r} - \alpha\omega G - \frac{\gamma}{r}G^2 \\ 
& & - \gamma'G - \half a_1 \omega r +
\frac{\Sigma'}{\Sigma}((G-1)(1+2G) + \gamma G ) + \eta'(\omega
r +\alpha G - \half a_1 r (1+2G) ) \\
& & + \sin (2\eta)G(\omega - \frac{\alpha}{r} + \half a_1 ) -
\frac{2\sin^2(\eta)}{r}G(G-1 + \gamma ) \Biggr]
\end{eqnarray*}
\chapter{The QED Flux Tube Phaseshifts}
\label{app:QED}
We explain how fermion phaseshifts and their Born series are computed in QED in the background of magnetic flux tubes.  These are ingredients in the fermion quantum corrections to the energy of flux tubes, as discussed in Sec.~\ref{sec:QED}.

\section{The Dirac Equation}
Consider QED in 2+1 dimensions with a four-component spinor.  In the radial gauge, vortex configurations of gauge fields give flux $F$ tubes:
\bea
A_0 & = & 0 \, , \nonumber \\
\vec{A} & = & \frac{\flux}{2 \pi \rho} f(\rho)\hat{\azAngle} \, . 
\eea  
For small $\rho, f(\rho) \propto \rho^2$, otherwise $B$ is singular.  For $\rho$ large compared to the width of the flux tube, $f(\rho)$ approaches 1.  In the case of no-net-flux embedding by considering a distant region of return flux,  $f(\rho)$ approaches 0 for $\rho$ beyond the region of return flux.  The energy $\omega$ eigenspinor, $\Psi$, obeys the time-independent Dirac equation
\be
H_D \Psi = \omega \Psi \, . 
\ee   
The Dirac Hamiltonian is
\be
H_D = - \gamma^0 \gamma^i \partial_i +\gamma^0 \left( m + V \right) \, , 
\ee
where the potential is
\be
V = \frac{e F}{2 \pi \rho} f(\rho) \left( - \gamma^1 \sin \azAngle + \gamma^2 \cos \azAngle \right) \, . 
\ee
Note that in the case of a flux tube with no return flux, the potential falls slowly like $1/\rho$ at large $\rho$, and conflicts with standard assumptions of scattering theory.  However, in the no-net-flux embedding, the potential is short-range.

Since the vortex configuration is cylindrically symmetric, the total angular momentum $J_z$ commutes with the above Hamiltonian.  The subscript $z$ is meaningless in 2+1 dimensions, but we include it because we are also interested in 3+1 dimensions in which case the z projection of the total angular momentum is a good quantum number.  For a given $J_z$ quantum number (say $M$, which is a half integer), the Dirac spinor has the form
\be
\Psi_M = \left( \begin{array}{c} i h_1(\rho) e^{i(M-\smallhalf)\azAngle} |+\rangle + h_2 (\rho) e^{i(M+\smallhalf)\azAngle} |-\rangle \\
i g_1(\rho) e^{i(M-\smallhalf)\azAngle} |+\rangle + g_2 (\rho) e^{i(M+\smallhalf)\azAngle} |-\rangle \end{array} \right) \, , 
\ee
where we have suppressed the $J_z$ label on the radial functions $f_i, g_i$.  The kets $|+\rangle , |-\rangle$ are spin up and spin down eigenkets of $S_z$.  In the chiral representation of the Dirac matrices (in which $\gamma_5$ is diagonal), the upper and lower components of the above spinor denote left and right handed projections respectively.

We use the first order Dirac equation to eliminate the left handed spinor and obtain second order differential equations for the radial functions of the right handed spinor:
\bea
0 & = & g_1'' + \frac{1}{\rho} g_1' + \left( k^2 - \frac{(M - \smallhalf + \res{F} f)^2}{\rho^2} \right) g_1 - \frac{\res{F} f'}{\rho} g_1 \, , \nonumber \\
0 & = & g_2'' + \frac{1}{\rho} g_2' + \left( k^2 - \frac{(M + \smallhalf + \res{F} f)^2}{\rho^2} \right) g_2 + \frac{\res{F} f'}{\rho} g_2 \, ,  
\eea
where $k^2 = \omega^2 - m^2$, the primes denote derivatives with respect to $\rho$, f is the profile function specifying the vortex configuration, and $\res{F} = \frac{e F}{2 \pi}$.

\section{Phaseshifts}
\subsection{Zero Flux Case}
First consider the no-net-flux embedding in which $f(\rho \rightarrow \infty) = 0$.  The asymptotic equations for large $\rho$ have outgoing wave solutions that are Hankel functions of the first kind.  We write the solutions in the presence of the background as a multiplicative modification of the free solutions: 
\be
g_1 (\rho) = e^{i\beta_1(\rho)}H_{|M-\smallhalf|}^{(1)} (k \rho) \, , g_2 (\rho) = e^{i\beta_2(\rho)} H_{|M+\smallhalf|}^{(1)} (k \rho) \, .
\ee  
Then, plugging in this form into the differential equations, we get the equations for $\beta_i$:
\bea
0 & = & i \beta_1'' + 2 i L_{M-\smallhalf}' \beta_1'  + \frac{i}{\rho} \beta_1' - (\beta_1')^2 - \frac{\res{F}}{\rho^2}\left[ 2 \left( M-\smallhalf \right)f + \rho f' + \res{F}f^2 \right] \, , \nonumber \\
0 & = & i \beta_2'' + 2 i \beta_2' L_{M+\smallhalf}' + \frac{i}{\rho} \beta_2' - (\beta_2')^2 - \frac{\res{F}}{\rho^2}\left[ 2 \left( M+\smallhalf \right)f - \rho f' + \res{F}f^2 \right] \, , 
\eea
where $L_\nu (k\rho) \equiv \ln H_{|\nu|} (k \rho)$.  The boundary conditions at infinity are that $\beta_i = \beta_i' = 0$.  Integrating the $\beta_i$ from $\infty$ to 0, the total phaseshift for $J_z=M$ is (summing over the two scattering channels and over the two signs of the energy)
\be
\delta_M (k) =  - 2 \mbox{Re} \left( \beta_1(0) + \beta_2(0) \right) \, .
\ee
(See \cite{Leipzig} for a derivation of the relationship between phaseshifts and $\beta$.)  

To obtain a Born series expansion of the phaseshifts, we introduce $\beta_i^{(j)}$, where $j$ labels the order in an expansion around $\vec{A}=0$.  The equations for the first two terms, $\beta_i^{(1)}, \beta_i^{(2)}$, are
\bea
0 & = & i (\beta_1^{(1)})'' + 2 i L_{M-\smallhalf}' (\beta_1^{(1)})'  + \frac{i}{\rho} (\beta_1^{(1)})' - \frac{\res{F}}{\rho^2}\left[ 2 \left( M-\smallhalf \right)f + \rho f'\right] \, , \nonumber \\
0 & = & i (\beta_2^{(1)})'' + 2 i L_{M+\smallhalf}' (\beta_2^{(1)})'  + \frac{i}{\rho} (\beta_2^{(1)})' - \frac{\res{F}}{\rho^2}\left[ 2 \left( M+\smallhalf \right)f - \rho f' \right] \, , \nonumber \\
0 & = & i (\beta_1^{(2)})'' + 2 i L_{M-\smallhalf}' (\beta_1^{(2)})'  + \frac{i}{\rho} (\beta_1^{(2)})' - \left[ (\beta_1^{(1)})' \right]^2 - \frac{\res{F}^2}{\rho^2}f^2 \, , \nonumber \\
0 & = & i (\beta_2^{(2)})'' + 2 i L_{M+\smallhalf}' (\beta_2^{(2)})'  + \frac{i}{\rho} (\beta_2^{(2)})' - \left[ (\beta_2^{(1)})' \right]^2 - \frac{\res{F}^2}{\rho^2}f^2 \, , 
\eea
with vanishing $\beta_i^{(j)}$ and $(\beta_i^{(j)})'$ as $\rho \rightarrow \infty$.  Then, the first two terms in the Born series expansion of the phaseshifts are obtained by integrating the $\beta_i^{(j)}$ to the origin:
\be
\delta_M^{(j)} = - 2 \mbox{Re} \left( \beta_1^{(j)}(0) + \beta_2^{(j)}(0) \right) \, .
\ee

\subsection{Non-zero Flux case}
In the case of non-zero flux, the profile function $f(\rho)$ goes to 1 at large $\rho$.  This gives a long-range scattering potential.  As before, expressing the radial functions $g_i (\rho)$ as multiplicative modifications of the asymptotic solutions (Hankel functions with their order shifted by the flux),
\be
g_1(\rho)  = e^{i\beta_1(\rho)}H_{|M-\smallhalf + \res{F}|}^{(1)} (k \rho) \, , g_2 (\rho) = e^{i\beta_2(\rho)} H_{|M+\smallhalf+\res{F}|}^{(1)} (k \rho) \, .
\ee  
Then, plugging in this form into the differential equations, we get the equations for $\beta_i$:
\bea
0 & = & i \beta_1'' + 2 i L_{M-\smallhalf+\res{F}}' \beta_1'  + \frac{i}{\rho} \beta_1' - (\beta_1')^2 \nonumber \\
 & & + \frac{\res{F}}{\rho^2}\left[ 2 \left( M-\smallhalf \right)(1-f) + + \res{F}(1-f^2) - f' \rho \right] \, , \nonumber \\
0 & = & i \beta_2'' + 2 i L_{M+\smallhalf+\res{F}}' \beta_2'  + \frac{i}{\rho} \beta_2' - (\beta_2')^2 \nonumber \\
 & & + \frac{\res{F}}{\rho^2}\left[ 2 \left( M+\smallhalf \right)(1-f) + + \res{F}(1-f^2) + f' \rho \right] \, , 
\eea
where, as before, $L_\nu (k\rho) \equiv \ln H_{|\nu|} (k \rho)$, and at infinity, $\beta_i = \beta_i' = 0$.  Integrating the $\beta_i$ from $\infty$ to 0, the total phaseshift for $J_z=M$ is (summing over the two scattering channels and over the two signs of the energy)
\be
\delta_{M, \rm ideal} (k) =  - 2 \mbox{Re} \left( \beta_1(0) + \beta_2(0) \right) \, .
\ee
However, as denoted by the subscript, the above phaseshift is relative to an ideal (0 width) flux tube, because the asymptotic equations do not correspond to the free equations.  So, as discussed in further detail in Sec.~\ref{sec:subtleties}, the phaseshift of the ideal vortex relative to the totally-trivial ($A_\mu=0$) configuration must be added to the above result:
\bea
\delta_M (k) & = & \delta_{M, \rm ideal} (k) + \delta_{M, \rm free} (k) \, , \nonumber \\
\delta_{M, \rm free} (k) & = & \pi \left( |M-\smallhalf|+|M+\smallhalf|-|M-\smallhalf+\res{F}|-|M+\smallhalf+\res{F}| \right) \, .
\eea

Although we could generate the Born series expansion of the phaseshift using an expansion of the $\beta_i$ as before, there is the complication that the asymptotic solutions are not free solutions because of the long-range potential.  This can be dealt with, of course.  However, we use the fact that the Born series expansion is an expansion in powers of the flux, $\res{F}$, and we compute the terms using numerical derivatives of the phaseshift with respect to the flux evaluated at $\res{F}=0$:
\bea
\delta_M^{(1)}(k) & = & \frac{d \delta_M (k)}{d \res{F}} \Biggr|_{\res{F}=0} \res{F} \, , \nonumber \\
\delta_M^{(2)}(k) & = & \half \frac{d^2 \delta_M (k)}{d \res{F}^2} \Biggr|_{\res{F}=0} \res{F}^2 \, .
\eea

Thus, we compute the phaseshifts and the first two terms in their Born series, which allows us to exactly compute the renormalized one-loop energies of magnetic flux tubes in QED in Sec.~\ref{sec:QED}.

\bibliography{main}
\bibliographystyle{unsrt}

\nocite{*}
\end{document}